\documentclass{article}
\usepackage{rotating}
\usepackage{verbatim}

\usepackage{placeins}
\usepackage{url}
\usepackage[noend]{algpseudocode}
\usepackage{algorithm}
\usepackage{color}
\usepackage[square,sort,comma]{natbib}
\usepackage{setspace}

\usepackage{amsthm,mathtools,amsfonts,amssymb}
\newtheorem{definition}{Definition}

\usepackage{bm}
\usepackage[nomarkers,notablist,nofiglist,tablesfirst,tabhead]{endfloat}

\newcommand\independent{\protect\mathpalette{\protect\independenT}{\perp}}
\def\independenT#1#2{\mathrel{\rlap{$#1#2$}\mkern2mu{#1#2}}}

\makeatletter
\newlength{\trianglerightwidth}
\algnewcommand{\LineComment}[1]{\State $\triangleright$ #1}
\makeatother

\makeatletter
\g@addto@macro{\UrlBreaks}{\UrlOrds}
\makeatother

\begin{document}
\title{Stochastic simulation framework for the Limit Order Book using liquidity motivated agents}
\date{\today}
\author{Efstathios Panayi\\ UCL, Department of Computer Science, WC1E 6BT, London, UK\\ efstathios.panayi.10@ucl.uk
        \and Gareth W. Peters \\UCL, Department of Computer Science, WC1E 6BT, London, UK\\ gareth.peters@ucl.ac.uk}
\maketitle


\begin{abstract}
In this paper we develop a new form of agent-based model for limit order books based on heterogeneous trading agents, whose motivations are liquidity driven. These agents are abstractions of real market participants, expressed in a stochastic model framework. We develop an efficient way to perform statistical calibration of the model parameters on Level 2 limit order book data from Chi-X, based on a combination of indirect inference and multi-objective optimisation. We then demonstrate how such an agent-based modelling framework can be of use in testing exchange regulations, as well as informing brokerage decisions and other trading based scenarios.
%
\end{abstract}


\section{Introduction}
In this paper, we develop a model that simulates trading activity in the Limit Order Book (LOB), the most common form of market mechanism, utilised in major stock exchanges to match the buying and selling interest in stocks \citet{jain2003institutional}. The LOB is a complicated, multivariate, event-driven stochastic process, resulting from the combination of buy and sell orders being grouped into a multi-level queueing framework, and \cite{gould2013limit} provides a characterisation of some of the main attributes. As an indication of the complexity of this process, one need only examine the attributes of the orders entering this set of LOB queues: Each order can be distinguished by order type, price, and size (in number of shares, number of contracts etc.). With regard to order type, there are limit orders, which enter at particular levels of the buy side (the bid) or the sell side (the ask) until executed or cancelled, or market orders, which are executed at the current best price. Time ordering is also important in establishing priority in a queue, and orders are typically given timestamps of millisecond or finer resolution by the trading venue. 

Our aim is to capture pervasive features of the LOB, which have been suggested to originate from the change in market structure over the last two decades. For example, the prominence of high frequency trading, which constitutes the majority of trading data today (approximately 73\% according to \cite{hendershott2011does}) has been suggested to be responsible for the a rapid decline in the number of orders that remain in the LOB until execution. Instead, orders are very frequently cancelled and resubmitted at different prices, either to gain priority, or to reduce the risk of adverse selection (getting `picked off' by a large trader). This induces dependencies between different event types (limit order submissions and cancellations, for example), which is very likely non-linear, and may be affected by prevailing market conditions.  

The dynamics that may arise from a LOB stochastic process are challenging to model, and doing so in a parsimonious manner is a particularly formidable task. \cite{rocsu2009dynamic} discusses the complexity of modelling the dynamics that emerge from the interaction of large numbers of anonymous traders, while \cite{large2007measuring} suggests that even when studying order replenishment alone, there are multiple dimensions to consider. 
Besides the trading interest itself, there are also numerous features that could also be incorporated into a model for intra-day trading on a financial exchange, which include particulars regarding the exchange mechanism that matches the trading interest in a particular asset, as well as exchange-specific rules governing the operation of the market under certain conditions. 

There have been two approaches that have prevailed in the LOB modelling literature. Firstly, agent-based frameworks, which typically involve a large number of economic agents interacting under a restricted set of agent attributes. \cite{cristelli2011critical} organises several such models according to their ability to interpret real market participant behaviours, as well as tractability, and finds that these two axes are very much at odds. As an example, the simplicity of the agent behaviours considered by \cite{maslov2000simple,farmer2005predictive}, makes their interpretation in terms of real market participant activity difficult. On the other hand, there have been efforts to introduce influences from real market behaviours, e.g. by \cite{arthur1996asset,chiarella2002simulation}, but several of these models have methodological problems related to empirical validation, discussed in \cite{windrum2007empirical}, or the calibration is not based on well understood simulation-based estimation frameworks.

The second approach to LOB modelling considers pure stochastic model frameworks, see for instance \cite{christensen2013rebuilding}. This approach abstracts away the market participant from the modelling process. Instead, a stochastic modelling approach is taken, where the complex trading dynamics are distilled into a set of statistical assumptions. These models can capture key empirical properties of the processes comprising the LOB stochastic structure \citep{cont2010stochastic,huang2012generalized}. They also give rise to LOB simulation frameworks which feature these same properties, see for instance \cite{daniels2003quantitative,christensen2013rebuilding}.

In this paper, we propose a third type of hybrid approach based on a selection of attributes from each of these methods. In particular, we develop a new form of agent-based model for limit order books based on liquidity motivated agents, in which the LOB price and volume dynamics are emergent features of the interaction between abstractions of real-world market participants. We develop two types of such agents in our framework, namely liquidity providers (market makers), and liquidity demanders, with the latter forming a stylised representation of algorithmic traders, noise traders, trend followers and other types of speculators. Their activity is expressed in a stochastic model framework, which is more detailed than typical simple agent models. This places our model part way between a traditional agent-based model and a pure limit order book stochastic model.

The model is structured to allow for efficient calibration under a rigorous statistical estimation framework. We introduce a new simulation-based estimation approach based on a combination of Indirect Inference and multi-objective optimisation. We calibrate our representative agent stochastic model to real high frequency data from Level 2 limit order book data from Chi-X. We show how such a procedure can be used to estimate the model, such that the resulting simulations approximate real data in more than one aspect, in our case the behaviour of the intra-day price and volume processes.

A practical benefit of the agent-based modelling approach we develop is that one can utilise it to estimate the effect of a regulatory intervention. In modern LOBs there have been proposals to introduce regulation in order to curb high frequency trading, in cases where it is seen to be harmful to market quality. Under the stochastic agent-based modelling framework we are able to evaluate the effect of a `quote-to-trade ratio' imposition, which has been discussed in this context. The empirical predictions of the model suggest that the imposition of such a ratio is, ceteris paribus, sufficient to limit extreme intra-day volatility in the price process.  

Our work contributes to the field of LOB modelling in a number of ways: Firstly, our model has structural components which are directly interpretable and easily understood in terms of market participants' behaviours. Compared to traditional agent-based models, which have considered the segmentation of the agent population into an element concerned with price fundamentals and another concerned with recent price fluctuations\footnote{The chartist and fundamentalist approach to agent-based modelling is covered in detail in Section \ref{subsec:abmlit}.}, our demarcation according to liquidity motivations is more reflective of current market behaviours. Secondly, it is able to capture key attributes of the observed LOB process, such as dynamics of asset price evolution, liquidity dynamics and volume process attributes. In addition, the model is able to capture the dependence in the intensity of limit order, market order and cancellation activity at different levels of the LOB, which has not been considered in previous models. Finally, as a contribution to the calibration of simulation models in general, the paper contributes a new statistical estimation framework for simulation models that is both rigorous and efficient. 

The rest of the paper is organised as follows. Section \ref{sec:lit} provides an overview of both the agent-based and stochastic LOB modelling literature, both of which this paper draws from. Section \ref{sec:model} presents a formal mathematical specification of each component of our stochastic representative agent-based model. Section \ref{sec:calib} introduces the estimation procedure employed in this paper, along with the features of the real data that we are interested in calibrating the model against. Section \ref{sec:results} presents the results from estimating various versions of the model of increasing complexity. Section \ref{sec:reg} presents a case study of the introduction of a quote-to-trade ratio in the simulated market. Section \ref{sec:conc} concludes.

\section{Related literature}
\label{sec:lit}

\subsection{Background on LOB simulation dynamics: Agent-based models }
\label{subsec:abmlit}
In the agent-based modelling literature for financial market simulations it is common practice to divide the trading population into fundamentalist and chartist traders. Early studies, such as that by \cite{taylor1992use}, undertook surveys on a number of London-based dealers to characterise the trading behaviours prevalent at the time. Such surveys refer to fundamentalist traders as deriving their views from an economic analysis of the traded asset. In the context of an agent-based model, fundamentalists traders distil their economic analysis into a single figure, the fundamental price of the asset, and trade accordingly. Being a chartist dealer, on the other hand, involved `providing forecasts or trading advice on the basis of largely visual inspection of past prices, without regard to any underlying economic or fundamental analysis'. Common chartist behaviours in an agent-based model include making decisions based on the price of the asset, compared to its moving average in a particular period, or assuming that a short move in a certain direction will continue in the near future (a momentum strategy).

The chartist \& fundamentalist literature in agent-based modelling began with the works of \cite{frankel1988chartists}, and \cite{kirman1993ants} and was then developed further by, for example, \cite{farmer2002price}, \cite{westerhoff2003nonlinearities} \cite{youssefmir1998bubbles} and \cite{vigfusson1997switching}, amongst others. As a first step in capturing heterogeneity in trading behaviours, this distinction in trading behaviours is important, and showed that there was a useful middle ground to explore between zero-intelligence-agent type approaches and perfect rationality models. However, markets have evolved, and the behaviours of participants have changed accordingly.  

A more relevant division of trading behaviours in modern markets is between the buy and sell side, with the latter providing liquidity to the former. More recent models of agent-based LOB activity have considered liquidity provision as a way to distinguish between the different types of agents (see, e.g. \cite{preis2007multi} for an example and \cite{lebaron2006agent} for a review of related work). On the one hand, we have liquidity providers, who may have quoting obligations (i.e. they are designated market makers\footnote{\url{https://www.nyse.com/market-model/dmm-case-studies}}), or not (which is usually the case with high frequency traders). On the other, we have liquidity demanders (or liquidity traders), whose need to trade is unrelated to the model, or to the price. Examples of these include fund managers in passive index funds.  

Our model also assumes the existence of these two types of agents, but it differs from most ABMs in that we do not model individual agents explicitly. We cannot claim to have precise knowledge of the strategies employed by any type of trader, and in any case, implementing even a small subset of such strategies would be a very difficult undertaking, due to the recent nature and complexity of a variety of high frequency trading firms strategies. We would expect, however, that the aggregation of the order flow from a class of agents would be more amenable to modelling. This provides the motivation for considering this agent activity in a stochastic modelling framework.

\subsection{Background on LOB simulation dynamics: Stochastic (non agent-based)}

The class of stochastic models is motivated more from a statistical perspective, where several components of market structure, as well as the details of market participant strategies, are abstracted away by a set of stylising statistical assumptions. The objective is usually to model a particular feature of the LOB process, such as the price or volume process, through a stochastic model. Particularly in response to empirical studies describing the change of market structure over time\footnote{\cite{hasbrouck2013low} and \cite{hendershott2011does} provide evidence of the increasing representation of high frequency trading firms in the market, although \citep{brogaard2014high} does not find that this increases institutional execution costs.}, such models have been used to help understand stock price dynamics at much shorter time intervals \citep{cartea2013modelling}. 

In terms of the approaches used in this context, several authors have considered the LOB as a set of queues of orders at each price, and for either side (bid and ask), and as a consequence, employ queue-type stochastic structures to perform LOB simulations. Examples of these include the queuing system proposed by \cite{cont2010stochastic} and, in a simpler specification, by \cite{cont2013price}. Under this model, the LOB is treated as a continuous time Markov chain, where all event types (limit orders at every level, cancellations, market orders) are mutually independent. They show that the assumption of a power law characterising the limit order intensity functions, as one moves from the best bid or ask, is a good match with empirical observations. They also obtain conditional probabilities of various LOB events that may be of interest in algorithmic trading. 

In an important extension of the Markov queueing system as an LOB simulator, \cite{huang2013simulating} consider the trading activity in unevenly spaced intervals in which the reference price (the mid price in this case) is constant. They also introduce some trivial dependence between activity at different levels of the LOB, in order to explain the consumption of liquidity beyond the first LOB level, when there is no resting volume at the first level. Simulations of the model using purely the event processes cannot closely reflect macro-level features of real markets, and some assumptions about the distribution of the resting volume beyond the first few levels is required for this.

The arrival process of limit orders, as well as market orders and cancellations is one of the most commonly modelled LOB aspects.
For example, in their effort to explain the concavity of the price impact function observed across stocks, \cite{smith2003statistical} considered limit order arrival rates on the bid and ask side as independent Poisson processes, and orders priced relative to the extant bid and offer prices. Their simplifications pertained to pricing on an infinite grid, constant size orders, and a constant cancellation rate. Later models, however, \citep{bowsher2007modelling,large2007measuring} observed clustering in trading activity, which is a feature of the LOB that cannot be captured by modelling order arrivals as independent Poisson processes. Instead, \cite{bowsher2007modelling} and \cite{large2007measuring} proposed the use of univariate and multi-variate Hawkes processes, to explain the clustering of trades and limit order arrivals after a trade (i.e. order replenishment), respectively. Recently \cite{huang2013simulating} also suggested a simple Markov queueing system to capture the dependence in the consumption of liquidity in the first two levels of the LOB.

Other stochastic models proposed for LOB simulations include that by \cite{rocsu2009dynamic} who introduced an LOB model which was intended to provide an alternative explanation for the submission of orders at different levels of the LOB, compared to the adverse selection risk favoured by the market microstructure theory. Instead, traders are assumed to have a higher expected utility from trading at a more favourable price, but lose utility proportionally to their waiting time when trading via limit orders. The model predicts that a competitive bid-ask spread can result from competition between liquidity providers, and that the possibility of large market orders can lead to a hump-shaped LOB. 

While the overall motivation for the agent behaviours we consider in our model comes from their liquidity impulses, the stochastic models we consider for their trading activity are related to the family of models described in this section. The assumption of independent, homogeneous Poisson processes for limit order arrivals is fairly simplistic, and we therefore incorporate dependence in the limit order arrival intensities at different levels of the LOB, as part of a flexible parametric model which we describe in the following section. 

\section{New perspective: Stochastic agent-based models for the LOB}
\label{sec:model}
In this section we present the formal mathematical specification for each component of our stochastic agent-based model. This includes the stochastic models for limit order placements and cancellations by a liquidity provider agent and the stochastic models for market order placements by liquidity demanding representative agents. The stochastic ABM framework can model the non-linear dependence in intra-day LOB activity, where the dependence is considered both between different types of events (e.g. limit and market orders), but also the same type of events, but at different levels (e.g. cancellations at level 2 and level 5 of the ask side of the LOB). We make extensive use of the flexible multivariate skew-t distribution, which is unique in enabling the modelling of heavy tails, tail dependence, skew and clustering of volatility \citep{demarta2005t,fung2010tail}.

\subsection{Limit Order Book simulation framework}
\label{sec:simframework}
We consider the intra-day LOB activity in fixed intervals of time $\ldots,[ t-1,t ),[ t,t+1 ),\ldots$. For every interval $[ t,t+1 )$, we allow the total number of levels on the bid or ask sides of the LOB to be dynamically adjusted as the simulation evolves. These LOB levels are defined with respect to two reference prices, equal to $p_{t-1}^{b,1}$ and $p_{t-1}^{a,1}$, i.e. the price of the highest bid and lowest ask price at the start of the interval. We consider these reference prices to be constant throughout the interval $[ t-1,t)$ and thus, the levels on the bid side of the book are defined at integer number of ticks away from $p_{t-1}^{a,1}$, while the levels on the ask side of the book are defined at at integer number of ticks away from $p_{t-1}^{b,1}$. 

\begin{figure}
\begin{center}
\includegraphics[width=0.59\textwidth ]{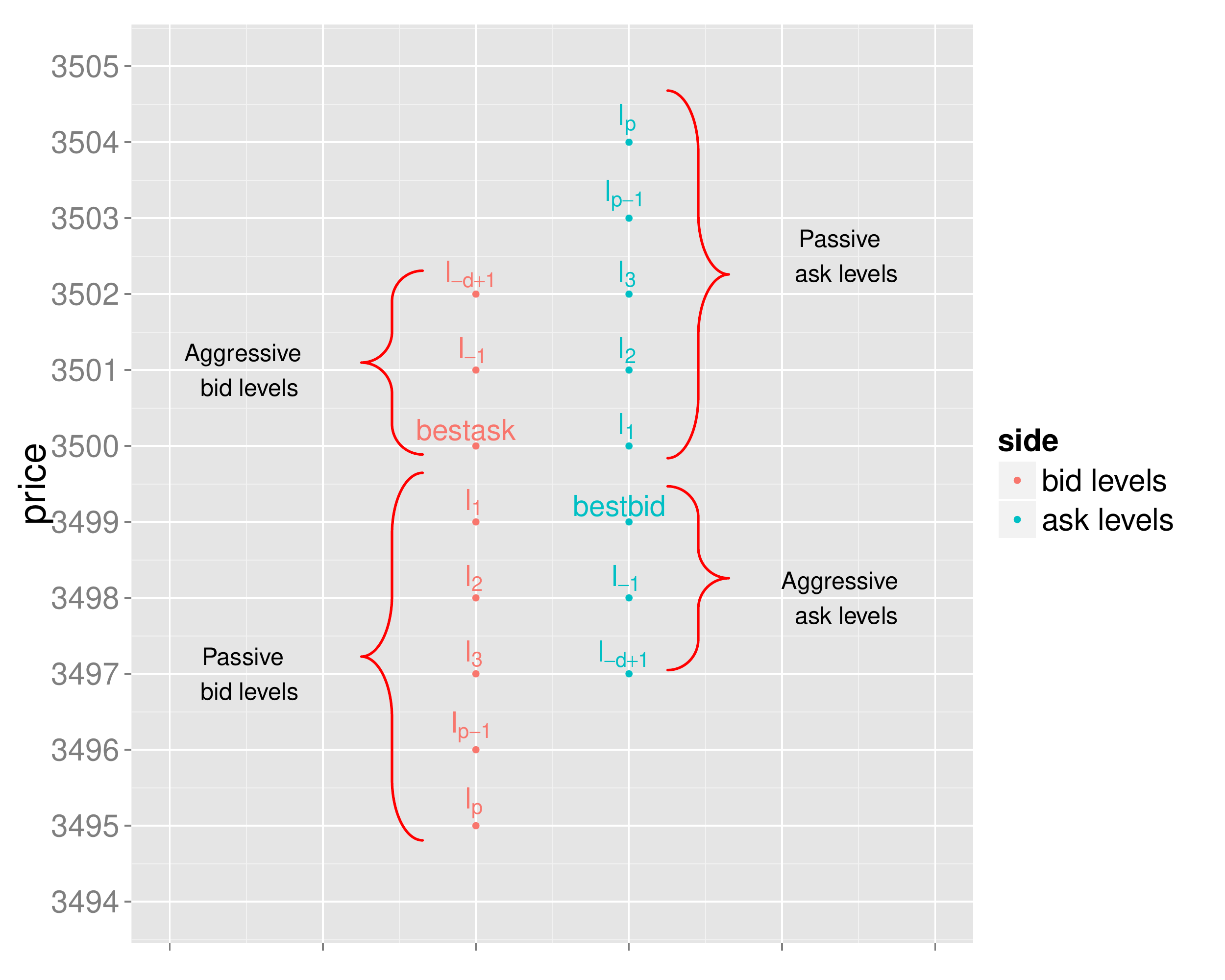}
\caption{The actively modelled levels of the LOB in the agent-based model presented in this paper. There are a total of $l_t$ levels on each side, where $l_p$ are passive levels and $l_d$ are direct, or aggressive levels (i.e. would lead to immediate execution). The levels of the ask are considered around the best bid price at the start of each interval, and likewise the levels of the bid side are considered around the best ask side at the start of each interval. In this figure, as in our model, we have $l_p=5$ and $l_d=3$.}
\label{fig:activeLOB}
\end{center}
\end{figure}

This does not mean that we expect the best bid and ask prices to remain constant, just that we model the activity (i.e. limit order arrivals, cancellations and executions) according to the distance in ticks from these reference prices during this period. We note that it is of course possible that the volume at the best bid price is consumed during the interval, and that limit orders to sell are posted at this price, which would be considered at 0 ticks away from the reference price. To allow for this possibility, we actively model the activity at $-l_d+1, \ldots,0,\ldots, l_p$ ticks away from each reference price. Here, the $p$ subscript will refer to passive orders, i.e. orders which would not lead to immediate execution, if the reference prices remained constant. $d$ refers to direct, or aggressive orders, where it is again understood that they are aggressive are with respect to the reference prices at the start of the period. Therefore, we actively model the activity at a total $l_t=l_p+l_d$ levels on the bid and ask, as indicated in Figure \ref{fig:activeLOB}.

\begin{figure}
\begin{center}
\includegraphics[width=0.49\textwidth ]{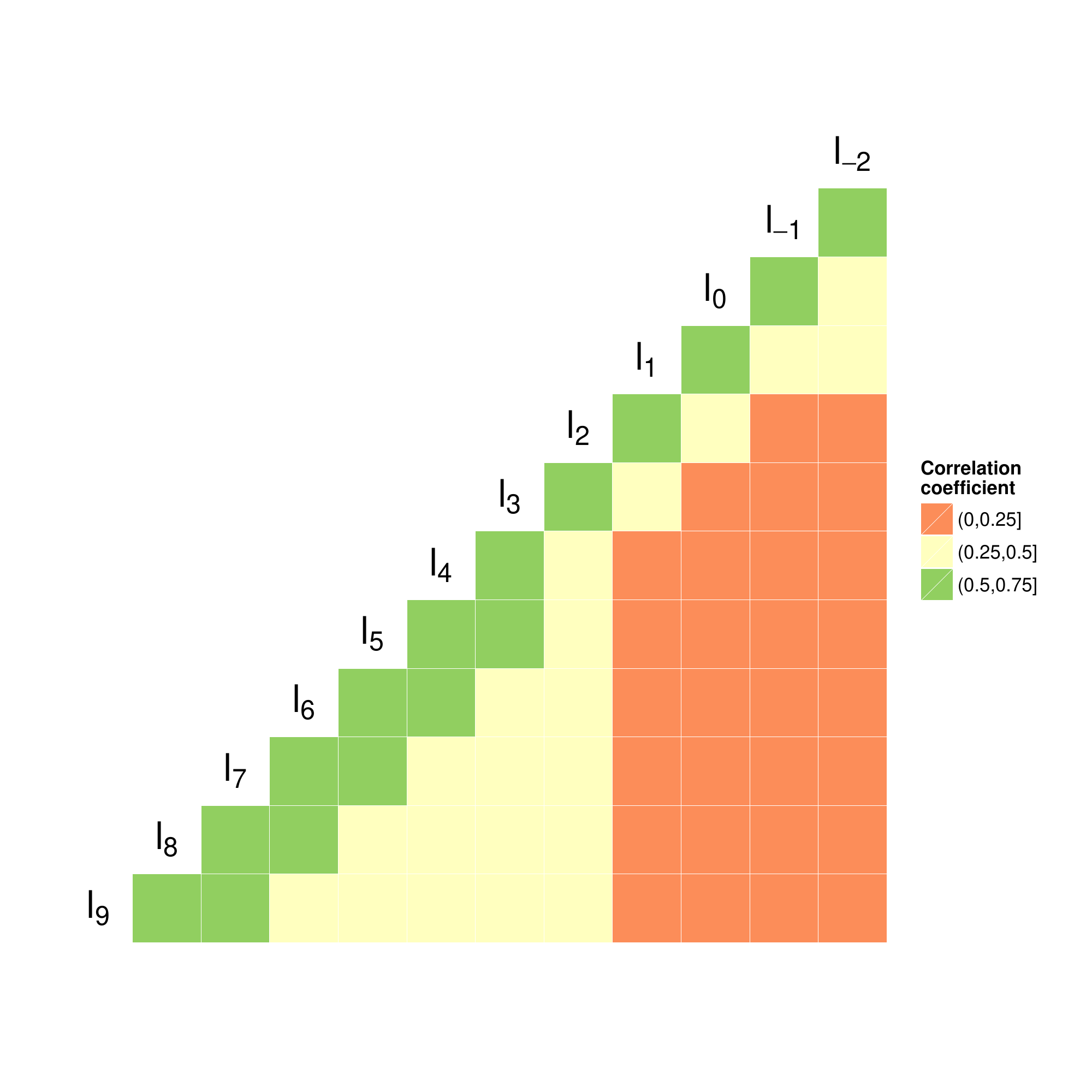}
\caption{Correlation in the LOB limit order submission intensities on the bid side of the LOB in 10 second intervals, with the levels defined with respect to the best ask price. $l_1$ to $l_9$ denote passive orders (i.e. priced above the reference price) and $l_0$ to $l_{-2}$ denote aggressive or direct orders (priced at or below the reference price, for immediate execution if the reference price had remained constant). The data set considered here is the daily LOB activity for stock BNP Paribas on 17/01/2012. }
\label{fig:lobintensitycor}
\end{center}
\end{figure}

We assume that activity that occurs further away is uncorrelated with the activity close to the top of the book (as is evident in Figure \ref{fig:lobintensitycor}), and therefore unlikely to have much of an impact on price evolution and the properties of the volume process. Therefore, the volume resting outside the actively modelled LOB levels ($-l_d+1, \ldots,0,\ldots, l_p$) on the bid and ask is assumed to remain unchanged until the agent interactions brings those levels inside the band of actively modelled levels.

To present the details of the simulation framework, including the stochastic model components for each agent, i.e. liquidity providers and liquidity demanders, we first define the following notation:
\begin{enumerate}
\item $\bm{V}_t^{a}=(V_t^{a,-l_d+1},\ldots,V_t^{a,l_p})$ - the random vector for the number of orders resting at each level on the ask side at time $t$ at the actively modelled levels of the LOB at time $t$
\item $\bm{N}_t^{LO,a}=(N_t^{LO,a,-l_d+1},\ldots,N_t^{LO,a,l_p})$ - the random vector for the number of limit orders entering the limit order book on the ask side at each level in the interval $[t-1,t)$
\item $\bm{N}_t^{C,a}=(N_t^{C,a,1},\ldots,N_t^{C,a,l_t})$ - the random vector for the number of limit orders cancelled on the ask side in the interval $[t-1,t)$
\item $N_t^{MO,a}$ - the random variable for the number of market orders submitted by liquidity demanders in the interval $[t-1,t)$
\end{enumerate}

We consider the processes for limit orders and market orders, as well as cancellations to be linked to the behaviour of real market participants in the LOB. In the following, we model the aggregation of the activity of 2 classes of liquidity motivated agents, namely liquidity providers and liquidity demanders. As we model LOB activity in discrete time intervals, we process the aggregate activity at the end of each time interval in the following order:
\begin{enumerate}
\item Limit order arrivals - passive - by the liquidity provider agent
\item Limit order arrivals - aggressive or direct - by the liquidity provider agent
\item Cancellations by the liquidity provider agent
\item Market orders by the liquidity demander agent.
\end{enumerate}

The rationale for this ordering is that the vast majority of limit order submissions and cancellations is typically accounted for by the activity of high-frequency traders, and many resting orders are cancelled before slower traders can execute against them. In addition, such an ordering allows us to condition on the state of the LOB, so that we do not have more cancellations at a particular level than the orders resting at that level. We do not see this as a limitation, as the time interval we consider can be made as small as desired for a given simulation. 

\subsection{Stochastic agent representation: liquidity providers and demanders}
We assume liquidity providers are responsible for all market-making behaviour (i.e. limit order submissions and cancellations on both the bid and ask side of the LOB). After liquidity is posted to the LOB, liquidity seeking market participants, such as mutual funds using some execution algorithm, can take advantage of the resting volume with market orders. For market makers, achieving a balance between volume executed on the bid and the ask side can be profitable; however, there is also the risk of adverse selection, i.e. trading against a trader with superior information, which may lead to losses if, e.g. a trader posts multiple market orders that consume the volume on several levels of the LOB. The risk of adverse selection as a result of asymmetric information is one of the basic tenets of market microstructure theory \citep{o1995market}. To reduce this risk, market makers cancel and resubmit orders at different prices and/or different sizes.

\begin{definition}[\textbf{Limit order submission process for the liquidity provider agent}]
Consider the limit order submission process of the liquidity provider agent to include both passive and aggressive limit orders on the bid and ask sides of the book, assumed to have the following stochastic model structure:
\begin{enumerate}
\item{Let the multivariate path-space random matrix $\bm{N}_{1:T}^{LO,k}\in \mathbb{N}_+^{l_t \times T}$ be constructed from random vectors for the numbers of limit order placements $\bm{N}_{1:T}^{LO,k} = \left(\bm{N}_1^{LO,k},\bm{N}_2^{LO,k}, \ldots,\bm{N}_T^{LO,k}\right)$. Furthermore, assume these random vectors for the number of orders at each level at time $t$ are each conditionally dependent on a latent stochastic process for the intensity at which the limit orders arrive, given by the random matrix $\bm{\Lambda}^{LO,k}_{1:T} \in \mathbb{R}_+^{l_t \times T}$ and on the path-space by $\bm{\Lambda}^{LO,k}_{1:T} = \left(\bm{\Lambda}_1^{LO,k},\bm{\Lambda}_2^{LO,k}, \ldots,\bm{\Lambda}_T^{LO,k}\right)$. In the following, $k \in \left\{a,b\right\}$ indicates the respective process on the ask and bid side.
}
\item{Assume the conditional independence property for the random vectors 
\begin{equation}
\left[\bm{N}_{s}^{LO,k}| \bm{\Lambda}^{LO,k}_{s}\right] \independent \left[\bm{N}_{t}^{LO,k}|\bm{\Lambda}^{LO,k}_{t} \right], \;\; \forall s \neq t,\;\; s,t \in \left\{1,2,\ldots,T\right\}.
\end{equation}
}
\item{For each time interval $[t-1,t)$ from the start of trading on the day, let the random vector for the number of new limit orders placed in each actively modelled level of the limit order book, i.e. the price points corresponding to ticks $(-l_d+1,\ldots,0,1,\ldots,l_p)$, as depicted in Figure \ref{fig:activeLOB}, be denoted by $\bm{N}_t^{LO,k}=(N_t^{LO,k,-l_d+1},\ldots,N_t^{LO,k,l_p})$, and assume that these random vectors satisfy the conditional independence property
\begin{equation}
\left[N_t^{LO,k,s}| \Lambda^{LO,k,s}_{t}\right] \independent \left[N_t^{LO,k,q}| \Lambda^{LO,k,q}_{t} \right], \;\; \forall s \neq q,\;\; s,q \in \left\{-l_d+1,\ldots,0,1,\ldots,l_p\right\}.
\end{equation}
}
\item{Assume the random vector $\bm{N}_t^{LO,k} \in \mathbb{N}_+^{l_t}$ is distributed according to a multivariate generalized Cox process with conditional distribution $\bm{N}_t^{LO,k} \sim \mathcal{GCP}\left(\bm{\lambda}^{LO,k}_t\right)$ given by 
\begin{equation} \label{EqnCntsDepLO}
\resizebox{.9\hsize}{!}{$
\mathbb{P}\text{r}\left(\left. N_t^{LO,k,-l_d+1} = n_1,\ldots,N_t^{LO,k,l_p} = n_{l_t}
\right|\bm{\Lambda}^{LO,k}_t = \bm{\lambda}^{LO,k}_t \right) = \prod_{s=-l_d+1}^{l_p} \frac{\left(\lambda_t^{LO,k,s}\right)^{n_s}}{n_s!}\,\exp\left[-\lambda_t^{LO,k,s}\right]
$}
\end{equation}
}
\item{Assume the independence property for random vectors of latent intensities unconditionally according to
\begin{equation}
\bm{\Lambda}_{s}^{LO,k} \independent \bm{\Lambda}_{t}^{LO,k}, \;\; \forall s \neq t,\;\; s,t \in \left\{1,2,\ldots,T\right\}.
\end{equation} 
}
\item{Assume that the intensity random vector $\bm{\Lambda}^{LO,k}_t \in \mathbb{R}_+^{l_t}$ is obtained through an element-wise transformation of the random vector $\bm{\Gamma}^{LO,k}_t \in \mathbb{R}^{l_t}$, where for each element we have the mapping 
\begin{equation} \label{EqnTransform}
\Lambda_t^{LO,k,s} = \mu_0^{LO,k,s} F\left(\Gamma_t^{LO,k,s}\right)
\end{equation}
where we have $s \in \left\{-l_d+1,\ldots,l_p\right\}$, baseline intensity parameters $\left\{\mu_0^{LO,k,s}\right\} \in \mathbb{R}_+$ and a strictly monotonic mapping $F:\mathbb{R} \mapsto [0,1]$.
}
\item{Assume the random vector $\bm{\Gamma}^{LO,k}_t \in \mathbb{R}$ is distributed according to a multivariate skew-t distribution $\bm{\Gamma}^{LO,k}_t \sim MSt(\bm{m}^k,\bm{\beta}^k,\nu^k,\Sigma^k)$ with location parameter vector $\bm{m}^k \in \mathbb{R}^{l_t}$, skewness parameter vector $\bm{\beta}^k \in \mathbb{R}^{l_t}$, degrees of freedom parameter $\nu^k \in \mathbb{N}_+$ and $l_t \times l_t$ covariance matrix $\Sigma^k$. Hence, $\bm{\Gamma}^{LO,k}_t$ has density function 

\begin{equation}
\resizebox{.9\hsize}{!}{$
f_{\bm{\Gamma}^{LO,k}_t}\left(\bm{\gamma}_t; \bm{m}^k,\bm{\beta}^k,\nu^k,\Sigma^k\right)= \frac{cK_{\frac{\nu^k+l_t}{2}}\left( \sqrt{(\nu^k+Q(\bm{\gamma}_t,\bm{m}^k))\left[\bm{\beta}^k\right]^T\left[\Sigma^k\right]^{-1}\bm{\beta}^k }\right)
\exp{(\bm{\gamma}_t-\bm{m}^k)}^T\left[\Sigma^k\right]^{-1}\bm{\beta}^k}
{\left( \sqrt{(\nu^k+Q(\bm{\gamma}_t,\bm{m}^k))\left[\bm{\beta}^k\right]^T\left[\Sigma^k\right]^{-1}\bm{\beta}^k} \right)^{-\frac{\nu^k+l_t}{2}} \left( 1+\frac{Q(\bm{\gamma}_t,\bm{m}^k)}{\nu^k} \right)^{\frac{\nu^k+l_t}{2}}} 
$}
\end{equation}

where $K_v(z)$ is a modified Bessel function of the second kind given by
\begin{equation}\label{eq:bessel}
K_{v}(z)=\frac{1}{2}\int_{0}^{\infty} y^{v-1}e^{-\frac{z}{2}(y+y^{-1})}dy
\end{equation}
and $c$ is a normalisation constant. We also define the function $Q(\cdot,\cdot)$ as follows:
\begin{align}
\label{eq:Q}
Q(\bm{\gamma}_t,\bm{m}^k)=(\bm{\gamma}_t-\bm{m}^k)^T \left[\Sigma^k\right]^{-1} (\bm{\gamma}_t-\bm{m}^k)
\end{align}
This model also admits skew-t marginals and a skew-t copula, see \cite{smith2012modelling} for details. Importantly, this stochastic model admits the following scale mixture representation,
\begin{equation} \label{EqnSimGam}
\bm{\Gamma}^{LO,k}_t \stackrel{d}{=} \bm{m}^k + \bm{\beta}^k W + \sqrt{W}\bm{Z}
\end{equation}
with Inverse-Gamma random variable $W \sim IGa\left(\frac{\nu^k}{2},\frac{\nu^k}{2}\right)$ and independent Gaussian random vector $\bm{Z} \sim N\left(\bm{0},\Sigma^k\right)$.
}
\item{Assume that for every element $N_t^{LO,k,s}$ of order counts from the random vector $\bm{N}_t^{LO,k}$, there is a corresponding random vector $\bm{O}_t^{LO,k,s} \in \mathbb{N}_+^{N_t^{LO,k,s}}$ of order sizes. We assume that the element $O_{i,t}^{LO,k,s}, i \in \left \{ 1,\ldots, N_t^{LO,k,s}\right \} $ is distributed as $O_{i,t}^{LO,k,s} \sim H(\cdot)$. Furthermore, we assume that order sizes are unconditionally independent $O_{i,t}^{LO,k,s} \independent O_{i',t}^{LO,k,s}$ for $i \neq i'$, $s \neq s'$ and $t \neq t'$.
}
\end{enumerate}
\end{definition}

We now define the second component of the liquidity provider agents, namely the cancellation process. The cancellation process has the same stochastic process model specification as the limit order submission process above, including a skew-t dependence structure between the stochastic intensities at each LOB level on the bid and ask. We therefore only specify the differences unique to the cancellation process relative to the order placement model definition in the below specification, to avoid repitition.

\begin{definition}[\textbf{Limit order cancellation process for liquidity provider agent}]
Consider the limit order cancellation process of the liquidity provider agent to have an identically specified stochastic model structure as the limit order submissions. The exception to this pertains to the assumption that the number of cancelled orders in each interval at each level is right-truncated at the total number of orders at that level.
\begin{enumerate}
\item{As for submissions, we assume for cancellations a multivariate path-space random matrix $\bm{N}_{1:T}^{C,k}\in \mathbb{N}_+^{l_t \times T}$ constructed from random vectors for the number of cancelled orders given by $\bm{N}_{1:T}^{C,k} = \left(\bm{N}_1^{C,k},\bm{N}_2^{C,k}, \ldots,\bm{N}_T^{C,k}\right)$. Furthermore, assume for these random vectors for the number of cancelled orders at each of the $l_t$ levels, the latent stochastic process for the intensity is given by the random matrix $\bm{\Lambda}^{C,k}_{1:T} \in \mathbb{R}_+^{l_t \times T}$ and given on the path-space by $\bm{\Lambda}^{C,k}_{1:T} = \left(\bm{\Lambda}_1^{C,k},\bm{\Lambda}_2^{C,k}, \ldots,\bm{\Lambda}_T^{C,k}\right)$.
}
\item{Assume that for the random vector $\tilde{\bm{V}}_t^{k}$ for the volume resting in the LOB after the placement of limit orders we have $\tilde{\bm{V}}_t^{k}=\bm{V}_{t-1}^{k}+\bm{N}_t^{LO,k}$, and that the random vector $\bm{N}_t^{C,k} \in \mathbb{N}_+^{l_t}$ is distributed according to a truncated multivariate generalized Cox process with conditional distribution $\bm{N}_t^{C,k}|\tilde{\bm{V}}_t^{k}=\underbar{v} \sim \mathcal{GCP}\left(\bm{\lambda}^{C,k}_t\right)\mathbb{I}(\bm{N}_t^{C,k}<\underbar{v})$ (with $\underbar{v}=(v_{-l_d+1},\ldots,v_{l_p})$) given by
\begin{align} \label{EqnCntsTruncCancel}
\mathbb{P}\text{r}\left(\left. N_t^{C,k,-l_d+1} = n_{-l_d+1},\ldots,N_t^{C,k,l_p} = n_{l_p}
\right|\bm{\Lambda}^{C,k}_t = \bm{\lambda}^{C,k}_t, \tilde{\bm{V}}_t^{k}=\underbar{v} \right) =\prod_{s=-l_d+1}^{l_p} \frac{\frac{(\lambda_t^{C,k,s})^{n_s}}{n_s!}}{\sum_{j=0}^{v_s} \frac{(\lambda_t^{C,k,s})^j}{j!}}.
\end{align}
}
\item{Assume that for the cancellation count $N_t^{C,k,s}$, the orders with highest priority are cancelled from level $s$ (which are also the oldest orders in their respective queue). Assume also that cancellations always remove an order in full, i.e. there are no partial cancellations. 
}
\end{enumerate}
\end{definition}

We complete the specification of the representative agents by considering the specification of the liquidity demander agent.

\begin{definition}[\textbf{Market order submission process for liquidity demander agent}]
Consider a representative agent for the liquidity providers to be composed of a \textbf{market order} component, which has the following stochastic structure:
\begin{enumerate}
\item{Assume a path-space random vector $N_{1:T}^{MO,k}\in \mathbb{N}_+^{1 \times T}$ for the number of market orders constructed from the random variables for the number of market orders in each interval $N_{1:T}^{MO,k} = \left(N_1^{MO,k},N_2^{MO,k}, \ldots,N_T^{MO,k}\right)$. Furthermore, assume that for these random variables the latent stochastic process for the intensity is given by random variable $\Lambda^{MO,k}_{1:T} \in \mathbb{R}_+^{l_t \times T}$, and given on the path-space by $\Lambda^{MO,k}_{1:T} = \left(\Lambda_1^{MO,k},\Lambda_2^{MO,k}, \ldots,\Lambda_T^{MO,k}\right)$.
}
\item{Assume the conditional independence property for the random variables 
\begin{equation}
\left[N_{s}^{MO,k}| \Lambda^{MO,k}_{s}\right] \independent \left[N_{t}^{MO,k}|\Lambda^{MO,k}_{t} \right], \;\; \forall s \neq t,\;\; s,t \in \left\{1,2,\ldots,T\right\}.
\end{equation}
}
\item{Assume that for the random variable $\tilde{R}_t^{k}$ for the volume resting on the opposite side of the LOB after the placement of limit orders and cancellations we have $\tilde{R}_t^{k}=\Sigma_{s=1}^{l_p} \left [ \tilde{V}_{t-\Delta t}^{k',s}-N_t^{C,k',s} \right ]$, where $k'=a$ if $k=b$, and vice-versa, and that the random variable $N_t^{MO,k} \in \mathbb{N}_+$ is distributed according to a truncated generalized Cox process with conditional distribution $N_t^{MO,k}|\tilde{R}_t^{k}=r \sim \mathcal{GCP}\left(\lambda^{MO,k}_t\right)\mathbb{I}(N_t^{MO,k}<r)$ given by
\begin{align} \label{EqnCntsTruncMO}
\mathbb{P}\text{r}\left(\left. N_t^{MO,k} = n \right|\Lambda^{MO,k}_t = \lambda^{MO,k}_t, \tilde{R}_t^{k}=r \right) = \frac{\frac{(\lambda_t^{MO,k})^{n}}{n!}}{\sum_{j=0}^{r} \frac{(\lambda_t^{MO,k})^j}{j!}}.
\end{align}
}
\item{Assume the independence property for random vectors of latent intensities unconditionally according to
\begin{equation}
\Lambda_{s}^{MO,k} \independent \Lambda_{t}^{MO,k}, \;\; \forall s \neq t,\;\; s,t \in \left\{1,2,\ldots,T\right\}.
\end{equation} 
}
\item{Assume that for each intensity random variable $\Lambda^{MO,k}_t \in \mathbb{R}_+$ there is a corresponding transformed intensity variable $\Gamma^{MO,k}_t \in \mathbb{R}$ and the relationship for each element is given by the mapping
\begin{equation} \label{EqnIntenMO}
\Lambda_t^{MO,k} = \mu_0^{MO,k} F\left(\Gamma_t^{MO,k}\right)
\end{equation}
for some baseline intensity parameter $\mu_0^{MO,k} \in \mathbb{R}_+$ and strictly monotonic mapping $F:\mathbb{R} \mapsto [0,1]$.
}
\item{Assume that the random variables $\Gamma^{MO,k}_t \in \mathbb{R}$, characterizing the intensity before transformation of the Generalized Cox-Process, are distributed in interval $[t-1,t)$ according to a univariate skew-t distribution $\Gamma^{MO,k}_t \sim St(m_t^{MO,k},\beta^{MO,k},\nu^{MO,k},\sigma^{MO,k})$. 
}
\item{Assume that for every element $N_t^{MO,k}$ of market order counts, there is a corresponding random vector $\bm{O}_t^{MO,k,s} \in \mathbb{N}_+^{N_t^{MO,k}}$ of order sizes. We assume that the element $O_{i,t}^{MO,k}, i \in \left \{ 1,\ldots, N_t^{MO,k}\right \} $ is distributed according to $O_{i,t}^{MO,k} \sim H(\cdot)$. Assume also that market order sizes are unconditionally independent $ O_{i,t}^{MO,k} \independent  O_{i',t}^{MO,k}$ for $i \neq i'$ or $t \neq t'$.
}
\end{enumerate}
\end{definition}

We denote the LOB state for the real dataset at time $t$ on a given day by the random vector $\bm{L}_{t}$, and this corresponds to the prices and volumes at each level of the bid and ask. Utilising the stochastic agent-based model specification described above, and given a parameter vector $\bm{\theta}$, which will generically represent all parameters of the liquidity providing and liquidity demanding agent types, one can then also generate simulations of intra-day LOB activity and arrive at the synthetic state $\bm{L}^{\ast}_{t}\left(\bm{\theta}\right)$. The state of the simulated LOB at time $t$ is obtained from the state at time $t-1$ and a set of stochastic components, denoted generically by $\bm{X}_t$, which are obtained from a single stochastic realisation of the following components of the agent-based models:

\begin{itemize}
\item{Limit order submission intensities $\bm{\Lambda}^{LO,b}_t$, $\bm{\Lambda}^{LO,a}_t$, order numbers $\bm{N}^{LO,b}_t$, $\bm{N}^{LO,a}_t$, and order sizes $\bm{O}_{i,t}^{LO,a,s},\allowbreak \bm{O}_{j,t}^{LO,b,s}$, where {$s=-l_d+1 \ldots l_p,i=1 \ldots N_t^{LO,a,s},j=1 \ldots N_t^{LO,b,s}$} }
\item{Limit order cancellation intensities $\bm{\Lambda}^{C,b}_t$, $\bm{\Lambda}^{C,a}_t$ and numbers of cancellations $\bm{N}^{C,b}_t$, $\bm{N}^{C,a}_t$ }
\item{Market order intensities $\bm{\Lambda}^{MO,b}_t$, $\bm{\Lambda}^{MO,a}_t$, numbers of market orders $\bm{N}^{MO,b}_t$, $\bm{N}^{MO,a}_t$,$\bm{V}^{MO,b}_t$,$\bm{V}^{MO,a}_t$ and market order sizes $\bm{O}_{i,t}^{MO,a},\bm{O}_{j,t}^{MO,b},i=1 \ldots N_t^{MO,a},j=1 \ldots N_t^{MO,b}$}
\end{itemize}
These stochastic features are combined with the previous state of the LOB, $\bm{L}^{\ast}_{t-1}\left(\bm{\theta}\right)$, to produce the new state $L^{\ast}_{t}\left(\bm{\theta}\right)$ for a given set of parameters $\bm{\theta}$, given by
\begin{equation}
\bm{L}^{\ast}_{t}\left(\bm{\theta}\right)=G(\bm{L}^{\ast}_{t-1}\left(\bm{\theta}\right),\bm{X}_t)
\end{equation}

$G(\cdot)$ is a transformation that maps the previous state of the LOB and the activity generated in the current step into a new step, much the same way as the matching engine updates the LOB after every event. As we model the activity in discrete intervals, however, the LOB is only updated at the end of every interval, and the incoming events (limit orders, market orders and cancellations) are processed in the order specified in Section \ref{sec:simframework}.  
Conditional then on a realization of these parameters $\boldsymbol{\theta}$, the trading activity in the LOB can be simulated according to the procedure described in Algorithm \ref{alg:sim}. 

\begin{algorithm}
\caption{Stochastic agent-based LOB simulation}
\label{alg:sim}
\begin{algorithmic}[1]
\Procedure{simulate}{$\bm{\theta},T$}
\For{$t=1\ldots T$}

\LineComment \textbf{\textsl{Simulate Liquidity Provider Limit Orders Bid/Ask.}}
	\For{$k=a,b$}
		\LineComment \textsl{Simulate dependent stochastic intensities for limit order submissions.}
        \State Sample $\bm{\Gamma}_{t}^{LO,k}=\bm{\gamma}_{t}^{LO,k} \sim MSt(\bm{m}^k,\bm{\beta}^k,\nu^k,\Sigma^k)$ via Equation \ref{EqnSimGam}.
		\State Apply transformation $\bm{\lambda}_{t}^{LO,k} = \bm{\mu}_0^k F(\bm{\gamma}_{t}^{LO,k})$ in Equation \ref{EqnTransform}.
		\LineComment \textsl{Simulate dependent limit order counts at each level bid/ask.}
        \State Sample $\bm{N}_t^{LO,k}=\bm{n}_t^{LO,k} \sim \mathcal{GCP}\left(\bm{\lambda}^{LO,k}_t\right)$ via Equation \ref{EqnCntsDepLO}.
		\LineComment \textsl{Simulate limit order sizes.}
		\For{$s=-l_d+1, \ldots l_p,i=1 \ldots N_t^{LO,k,s}$}
	        \State $\bm{O}_{i,t}^{LO,k,s} \sim H(\cdot)$
	    \EndFor
    \EndFor
\LineComment \textbf{\textsl{Simulate Liquidity Provider Cancelled Limit Orders Bid/Ask.}}
    \For{$k=a,b$}
				\LineComment \textsl{Evaluate total volumes at each level bid and ask.}
    		\State $\tilde{\bm{V}}^{LO,k}_t = \bm{V}^{LO,k}_{t-1}+\bm{N}^{LO,k}_{t}=\tilde{\bm{v}}^{LO,k}_t$
				\LineComment \textsl{Simulate dependent stochastic intensity for bid and ask cancellation counts.}
        \State Sample $\bm{\Gamma}_{t}^{C,k}=\bm{\gamma}_{t}^{C,k} \sim MSt(\bm{m}^{C,k},\bm{\beta}^{C,k},\nu^{C,k},\Sigma^{C,k})$ via Equation \ref{EqnSimGam}.
        \State Apply transformation $\bm{\lambda}_{t}^{C,k} = \bm{\mu}_0^{C,k} F(\bm{\gamma}_{t}^{C,k})$ in Equation \ref{EqnTransform}.
				\LineComment \textsl{Simulate dependent limit order cancellation counts at each level of the bid/ask.}
        \State Sample $\bm{N}_t^{C,k}=\bm{n}_t^{C,k} \sim \mathcal{GCP}\left(\bm{\lambda}^{C,k}_t\right)\mathbb{I}(\bm{N}_t^{C,k}<\tilde{\bm{v}}^{LO,k}_t)$ via Equation \ref{EqnCntsTruncCancel}.
    \EndFor
\LineComment \textbf{\textsl{Simulate Liquidity Demander Market Orders.}}
    \For{$k=a,b$}
				\LineComment \textsl{Evaluate the current resting volumes on each level of the bid/ask.}
    		\State $\tilde{R}^{LO,k}_t = \Sigma_{s=1}^{l_p} \left [ \tilde{V}^{LO,k',s}_t-N^{C,k',s}_{t} \right ]=\tilde{r}^{LO,k}_t$
				\LineComment \textsl{Simulate stochastic intensities for market order submissions.}
				\State Sample $\gamma^{MO,k} \sim St(m_t^{MO,k},\beta^{MO,k},\nu^{MO,k},\sigma^{MO,k})$ from skew-t distribution.
				\State Evaluate transformation $\lambda_{t}^{MO,k} = \mu_0^{MO,k} F(\gamma_{t}^{MO,k})$ in Equation \ref{EqnIntenMO}.
				\LineComment \textsl{Simulate market order counts.}
        \State Sample $N_{t}^{MO,k}|\tilde{r}^{LO,k}_t \sim \mathcal{GCP}\left(\bm{\lambda}^{MO,k}_t\right)\mathbb{I}(\bm{N}_t^{MO,k}<\tilde{r}^{LO,k}_t)$ via Equation \ref{EqnCntsTruncMO}.				
				\LineComment \textsl{Simulate market order sizes.}
        \For{$i=1 \ldots N_t^{MO,k}$}
	        \State $\bm{O}_{i,t}^{MO,k} \sim H(\cdot)$
	    \EndFor
    \EndFor
         \State $L_t \gets$ \Call{G}{$L_{t-1},\bm{N}_t^{LO,a},\bm{N}_t^{LO,b},\bm{N}_t^{C,a},\bm{N}_t^{C,b},N_t^{MO,a},N_t^{MO,a},\bm{O}_{t}^{LO,a},\bm{O}_{t}^{LO,b},\bm{O}_{t}^{MO,a},\bm{O}_{t}^{MO,b}$}
      \EndFor
\Return $\bm{L}=\left \{ L_1,\ldots,L_T \right \}$
\EndProcedure
\end{algorithmic}
\end{algorithm}

\section{Simulation based likelihood calibration} 
\label{sec:calib}
A common attribute of all agent-based modelling frameworks is that they are able to generate realisations of the stochastic process they represent, in our case the LOB process. That is, given a set of specifications for the parameters of the agents, the simulation of the agent model is trivial and efficient. However, it is also commonly the case that there is either no direct tractable (to evaluate pointwise) likelihood model or the likelihood model is complex and computationally costly to evaluate. In these cases, traditional parameter estimation methods based on likelihood inference are not directly applicable, when calibrating such models to observed LOB data. There are, however, a range of methods, which have yet to be utilised widely in the agent-based modelling literature, that allow one to still perform calibration of models, i.e. parameter estimation, for models specified in a simulation based format. 

The structure of our model ensures that we can capture features such as the non-linear dependencies between the activity at different LOB levels. This activity includes limit order submissions that can be passive or aggressive, cancellations and market orders, and can arise from two different classes of agents. Given this complexity, obtaining the distributional form of the likelihood will be impossible. We therefore propose estimating the model via a simulation-based method called Indirect Inference. In particular, we develop a novel extension to one of these classes of statistical simulation based likelihood inference procedures known as Indirect Inference.

\subsection{Background on Indirect Inference}
There is a substantial body of academic work related to simulation-based likelihood inference, and we focus on the subclass known as Indirect Inference, introduced by \citet{smith1990three,smith1993estimating} and \citet{gourieroux1993indirect} and covered extensively in \cite{gourieroux2010indirect,gallant1996moments} and the book length coverage in \cite{gourieroux1997simulation}. At its most fundamental level, Indirect Inference is a technique for parameter estimation in simulation based stochastic models. These are models for which one cannot evaluate the density for the data generating model, but for which one can generate data given a set of parameters. One can then compare the simulated data with the observed data, and obtain a measure of fitness for a set of parameters based on this comparison.

To achieve this via Indirect Inference, one introduces a new model, called the `auxiliary model', which is mis-specified and typically not even generative, but can generally be estimated easily via for instance maximum likelihood estimation. This auxiliary model has its own parameter vector $\bm{\beta}$, with point estimator $\widehat{\bm{\beta}}$. These parameters of the auxiliary model describe aspects of the distributions of the observations. The idea of Indirect Inference is then to simply try to match aspects of the estimated auxiliary model parameters on the observed data $\bm{y}$, given by $\widehat{\bm{\beta}}(\bm{y})$, and the estimated auxiliary model parameters on the simulated data $\bm{y}^{\ast}(\bm{\theta})$, which is obtained through simulation using parameters of the actual model $\bm{\theta}$, given by $\widehat{\bm{\beta}}(\bm{y}^{\ast}(\bm{\theta}))$. 

One sees that Indirect Inference only requires that the model one wants to estimate can be simulated, and proceeds by fitting a simpler auxiliary model to both the simulated and the real data. Estimates of the model parameters are then obtained by minimising the difference between the parameter vectors of the auxiliary model fit to the simulated data and the real data. 

When considering the choice of an auxiliary model, the simplest form one may consider involves a comparison formed between a single summary statistic calculated on the real observed data, say $\bm{y}$ and also on the simulated synthetic data $\bm{y}^{\ast}$. Alternatively, one may consider methods that consider the use of a vector of summary auxiliary parameters, such as in \citet{winker2007objective} who consider minimization of a weighted L2 quadratic error function between the real data vector of estimated moments and the synthetic simulated data equivalents. Others who have adopted such methods include \citet{mcfadden1989method} and \citet{pakes1989simulation} who each proposed a modification of the method of moments estimator, called the Method of Simulated Moments (MSM). Other, alternative simulation-based estimation techniques include the simulated maximum likelihood (SML) and the method of simulated scores (MSS). Such techniques have been used in the estimation of a number of economic models, for example dynamic stochastic general equilibrium (DSGE) models \citet{ruge2007methods} and Markov models of asset pricing \citet{duffie1993simulated}. 

In this paper, the auxiliary models we consider are based on aspects of the LOB stochastic process that we analyze. The key features we consider include the variation in the price and the volume resting in the LOB. In particular, we would like to capture the clustering of volatility in intra-day log returns and the dynamic behaviour of total volume in the first $n$ levels of the LOB. 

In detail, the sequence for obtaining the Indirect Inference estimator is as follows:
\begin{enumerate}
\item Take the observed sequence of LOB states $\bm{L}_{1:T}$ and transform them to auxiliary model data set \linebreak $\bm{y} = \mathcal{T}\left(\bm{L}_{1:T}\right)$.
\item Using observed auxiliary model data $\bm{y}$, estimate auxiliary model parameters $\widehat{\bm{\beta}}\left( \bm{y}\right)$. 
\item Initialize parameter vector of stochastic agent LOB model, in our case liquidity provider and liquidity demander agent models parameters $\bm{\theta}^{(0)}$. Then simulate a synthetic realization of the LOB model $L^{\ast}_{1:T}\left(\bm{\theta}^{(0)}\right)$ from the stochastic agent model. \label{step:start}
\item Take the synthetic sequence of LOB states $L^{\ast}_{1:T}\left(\bm{\theta}^{(0)}\right)$ and transform them to auxiliary model data set $\bm{y}^{\ast}(\bm{\theta}^{(0)}) = \mathcal{T}\left(L^{\ast}_{1:T}\left(\bm{\theta}^{(0)}\right)\right)$.
\item Using synthetic auxiliary model data $\bm{y}^{\ast}(\bm{\theta}^{(0)})$, estimate auxiliary model parameters $\widehat{\bm{\beta}}_{0}\left( \bm{y}^{\ast}(\bm{\theta}^{(0)})\right)$. 
\item Estimate Mahalanobis distance or Euclidean distances between auxiliary parameter vectors \linebreak $\mathcal{D}\left(\widehat{\bm{\beta}}\left( \bm{y}\right), \widehat{\bm{\beta}}_{0}( \bm{y}^\ast(\bm{\theta}^{(0)})) \right)$
\item Set optimal parameter vector $\widehat{\bm{\theta}}^{opt} = \bm{\theta}^{(0)}$ with distance $\mathcal{D}_{\text{min}} = \mathcal{D}\left(\widehat{\bm{\beta}}\left( \bm{y}\right),\widehat{\bm{\beta}}_{0}( \bm{y}^\ast(\bm{\theta}^{(0)})) \right)$. \label{step:end}
\item Repeat steps \ref{step:start} to \ref{step:end} with proposed parameter vector $\bm{\theta}^{(j)}$ until convergence or for $J$ total iterations, with step (vii) applied conditionally on the event \\ $\mathcal{D}_{\text{min}} > \mathcal{D}\left(\widehat{\bm{\beta}}\left( \bm{y}\right),\widehat{\bm{\beta}}_{j}\left( \bm{y}^{\ast}\left(\bm{\theta}^{(j)}\right)\right) \right)$
\end{enumerate}

Several theoretical properties are known about the estimators obtained from such a data generative procedure, see discussions in \cite{smith2008indirect} and \cite{genton2003robust}. Under certain assumptions it can be shown that the Indirect Inference procedure produces a point estimator of the model parameters which is both consistent and asymptotically Normal under fairly unrestrictive regularity conditions (\cite{gourieroux1997simulation}):
\begin{enumerate}
\item The likelihood, which we maximise, in order to estimate the auxiliary model parameters $\bm{\beta}$, tends asymptotically to a non-stochastic limit.
\item This limit is continuous in the simulation model parameters $\bm{\theta}$.
\item The so-called \textsl{binding function} linking the parameters of the auxiliary model to the parameters of the actual model we are trying to estimate is one-to-one and its derivative with respect to the auxiliary model parameters is of full column rank.
\end{enumerate}
In addition, Indirect Inference can be shown to be asymptotically efficient when the model is correctly specified for the observed data.  

\subsection{Multi-objective Indirect Inference for simulation-based model calibration}
To perform estimation of our agent stochastic model, we develop a novel extension of simulation-based estimation procedures which combines two key ideas: simulation-based likelihood inference based on Indirect Inference, and multi-objective optimisation methods, typically utilised in genetic search algorithms. We denote the resulting class of estimation methods as Multi-objective-II. The proposed Multi-objective-II estimation framework, unlike standard indirect inference, is designed to allow one to utilise multiple auxiliary models, each capturing different features of the LOB stochastic process. In this sense, this is a multi-objective extension of standard Indirect Inference procedures, which will naturally allow us to explore relevant features of the target stochastic process given by the LOB.

To proceed with the specification of the multi-objective-II estimation methodology, in addition to the LOB simulation framework described in Section \ref{sec:model}, we need to specify 
\begin{itemize}
\item The auxiliary model(s), each parameterised by a set of parameter vectors, generically denoted by $\bm{\beta}$, which are determined according to the features of the observed data stochastic process we would like to approximate with our model.
\item The objective function quantifying the difference in the auxiliary model(s) parameters fit to the real data (for which we will use the shorthand $\widehat{\bm{\beta}}$ to represent 
$\widehat{\bm{\beta}} \left(\bm{y}\right)$) and the auxiliary model(s) 
parameter fits to the synthetically generated data (where we will use the shorthand $\widehat{\bm{\beta}}^{\ast}(\bm{\theta})$ for $\widehat{\bm{\beta}}( \bm{y}^\ast(\bm{\theta})$)
\item The search method that will explore the parameter space of the stochastic agent-based model when performing simulation based optimization for stochastic agent LOB model calibration.
\end{itemize}

\subsubsection{The auxiliary models}
The auxiliary model(s), sometimes known as the estimating function(s), serve to capture aspects of the real data that we want reflected in our simulation, i.e. they do not necessarily have to correspond closely to the data generating process, but each should capture some relevant features that will inform estimation of the stochastic simulation model parameters. In standard Indirect Inference methods, there is only one auxiliary model utilised which usually comes from a relatively simple class of models, for guidelines relating to selection see \cite{heggland2004estimating}.

In our framework, for a given candidate parameter vector $\bm{\theta}$ we generate $M$ realisations of trajectories of the LOB process, i.e. $\left\{\bm{L}^{\ast,m}_t(\bm{\theta})\right\}_{t>0, m \in \left\{1,2,\ldots,M\right\}}$, from the stochastic agent-based LOB model. Then for each auxiliary model, parameterised by some vector, generically denoted by $\bm{\beta}$, we utilise the simulated data to obtain estimates of the auxiliary model parameters, for instance via a maximum likelihood framework:
\begin{equation}\label{EqnMLEAux}
\widehat{\bm{\beta}}^{\ast}\left(\bm{\theta}\right)=\arg \underset{\bm{\beta}}{\max} \sum_{m=1}^{M} \sum_{i=1}^{T} \log (f(\mathcal{T}(\bm{L}_t^m(\bm{\theta}))|\mathcal{T}(\bm{L}_{t-1}^{\ast,m}(\bm{\theta}));\bm{\beta})).
\end{equation}

In principle, one can adopt as many auxiliary models as is deemed desirable for a particular application. However, several authors have explored the effect of the number of objective functions $K$ on the estimation performance under a multi-objective optimization framework. For instance, \cite{purshouse2003evolutionary} and \cite{hughes2005evolutionary} suggest that Pareto-ranking based methods, such as the one used in this paper, scale poorly with the number of objectives. \cite{koppen2005fuzzy} explains that an increase in the number of objectives may have a detrimental effect on the optimisation because the probability of dominance in a Pareto optimality based multi-objective framework will go to zero. A second issue with having a large number of objectives is the difficulty in comparing the results qualitatively, since in a task with $K$ objectives, a set of solutions lies in a $K-1$ hyperspace. Based on this guidance, we focus on capturing two core features of LOB stochastic process, related to the evolution of the price and the properties of the volume resting near the top of the book.

\begin{figure}
\begin{center}
\includegraphics[width=0.7\textwidth]{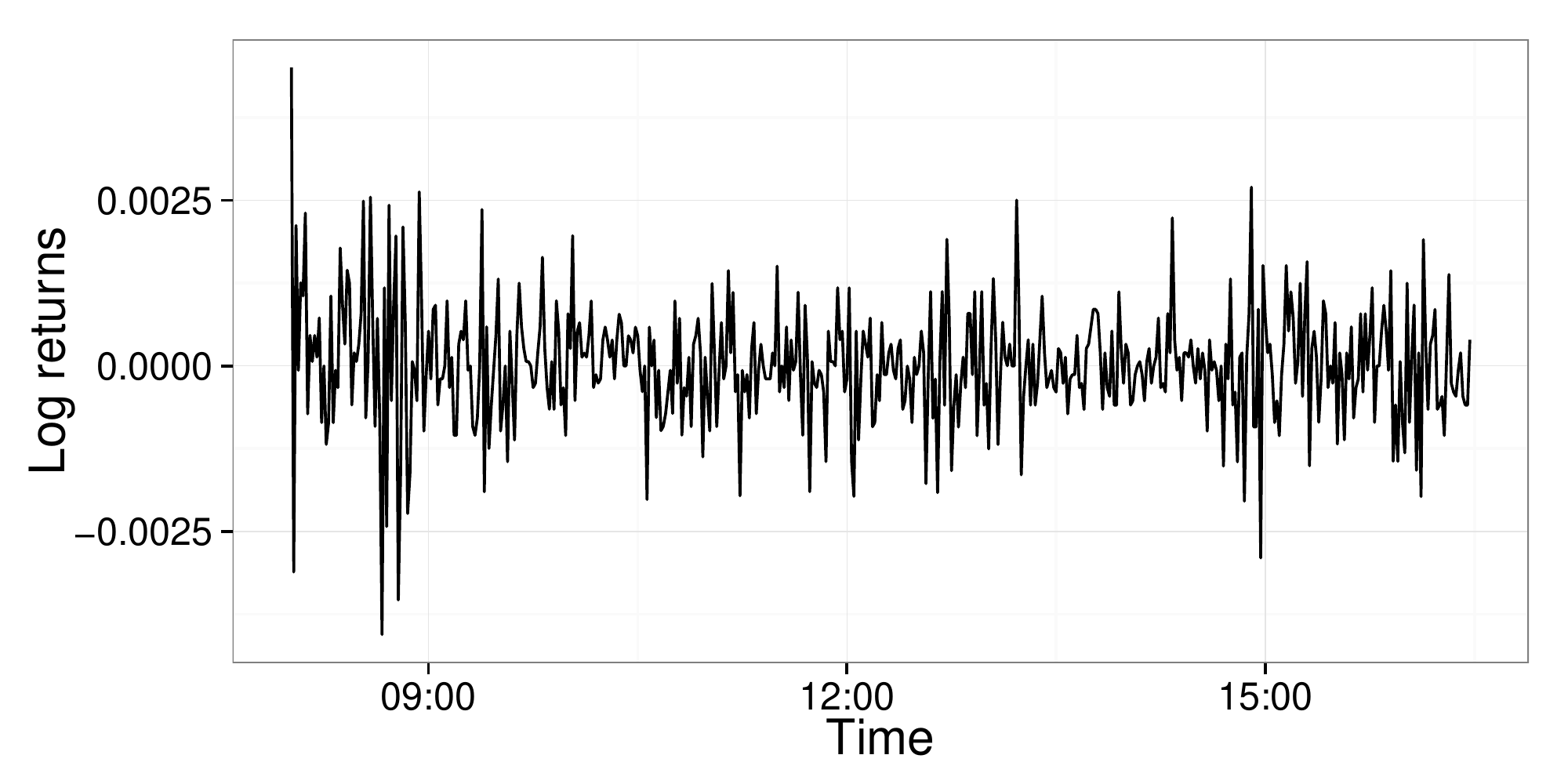}
\caption{One-minute log returns for stock BNP Paribas on a typical day.}
\label{fig:oneminutelogreturns}
\end{center}
\end{figure}

\noindent \textbf{Auxiliary Model 1 - Price features:} If we denote the mid-price as $p^{mid}_t=\frac{p^{a,1}_t+p^{b,1}_t}{2}$ then the log return is defined as 
\begin{equation*}
r_t=\ln \frac{p^{mid}_t}{p^{mid}_{t-\Delta_t}}
\end{equation*}
where $\Delta_t$ is a suitable interval, in our case 1 minute. The timeseries of log returns for a typical day for an illustrative stock GDF Suez is presented in Figure \ref{fig:oneminutelogreturns}. 

This illustrative timeseries displays typical features of mid price dynamics, such as heteroskedasticity. The presence of ARCH effects was formally confirmed by an ARCH-LM test. Hence, the volatility \linebreak 
$\sigma_t=\sqrt{Var(r_t|r_{t-1}, \ldots)}$ is not constant, and can be captured with a generalised autoregressive conditionally heteroskedastic model, or GARCH(p,q) model, where with $r_t=\sigma_t \eta_t$ and $\eta_t\sim N(0,1)$, we have for the squared volatility 
\begin{equation*}
\sigma^2_t = a_0 + a_1 r_{t-1}^2+ \ldots + a_p r_{t-p}^2 + b_1 \sigma^2_{t-1} + \ldots + b_q \sigma^2_{t-q}
\end{equation*} 
where $a_i\geq 0$, $b_j\geq 0$ for all $i \in \left\{1,\ldots,p\right\}$ and $j \in \left\{1,\ldots,q\right\}$. For simplicity of the auxiliary model we utilise a GARCH(1,1) model for this aspect of the data, parameterized by $\bm{\beta}_1=(a_0, a_1,b_1$).

\begin{figure}
\begin{center}
\includegraphics[width=0.3\textwidth]{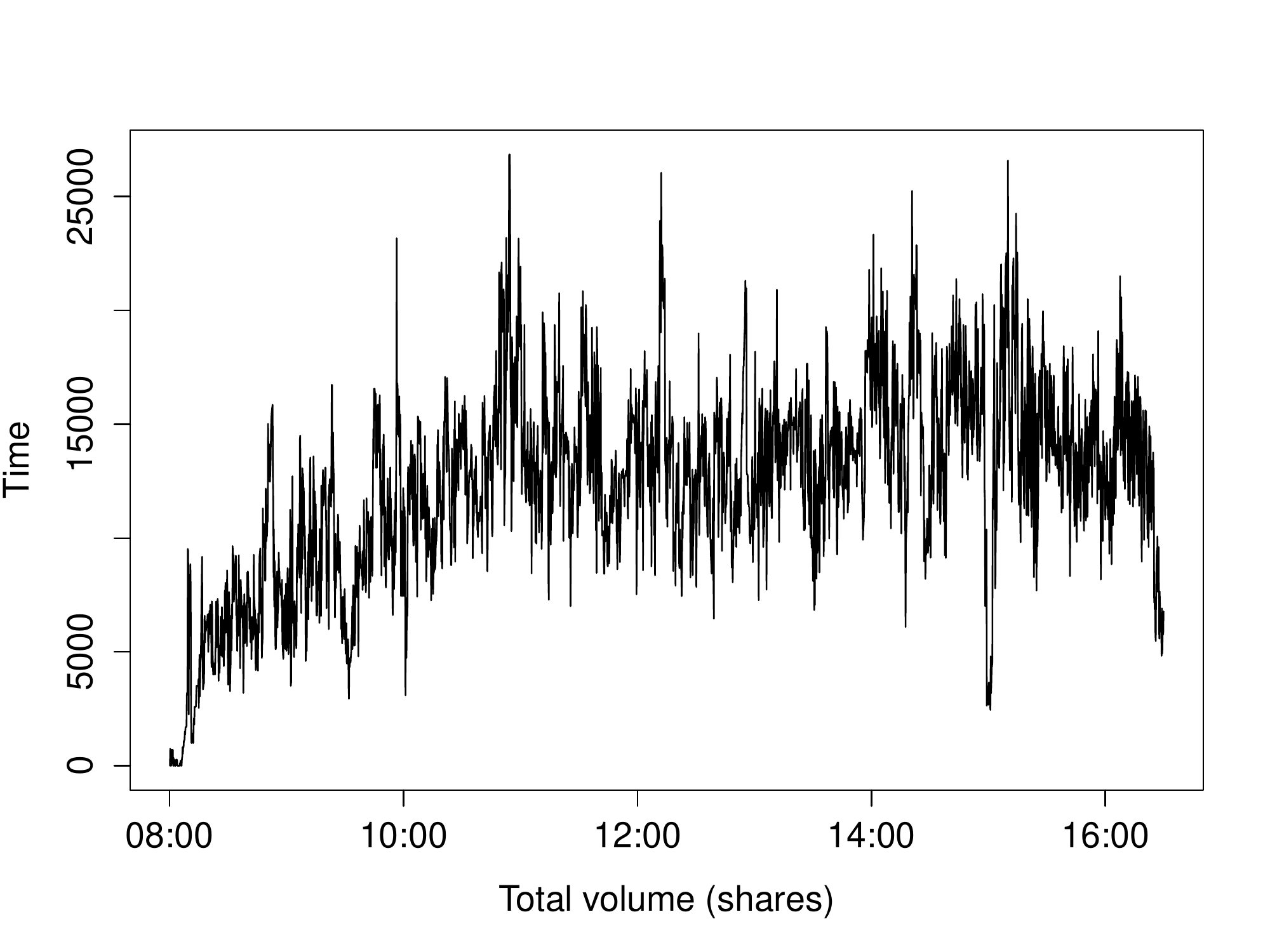}
\includegraphics[width=0.3\textwidth]{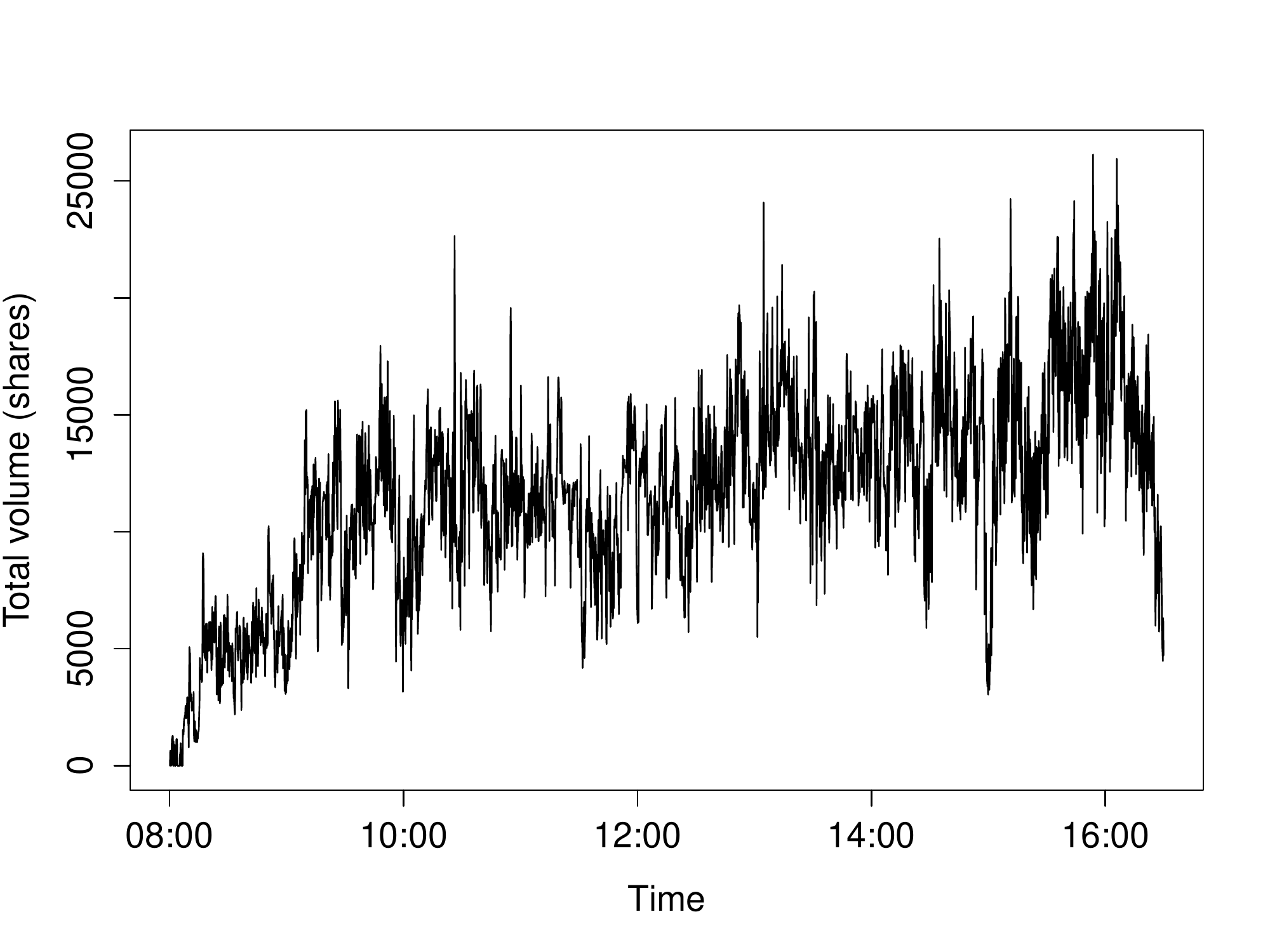}\\
\includegraphics[width=0.3\textwidth]{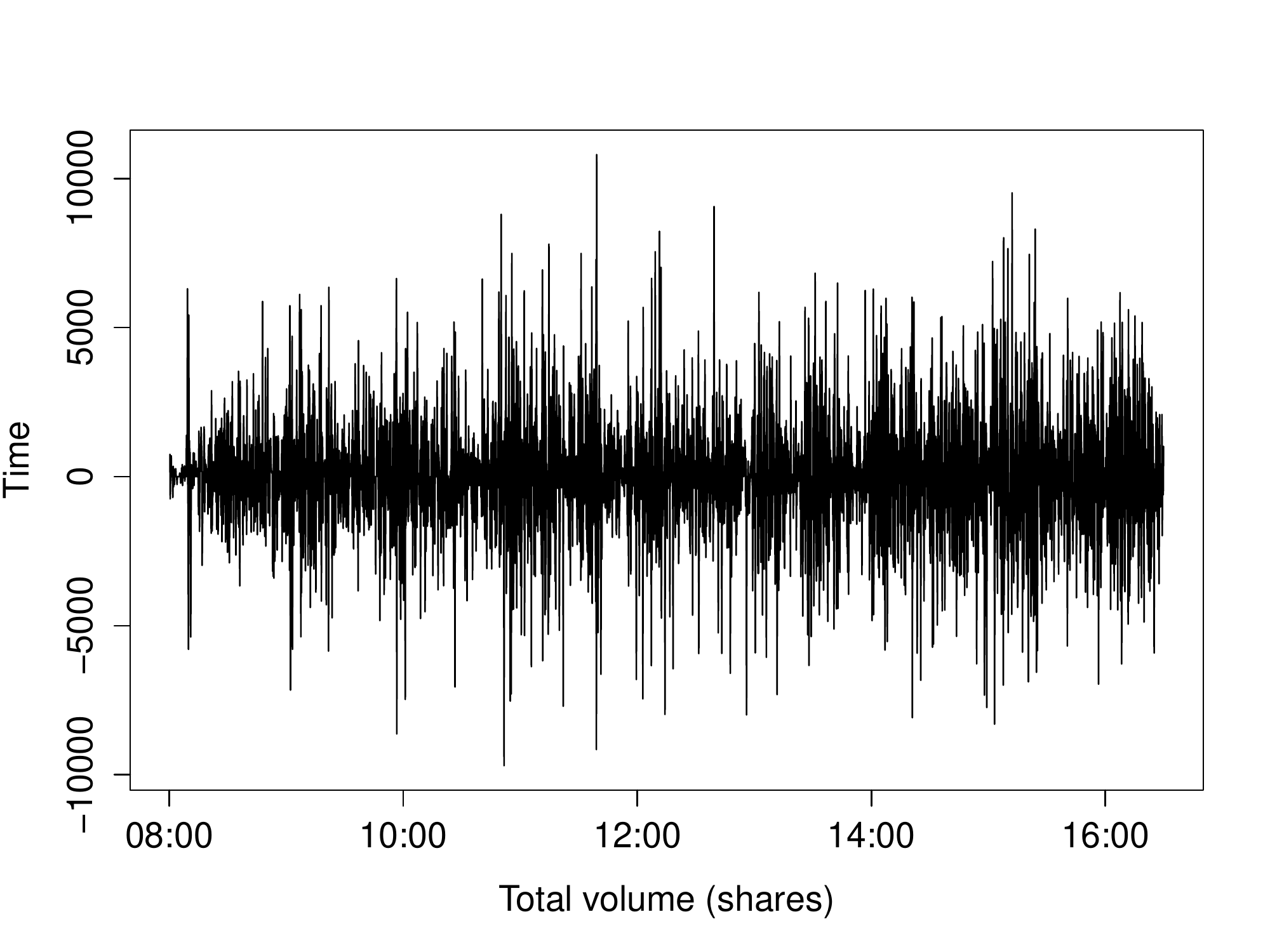}
\includegraphics[width=0.3\textwidth]{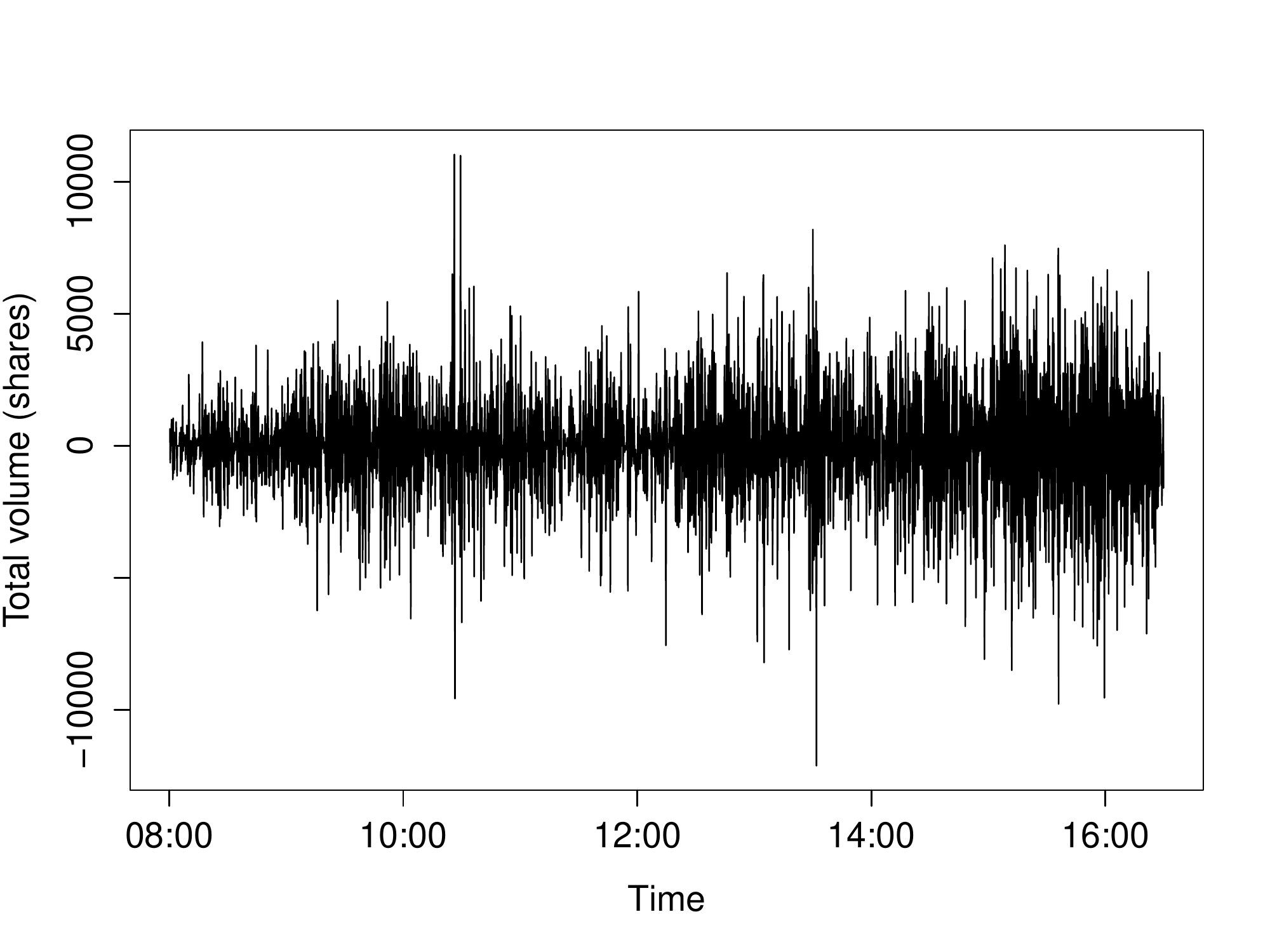}\\
\includegraphics[width=0.3\textwidth]{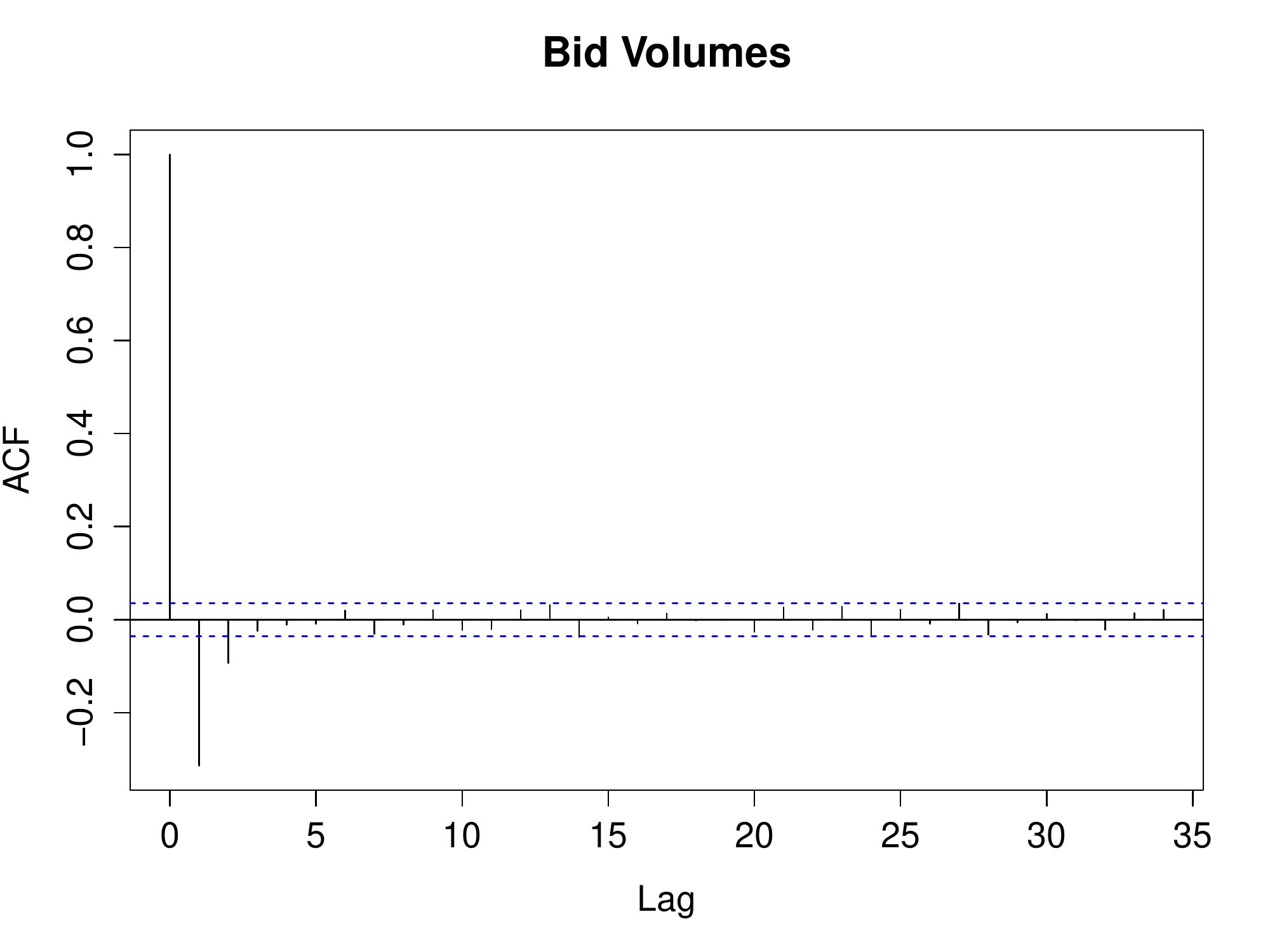}
\includegraphics[width=0.3\textwidth]{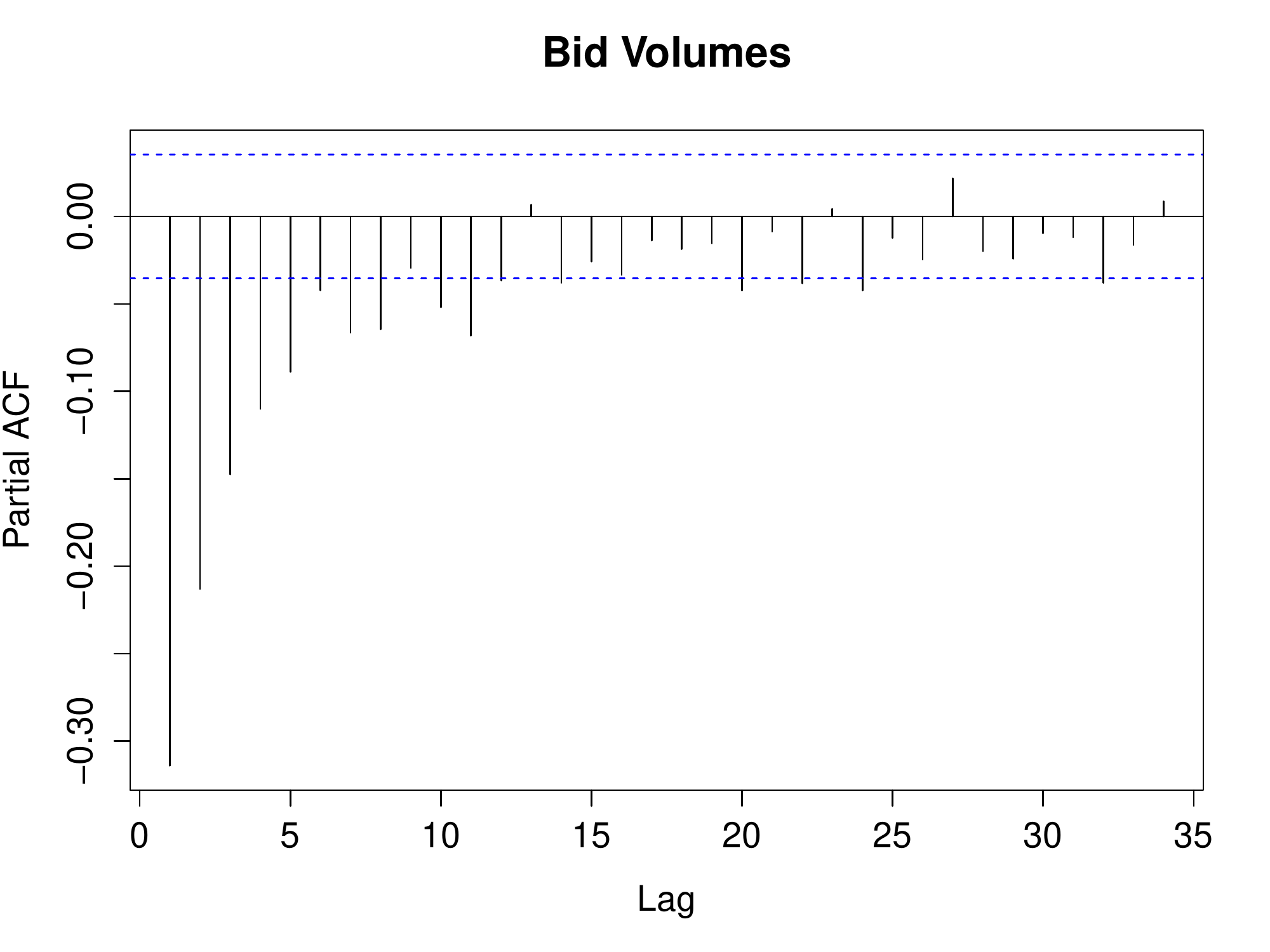}
\caption{Top row subplots: Total sell volume resting in the LOB in the first 5 ticks away from the best bid (left), total buy volume resting in the first 5 ticks away from the best ask (right) for stock GDF Suez on a typical day. Middle Row Subplots: First differences of figures above. Bottom Row Subplots: Sample ACF and PACF.}
\label{fig:bidaskvolumesdiff}
\end{center}
\end{figure}

\noindent \textbf{Auxiliary Model 2 - Volume features:} In Figure \ref{fig:bidaskvolumesdiff} we demonstrate an example of the volume on the bid and ask side for a typical day for stock GDF Suez. We fit an ARIMA model to this data, in order to capture the time series structure of the LOB volumes. We will err on the side of parsimony during model identification, as we would like to obtain an auxiliary model with few parameters in our Indirect Inference procedure. 

We first remove observed linear trends present in the LOB volume timeseries throughout the day by taking first differences, see Figure \ref{fig:bidaskvolumesdiff}. The resulting sample ACF and PACF is given in Figure \ref{fig:bidaskvolumesdiff} and it indicates that an MA(1) model is appropriate. Hence, we fit an ARIMA(0,1,1) model to the volume data. 

\subsubsection{Combining multi-objective optimisation and Indirect Inference}
Thus far, for a given set of parameters in our stochastic LOB agent model, we have simulated the order book process. This simulated data was then utilised to construct a framework in which we obtained multiple fitted parameter vectors, one for each auxiliary model considered. We now need to consider how to judge the suitability of the model parameter vector in capturing the true observed LOB stochastic process dynamics.

In standard Indirect Inference based frameworks, one would concatenate all the auxiliary model output parameter vector estimates into a single vector of auxiliary model parameters, in order to produce a single distance measure or discrepancy between the simulated data and actual data. This concatenation induces a loss in information, as for instance some auxiliary parameter model discrepancies may be on different scales to others. Therefore, if a na\"{\i}ve concatenation is applied, this often results in domination of a select few criteria, rather than considering each component in its own right.

We overcome this issue through the introduction of a multi-objective optimization framework. Such methods naturally adapt the simulation-based estimation to allow for competing criteria when assessing the suitability of the stochastic agent LOB model parameters via a collection of auxiliary model fits. The multi-objective optimisation method thus enables us to consider multiple distance measures, of discrepancy scores, as separate objective functions. 

In this framework, the fitness, or suitability, of a parameter vector $\bm{\theta}$ for the stochastic agent LOB model is measured by simulating from the generative model and quantifying the difference between each auxiliary model's parameters. Each auxiliary model is fit to both the tranformation of the observed data to obtain ($\widehat{\bm{\beta}}_k$) and to the transformation of the simulated LOB data ($\widehat{\bm{\beta}}_k^{\ast}\left(\bm{\theta}\right)$), for which a discrepancy score is calculated by measuring the distance between the two. The most commonly utilised distance measures are based on some form of weighted or unweighted norm, such as the $L_p$-norm, or Minkowski distance of order $p$, of which the $L_\infty$-norm, the $L_1$-norm and the $L_2$-norm 
are frequently used in practice. We adopted the $L_2$-norm to measure discrepancies for both the price and volume-based auxiliary models we considered, generically given for the $k$-th auxiliary model by
\begin{equation}
\label{eq:objdiff}
\mathcal{D}_k(\bm{\theta}) = \mathcal{D}\left(\widehat{\bm{\beta}}_k , \widehat{\bm{\beta}}^{\ast}_k\left(\bm{\theta}\right)\right) = \sum_{i=1}^{q_k} \left(\left[\widehat{\bm{\beta}}_k\right]_i - \left[\widehat{\bm{\beta}}^{\ast}_k\left(\bm{\theta}\right)\right]_i\right)^2.
\end{equation}
for each $q_k$-dimensional auxiliary model, $k=1, \ldots , K$. 

\subsubsection{Multi-objective optimisation and the role of Pareto optimality}
When our search is for an optimal parameter vector $\bm{\theta}$ that should satisfy multiple objective functions, in a vector $\bm{\mathcal{D}}(\bm{\theta}):=\left[\mathcal{D}_1(\bm{\theta}),\ldots,\mathcal{D}_K(\bm{\theta})\right]$ to be minimised, there are many cases where there will not be a global minimum with respect to each individual objective. In this case, one can consider as an alternative to the single optimal value produced by an optimisation method, the notion of \textsl{Pareto optimality}, in reference to the Pareto efficient frontier. Informally, this is the search for solutions such that there is no solution in the search space that can unilaterally improve a single criterion (objective function) without worsening another criterion, and this is formally defined in Definition \ref{DefnPO} for the case of our estimation framework.

\begin{definition}[Pareto Optimal Dominance of Parameter Solutions] \label{DefnPO}
Consider the set of $K$ auxiliary models producing parameter vectors $\left\{\bm{\beta}_k\right\}_{k \in \left\{1,2,\ldots,K\right\}}$, each based on an underlying parameter vector $\bm{\theta}\in \Omega$, that produce, for selected objective functions, the values $\bm{\mathcal{D}}(\bm{\theta}):=\left[\mathcal{D}_1(\bm{\theta}),\ldots,\mathcal{D}_K(\bm{\theta})\right]$. 
Then the selection of $\bm{\theta} \in \Omega$ is called Pareto-optimal or (non-dominated) with respect to the set of solutions in the feasible region $\Omega$, if 
\begin{equation}
\nexists \widetilde{\bm{\theta}} \in \Omega \;\; \text{s.t} \;\; \bm{\mathcal{D}}(\widetilde{\bm{\theta}}) \prec \bm{\mathcal{D}}(\bm{\theta}),
\end{equation}
where we say that $\bm{\mathcal{D}}(\bm{\theta})$ dominates $\bm{\mathcal{D}}(\widetilde{\bm{\theta}})$, denoted by $\bm{\mathcal{D}}(\bm{\theta}) \prec \bm{\mathcal{D}}(\widetilde{\bm{\theta}})$, if 
\begin{equation}
\mathcal{D}_k(\bm{\theta}) \leq \mathcal{D}_k(\widetilde{\bm{\theta}}) \;\; \forall k \in \left\{1,2,\ldots,K\right\} \;\; \text{and} \;\; \exists k \;\; \text{s.t.} \;\; \mathcal{D}_k(\bm{\theta}) < \mathcal{D}_k(\widetilde{\bm{\theta}}).
\end{equation}
\end{definition}

From this, we can then state the overall objective, incorporating all $K$ auxiliary models and a common selection of $L2$-norm objective functions for the parameter vector $\bm{\theta}$ of the stochastic agent-based model as follows
\begin{equation}
\begin{split}
\widehat{\bm{\theta}}&= \arg \underset{\bm{\theta}\in \Omega}{\min} \left[\mathcal{D}_1(\bm{\theta}),\ldots,\mathcal{D}_K(\bm{\theta})\right]\\
&= \arg \underset{\bm{\theta}\in \Omega}{\min} \left\{ \mathcal{D}\left(\hat{\bm{\beta}}_1, \hat{\bm{\beta}}^{\ast}_1\left(\bm{\theta}\right)\right),\ldots, \mathcal{D}\left(\hat{\bm{\beta}}_K, \hat{\bm{\beta}}^{\ast}_K\left(\bm{\theta}\right)\right) \right\}\\
&= \arg \underset{\bm{\theta}\in \Omega}{\min} \left\{ 
\sum_{i=1}^{q_1} \left(\left[\hat{\bm{\beta}}_1\right]_i - \left[\hat{\bm{\beta}}^{\ast}_1\left(\bm{\theta}\right)\right]_i\right)^2, \ldots, 
\sum_{i=1}^{q_K} \left(\left[\hat{\bm{\beta}}_K\right]_i - \left[\hat{\bm{\beta}}^{\ast}_K\left(\bm{\theta}\right)\right]_i\right)^2 \right\}\\
&\;\;\;\;  \text{subject to } \;\; \theta_{1_{L}}\leq \theta_{1}\leq \theta_{1_{U}},...,\theta_{n_{L}}\leq \theta_{n}\leq \theta_{n_{U}}
\end{split}
\end{equation}
where $[\theta_{i_{L}},\theta_{i_{U}}]$, for all $i$, which denote the boundaries of the feasible region $\Omega$.

To complete the specification of the multi-objective Indirect Inference simulation based estimation framework we propose, we require a method to search the constrained parameter space $\Omega$ for feasible and Pareto optimal solutions. A variety of stochastic search methods are available for use in this context, see discussion in \cite{coello2007evolutionary}.

We propose the use of an evolutionary genetic search method for this purpose, known in the literature as Multi-Objective Evolutionary Algorithms (MOEAs). We develop a version of such a stochastic search framework which combines the widely utilised NSGA-II genetic search algorithm by \cite{deb2002fast}, which is a Pareto-ranking based  method, with an additional mutation kernel we designed specifically for a covariance matrix mutation operator, based on the framework developed in \cite{peters2012copula}. This additional mutation component is combined with the framework of NSGA-II, to ensure that the proposed covariance matrices in the stochastic agent LOB model, which are proposed at each step of the search, remain positive definite and symmetric. Details of this genetic search algorithm are provided in Appendix \ref{app:evosearch}.

\section{Stochastic agent LOB model assessment and calibration to real LOB data}
\label{sec:results}
We have provided a description of the stochastic agent-based LOB model we developed for modelling trading interactions and their dependency. In addition, we have developed a method for the calibration of model parameters to observed LOB data. In this section, we illustrate the results of this calibration on real data, through a sequence of studies which aim to practically assess the importance of each component of the stochastic agent LOB model specification. To achieve this, we make a number of model simplifications and progressively relax these simplifying assumptions, in order to provide an understanding of the role each feature of our proposed model plays in the simulation framework. The \textbf{reference model} is the basic framework against which we compare the more detailed versions of the model, as detailed below.

\subsection{Developing a baseline simplified reference stochastic agent LOB model}

In the stochastic agent-based LOB model, the liquidity provider agent has limit order submission and cancellation components which each require the specification of four independent $l_t$-dimensional multivariate skew-t distributions for the bid and ask sides, with $l_p=5$ `passive' levels and $l_d=3$ `direct', or aggressive levels for a total of $l_t=8$ actively modelled levels for each side of the book. For each of these stochastic model components we require the estimation of the parameters: $\bm{m} \in \mathbb{R}^d$, the location for the mean intensity vector; $\bm{\gamma} \in \mathbb{R}^d$, the skewness of the stochastic intensity vector; $\nu \in \mathbb{R}^{+}$ which directly influences the heavy-tailedness of the stochastic intensity vector and $\Sigma \in \mathbb{R}^{d\times d}$ the covariance matrix of the stochastic intensity vector for order arrivals. We consider aggregate activity in 10 second intervals, and for the 8.5 hour trading days for the asset under consideration here, we have $T=3060$ intervals in the day. The basic reference model is characterised by the following model assumptions:
\begin{itemize}
\item{We assume that the associated limit order submission distributions for the bid and ask have common parameter value settings. In addition, market order submission distributions for the bid and ask are also assumed to have common parameter value settings. This is reasonably consistent with empirical observations for a large number of assets when observing the submission activity on either side of the LOB throughout the trading day.
}
\item{Since the vast majority of orders get cancelled prior to execution, we consider the parameters of the distribution of cancellations to also match the distribution of limit order placements.}
\item{We also set $\bm{m}=\bm{0}$ and consider the skewness vector, $\bm{\gamma}$, to take a common value in all levels of the bid and ask such that $\bm{\gamma}=\gamma_0 \bm{1}$, where $\bm{1}$ is a vector of ones.   
}
\item{The monotonic mapping $F(\cdot)$, transforming the random variables $\Gamma^{LO,k,s},\Gamma^{C,k,s},\Gamma^{MO,k}$ into intensity random variables $\Lambda^{LO,k,s},\Lambda^{C,k,s},\Lambda^{MO,k}$ is set as the CDF of the standard Normal. This transformation is necessary in order to ensure that intensities are positive, and to bound the event counts.}
\item{For the baseline intensities of limit order activity at each level, we assume that they will be the same for the `passive' limit orders on both sides, i.e. $\mu_0^{LO,a,1}=\ldots=\mu_0^{LO,a,l_p}=\mu_0^{LO,b,1}=\ldots=\mu_0^{LO,b,l_p}=\mu_0^{LO,p}$, while `aggressive' limit orders will have a different limit order intensity, i.e. $\mu_0^{LO,a,0}=\ldots=\mu_0^{LO,a,-l_d+1}=\mu_0^{LO,b,0}=\ldots=\lambda_0^{b,-l_d+1}=\mu_0^{LO,d}$. Market order baseline intensities are also equal on either side, i.e. $\mu_0^{MO,a}=\mu_0^{MO,b}=\mu_0^{MO}$. The cancellation baseline activity will be the same as the submission baseline activity. 
}
\item{Finally, we assume constant order sizes, i.e. $O_{i,t}^{LO,k,s}=c=O_{j,t}^{MO,k}$ for all 
$i\in \left\{1,\ldots, N_t^{LO,k,s}\right\}$, $j\in \left\{1,\ldots, N_t^{MO,k}\right\}$, $k \in \left\{a,b \right\}$, $s \in \left\{-l_d+1, \ldots, l_p \right\}$ and $t \in \left\{1,\ldots,T \right\}$.
}
\end{itemize}
Hence, the basic reference model has the following parameter vector $\left \{ \mu_0^{LO,p},\mu_0^{LO,d}, \mu_0^{MO},\gamma_0,\nu, \sigma^{MO} \right \}$, as well as the covariance matrix $\Sigma$ to be estimated. 

The cancellations are modelled by a dynamically evolving volume process, i.e. the Cox process is truncated to the available number of orders at each level, as specified in the model by $N_t^{C,k,s}|\left\{\overset{\sim}{V}^{LO,k,s}_{t} =v\right\} \sim Po(\lambda_t^{C,k,s})\mathbb{I}(N_t^{C,k,s}<v)$ where we denote by $V^{LO,k,s}_{t-1}$ the volume at level $L_i$ at the start of the $[t-1,t)$ interval and $\overset{\sim}{V}^{LO,k,s}_{t}$ is the volume available after the arrival of the limit orders at time $t$, but before the cancellations and executions. One can simulate from the model, in order to obtain the state of the LOB at time $t$, $L^*_t$, and thus the available volume $v$, so that one can then draw from a truncated Poisson distribution with a truncation limit of $v$. 

\begin{figure}
\begin{center}
\includegraphics[width=0.48\textwidth]{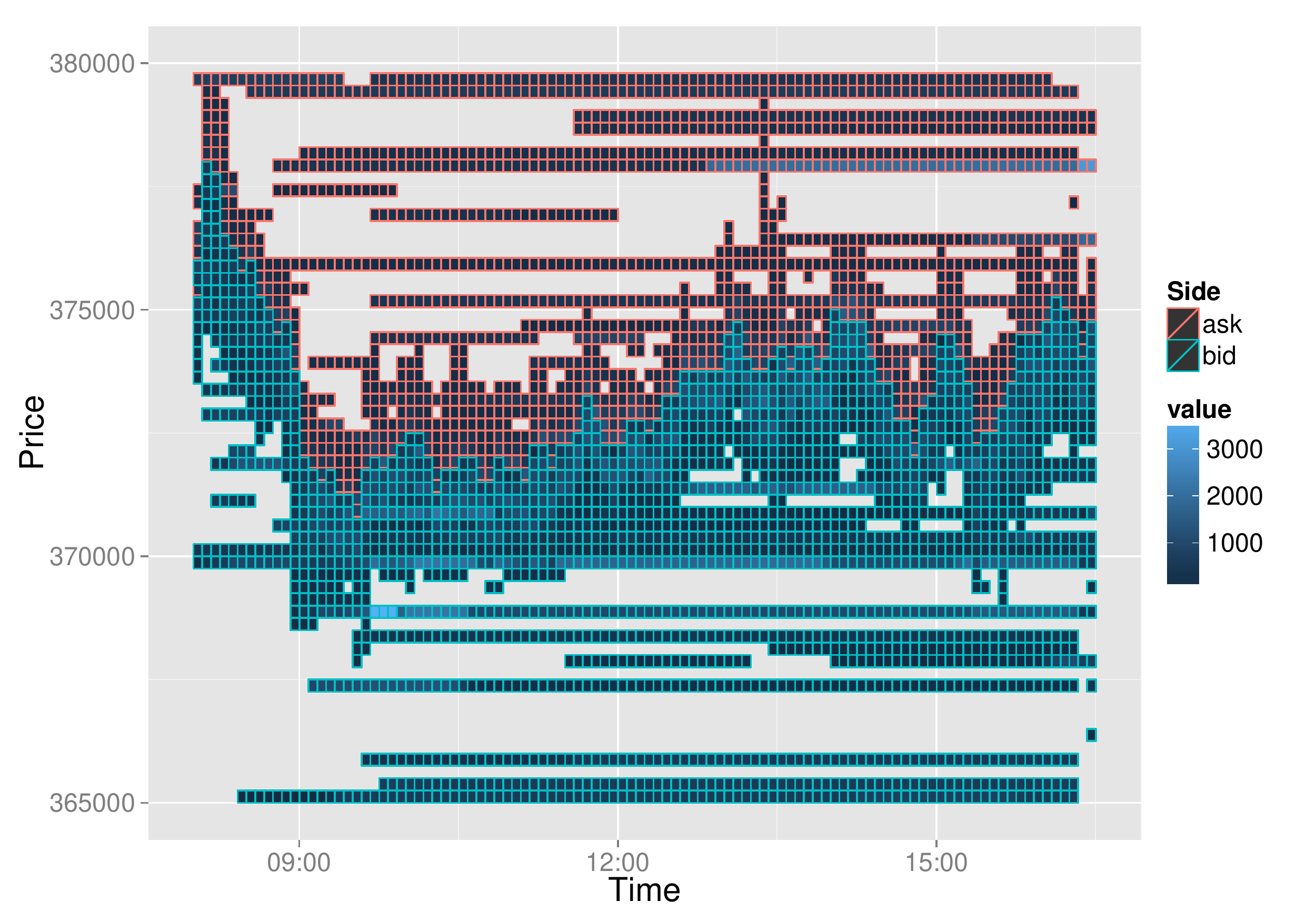}
\includegraphics[width=0.48\textwidth]{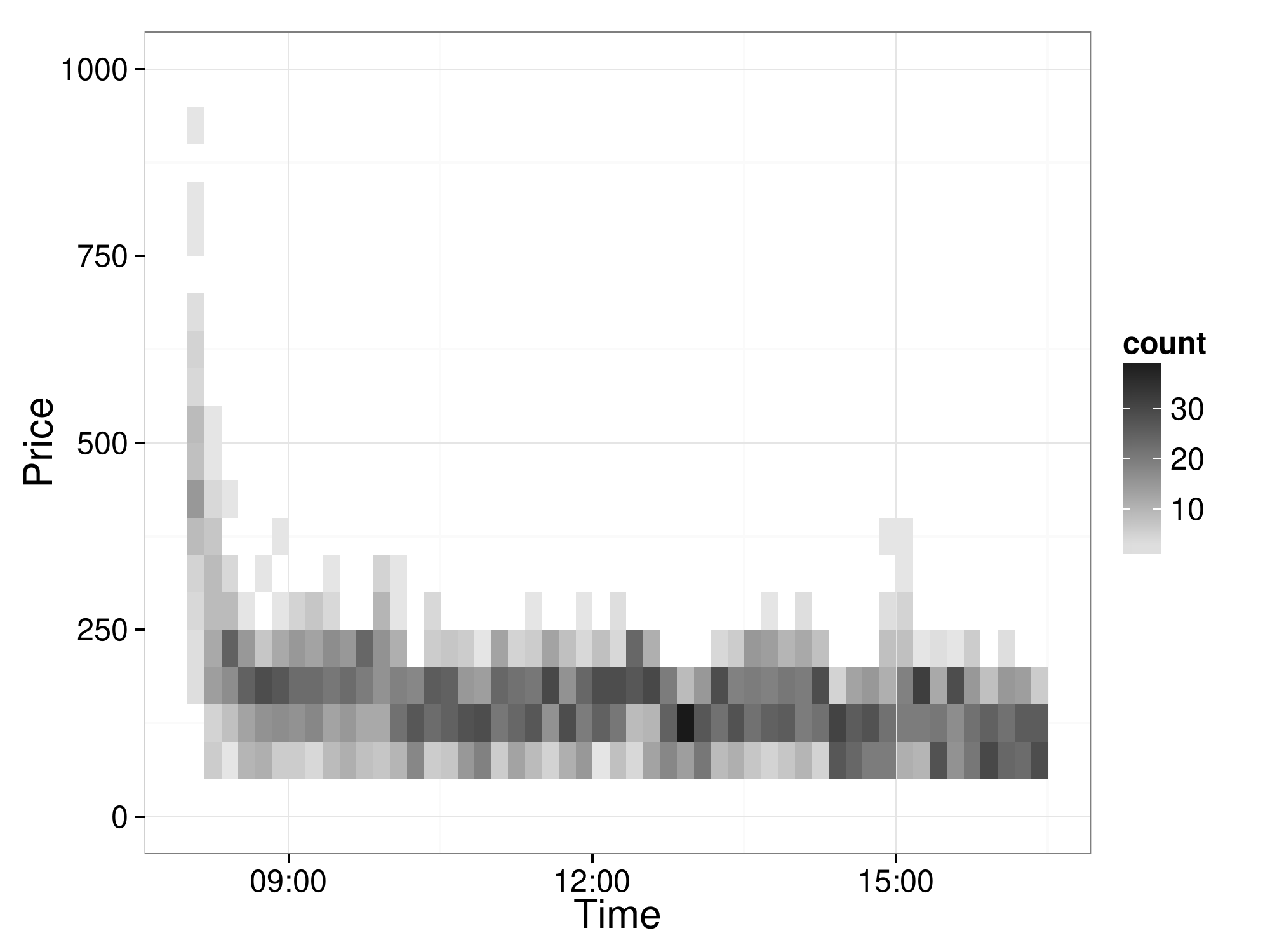}
\caption{For stock BNP Paribas, the intensity of the volume process on either side, where the shading of each bin indicates the average number of shares available at those prices in that period. The plot on the right shows the evolution of the spread throughout the trading day. }
\label{fig:realdata}
\end{center}
\end{figure}

Before we begin the study of the stochastic agent-based LOB model and its calibration and simulation behaviour, we first show for a representative trading day, the evolution of the spread, as well as the intensity of the volume process around the top of the book, for one of the most liquid stocks in the CAC40, namely BNP Paribas, in Figure \ref{fig:realdata}. This provides an illustration of the LOB dynamics we should aim to recover with the model once accurately calibrated. We estimate the model on the data from this day, as an illustration of the calibration procedure. 

\subsection{Reference model: Calibration}
\label{subsec:calib}
\begin{figure}
\begin{center}
\includegraphics[width=0.6\textwidth]{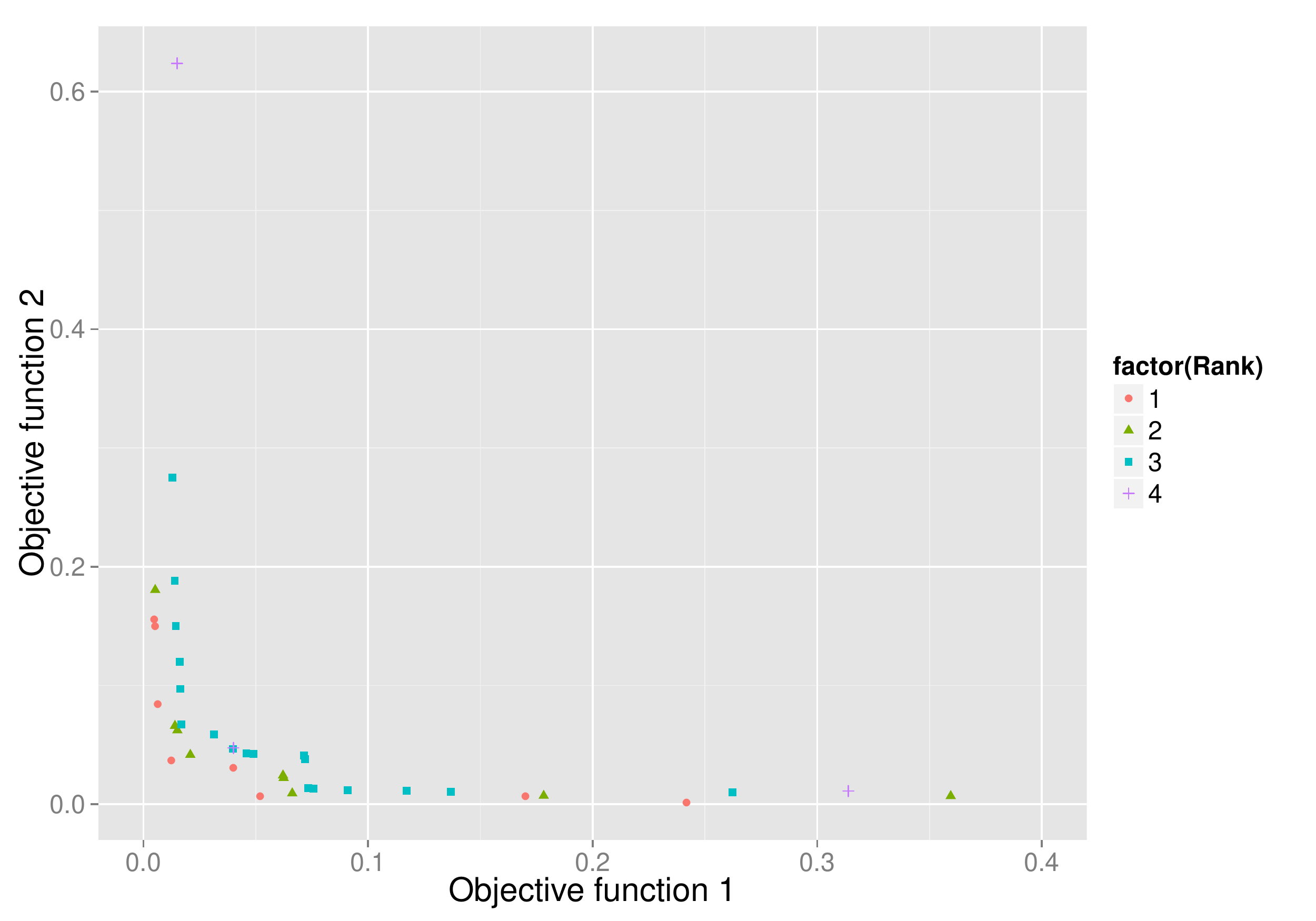}
\caption{Objective function values for the parameter vectors produced by the multi-objective II method. These are grouped by non-domination rank, with a rank of 1 indicating non-dominated vectors, a rank of 2 indicating vectors dominated only by a single other vector and so on. Note that the points in each group form a Pareto front, a feature of the optimisation.}
\label{fig:basicobjvalue}
\end{center}
\end{figure}

We present in Table \ref{tab:estbasic} the results of the estimation using the multi-objective II approach proposed in this paper. There are 8 non-dominated solutions spread out accross the Pareto optimal front, each of which also has an associated covariance matrix, which has not been included here due to space considerations, instead we provide the trace as a summary. In the table, we also present a further 4 solutions with a non-domination rank of 2, i.e. parameter vectors which were dominated in both objective functions by only one other parameter vector. We present the non-domination rank, as well as the objective function values of the entire final parameter population in Figure \ref{fig:basicobjvalue}. We note that in terms of the 2 objective function values associated with these parameter vectors, these are spread out across the Pareto front.

\begin{figure}
\begin{center}
\includegraphics[width=0.49\textwidth]{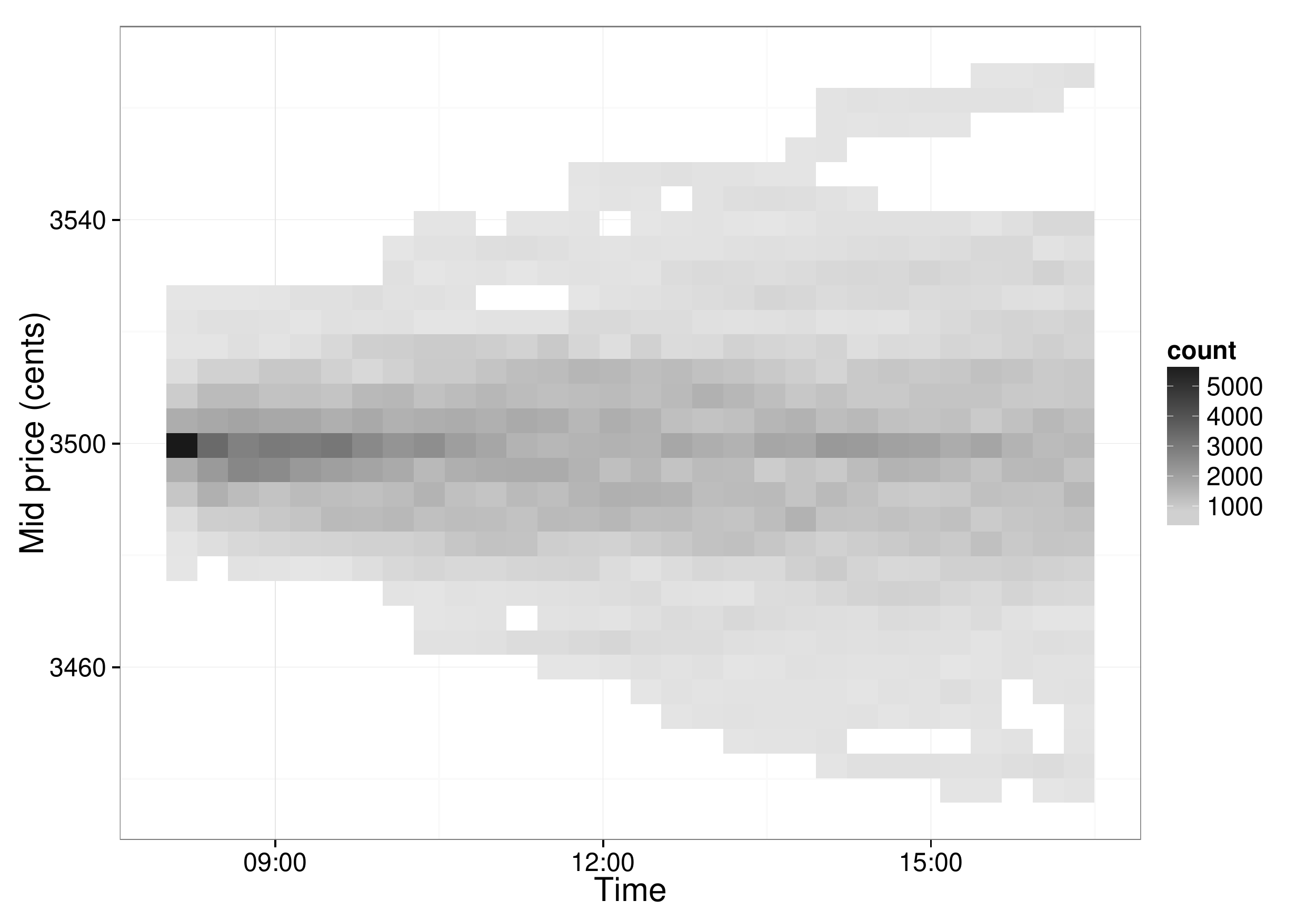}
\includegraphics[width=0.49\textwidth]{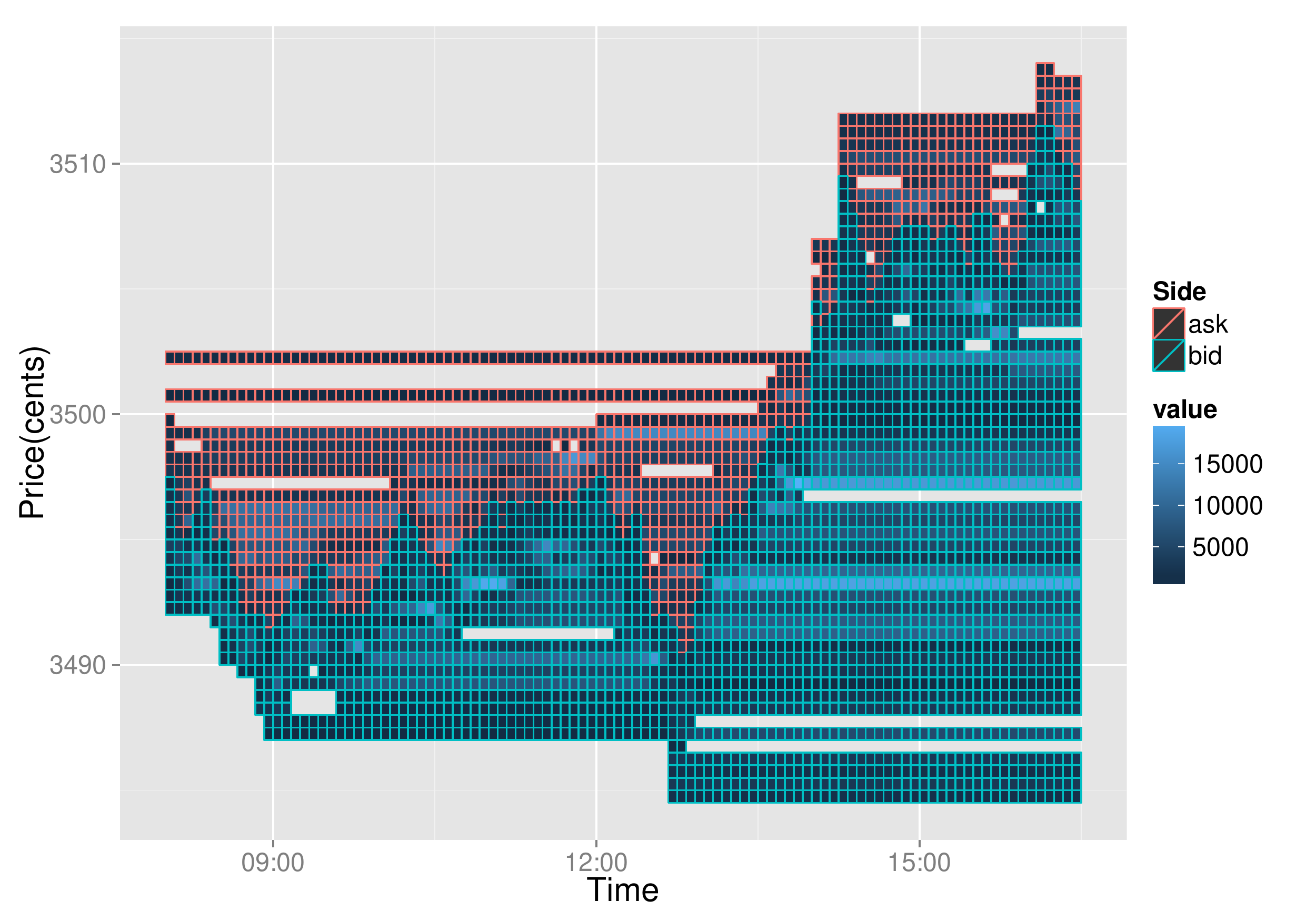}
\includegraphics[width=0.49\textwidth]{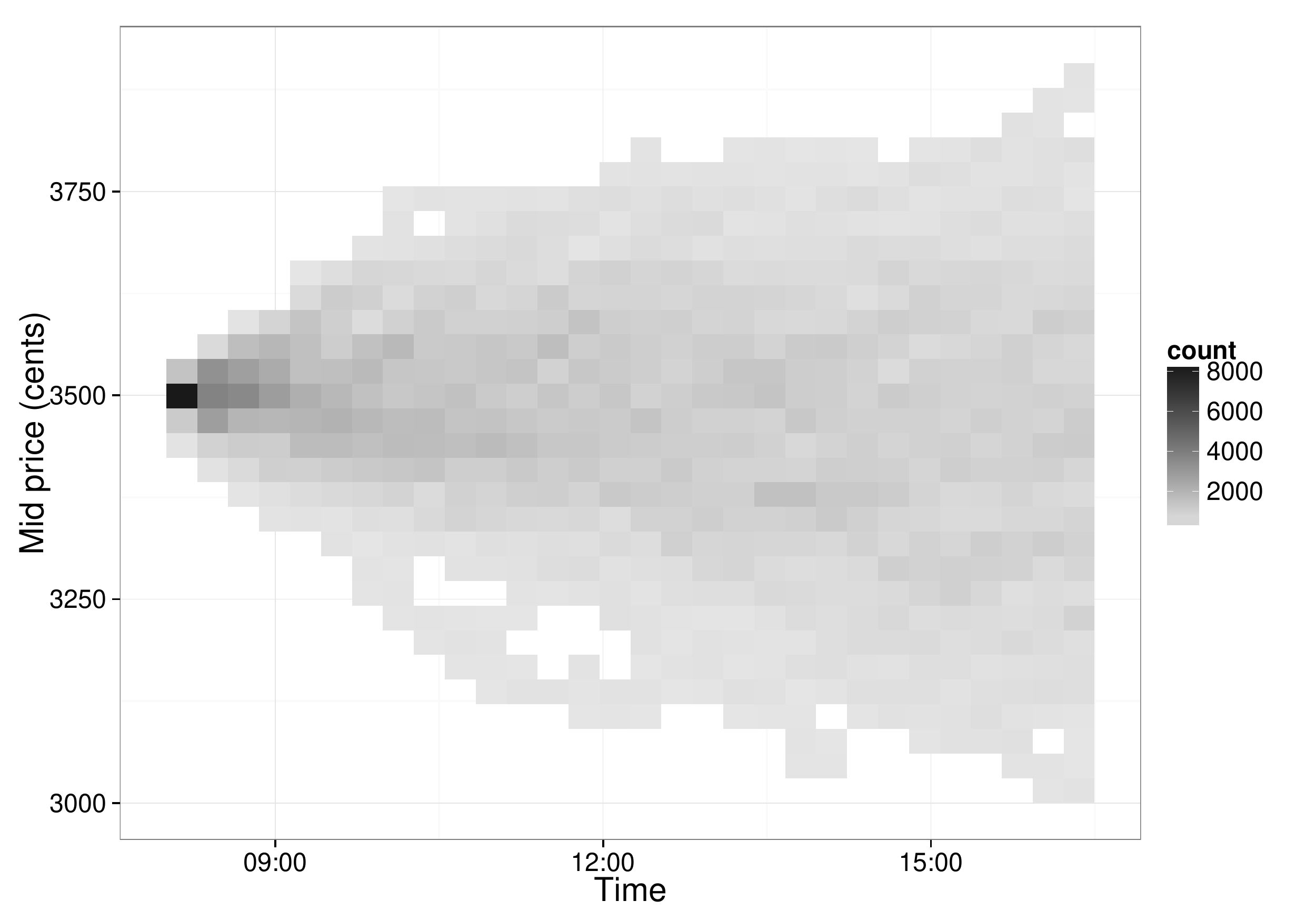}
\includegraphics[width=0.49\textwidth]{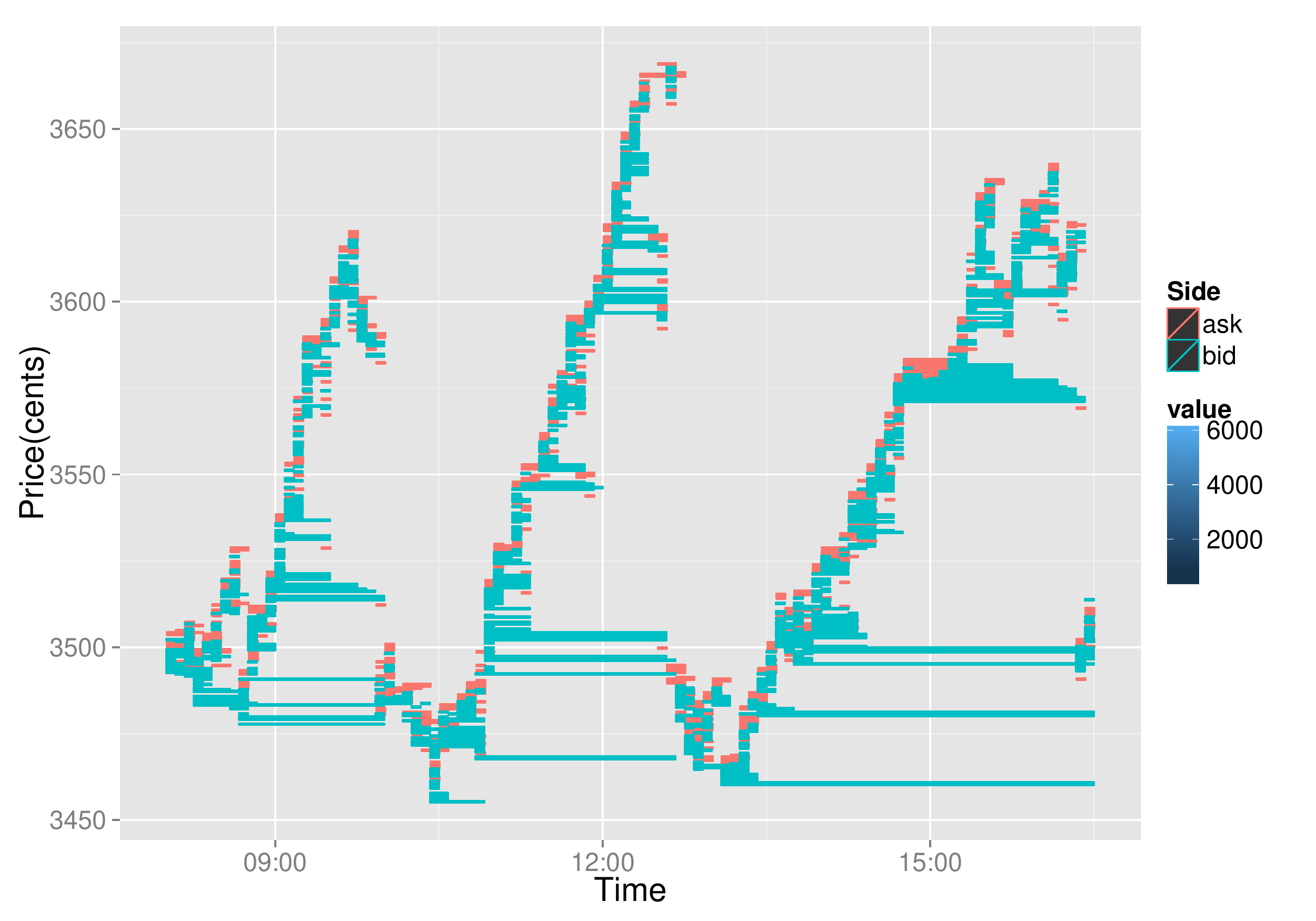}
\caption{Simulations using 2 of the non-dominated parameter vectors resulting from estimating the basic model with NSGA-II. The figures on the left are heatmaps of the asset mid price over 100 simulations, while the figures on the right represent the state of the LOB over a single simulation.}
\label{fig:simsbasic}
\end{center}
\end{figure}

We assess the fit by a qualitative comparison of the simulations produced with the estimated parameters. In Figure \ref{fig:simsbasic} we present, for the first 2 Pareto optimal solutions of the parameter vectors in Table \ref{tab:estbasic}, summaries of the price process for repeated simulations, as well as an example of the LOB evolution throughout the day. We see that the two Pareto optimal solution parameter vectors produce a broad variety of different price trajectories over repeated simulations. In particular, some points on the Pareto front of solutions for this basic reference model produce a time series of simulated prices which replicates a trading day with relatively volatile trade activity, whilst other points on the Pareto front favour more constrained trading simulated price activities. To understand how this may occur, we note that this is likely to be due to the relatively high baseline rate of market orders compared to baseline limit order rates in the first set of Pareto optimal solutions, compared to the second. 

\begin{table}[ht]
\caption{Non-dominated solutions after 40 iterations, with a population size of 40.}
\centering
\begin{tabular}{rrrrrrrrr}
  \hline
  & $\mu_0^{LO,p}$ & $\mu_0^{LO,d}$ & $\mu_0^{MO}$ & $\gamma_0$ & $\nu_0$ & $\sigma^{MO}_0$ & $\mathrm{Tr}(\Sigma)$ \\ 
  \hline
1 & 30.84 & 8.16 & 4.75 & -0.18 & 33.70 & 1.78  & 7.11 \\ 
  2 & 31.16 & 5.13 & 4.41 & 9.96 & 28.07 & 9.95 & 4.60 \\ 
  3 & 31.16 & 5.13 & 4.41 & 9.96 & 21.74 & 9.95 & 5.52 \\ 
  4 & 29.82 & 5.24 & 4.45 & -0.52 & 20.57 & 4.70 & 4.26 \\ 
  5 & 46.87 & 7.42 & 4.77 & 0.64 & 28.34 & 8.81 & 6.82 \\ 
  6 & 22.05 & 8.18 & 8.13 & -1.68 & 24.65 & 1.83 & 5.70 \\ 
  7 & 19.83 & 5.41 & 0.68 & -0.28 & 28.85 & 2.15 & 5.25 \\ 
  8 & 12.95 & 3.12 & 2.93 & 2.35 & 35.25 & 3.58 & 6.01 \\ 
  9 & 30.84 & 8.16 & 4.75 & -0.18 & 33.70 & 1.78 & 5.14 \\ 
  10 & 31.16 & 5.13 & 4.41 & 9.96 & 28.07 & 9.95 & 5.20 \\ 
  11 & 31.16 & 5.13 & 4.41 & 9.96 & 21.74 & 9.95 & 7.30 \\ 
  12 & 29.82 & 5.24 & 4.45 & -0.52 & 20.57 & 4.70 & 4.32 \\ 
   \hline
\end{tabular}
\label{tab:estbasic}
\end{table}

In Appendix \ref{app:results}, we provide further calibration results for the reference model, for multiple assets, over an extended period of 15 trading days. Summarising these results, we show that within the set of solutions produced by our estimation procedure, there is very commonly a subset which produce simulations which are similar to real trading observations in terms of their price and volume behaviour, which are the summaries of the LOB which our auxiliary models related to.

\subsection{Relaxing assumptions of the reference stochastic agent LOB model}
\label{subsec:relax}
The baseline model results are encouraging, however we still need to determine what influence the simplifying statistical model assumptions made in the reference model specification have on the calibration performance. This will now be assessed by progressively relaxing the assumptions and making less restrictive model assumptions. Our criterion for improvement relative to the reference model will be a reduction in the values of the objective functions of the solutions on the Pareto optimal front. We will only suggest that particular features should be relaxed if we observe such an improvement.

\subsubsection{Incorporating an order size distribution}
\begin{figure}
\begin{center}
\includegraphics[width=0.45\textwidth]{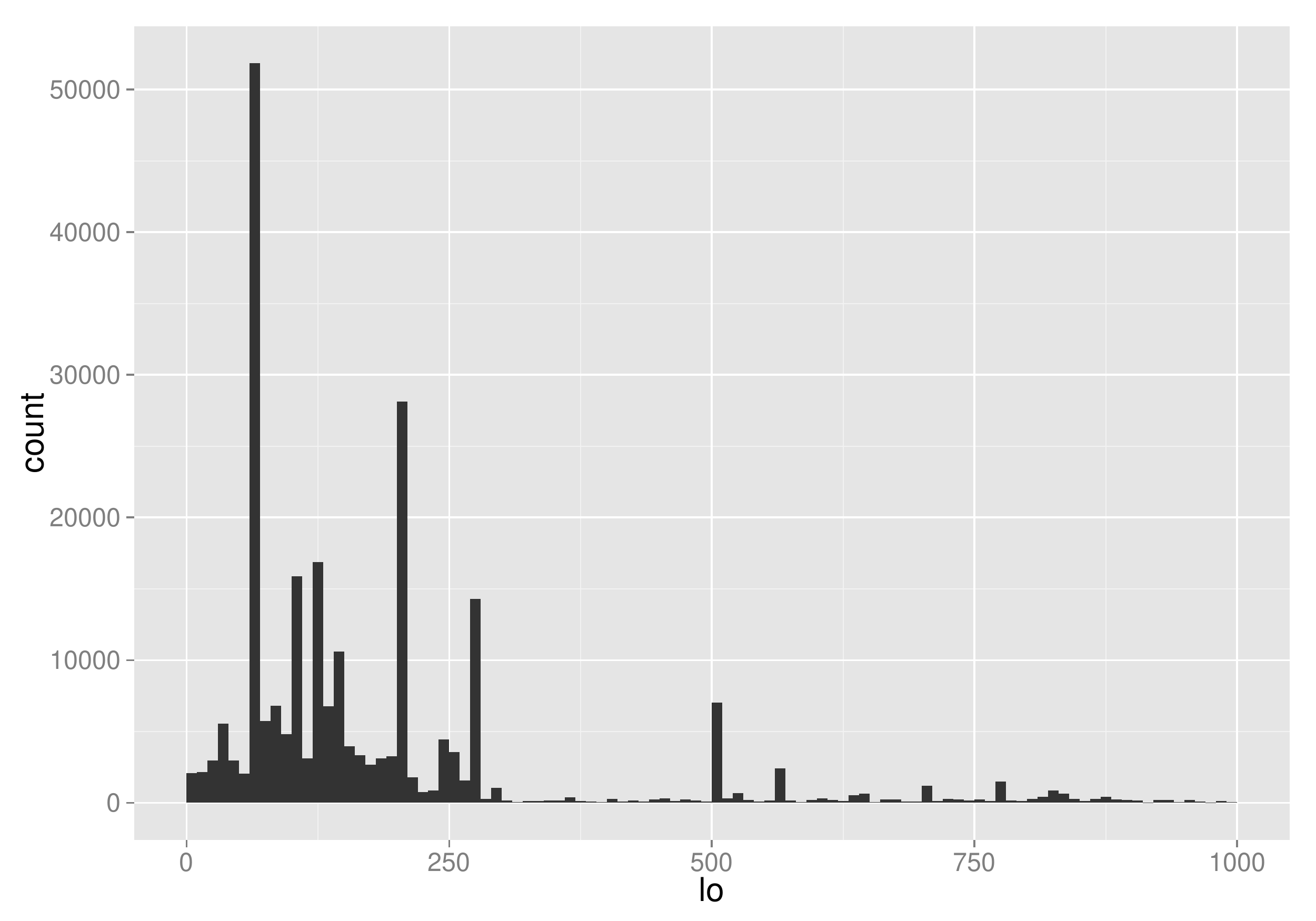}
\includegraphics[width=0.45\textwidth]{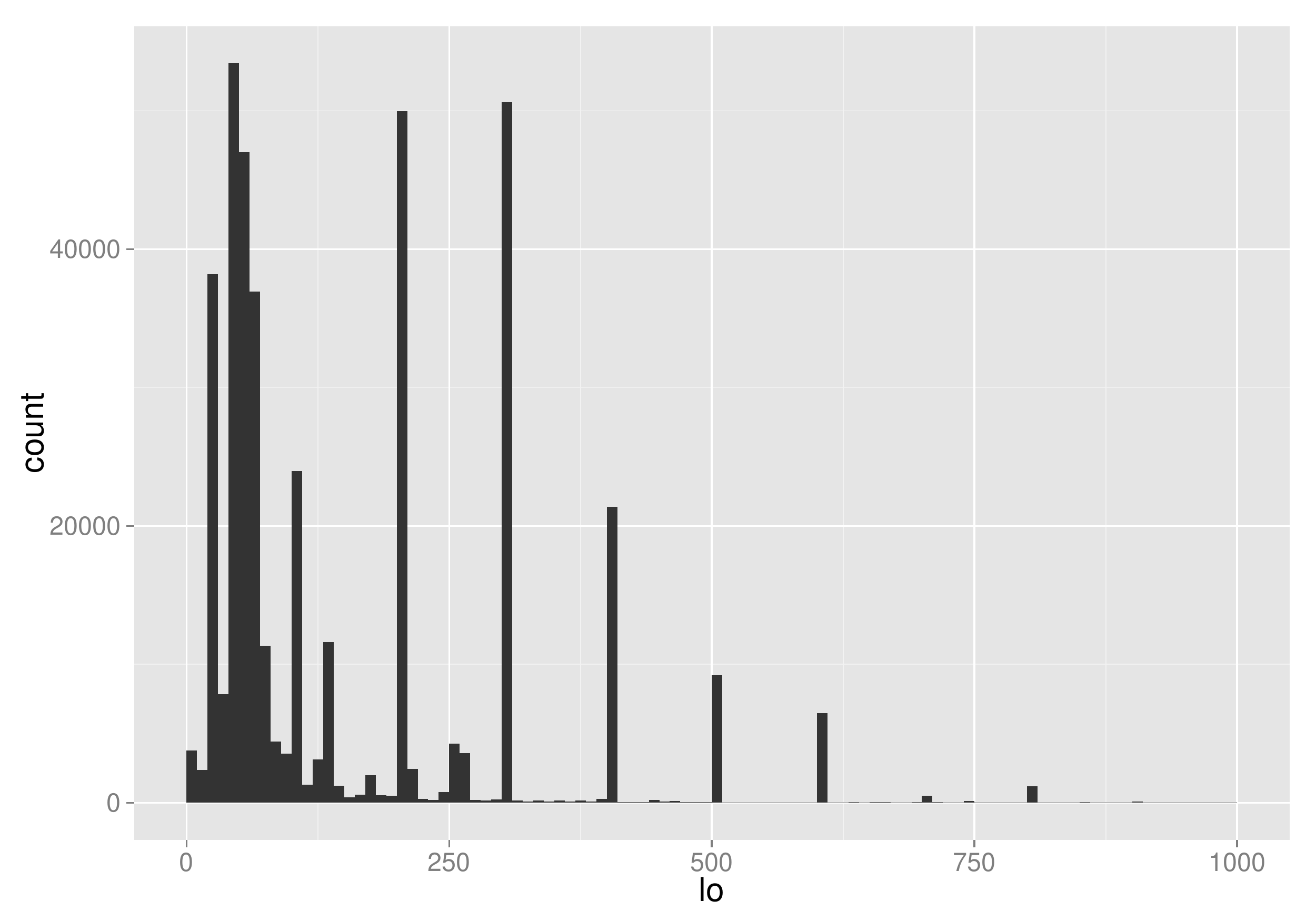}
\includegraphics[width=0.45\textwidth]{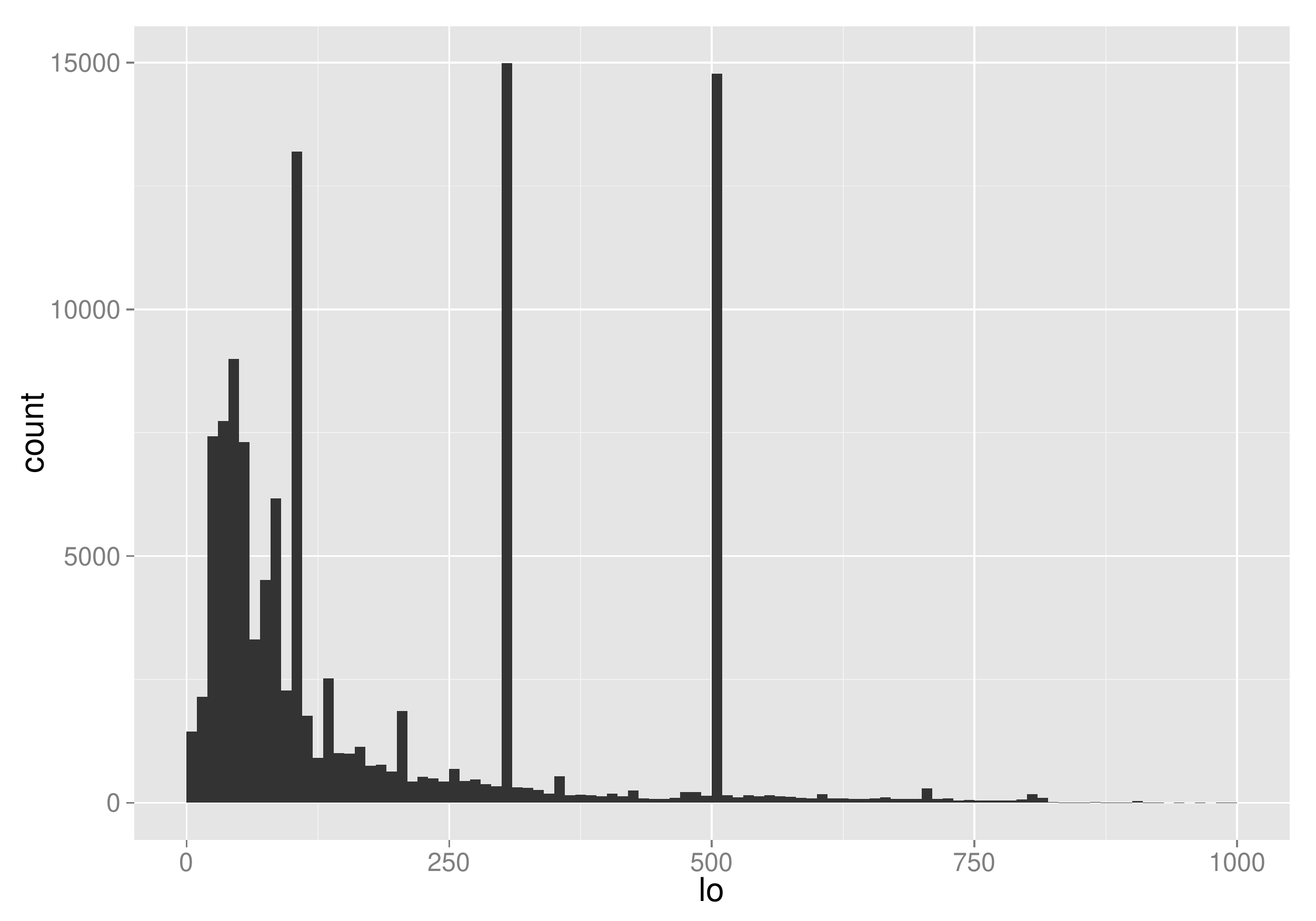}
\includegraphics[width=0.45\textwidth]{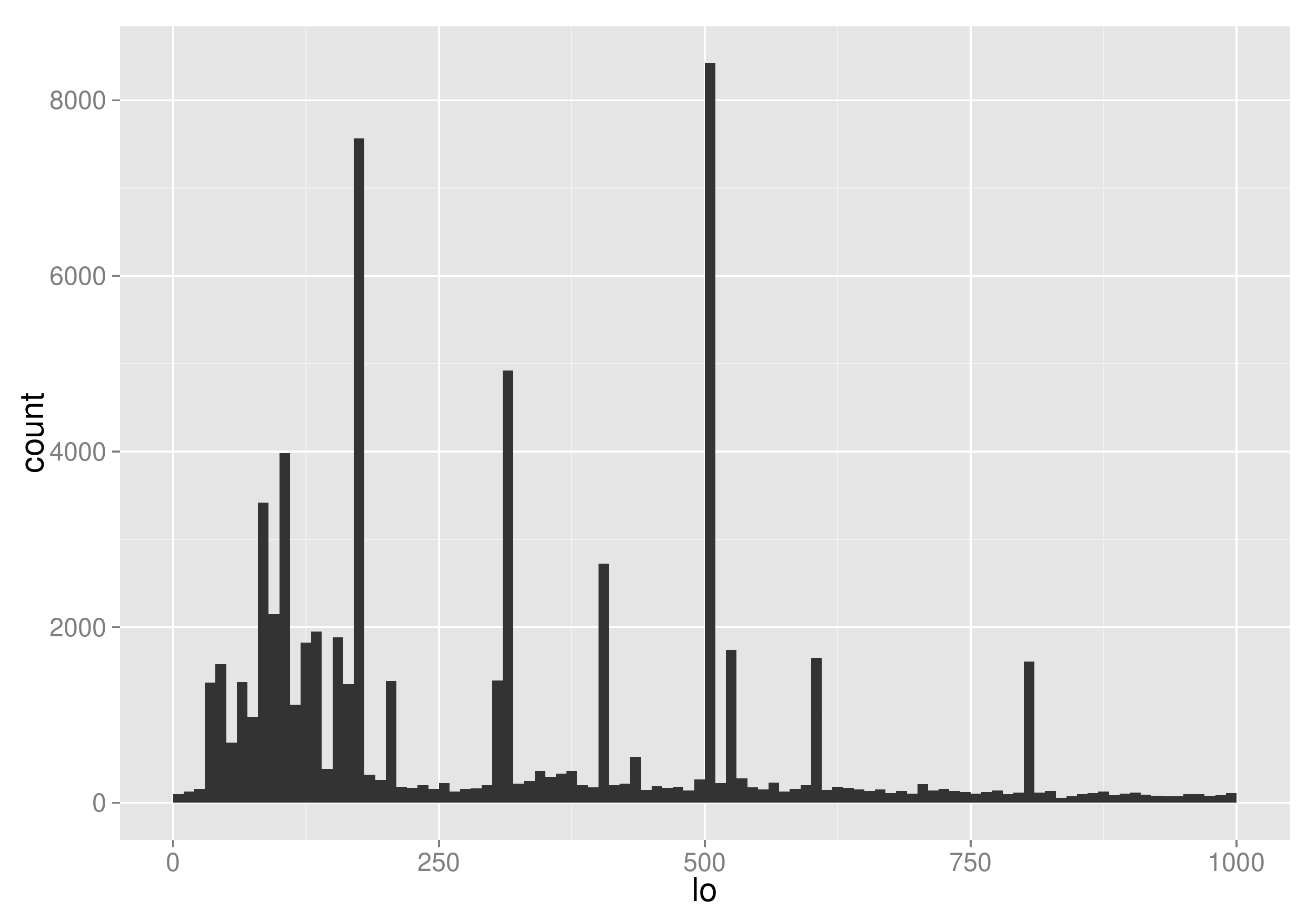}
\caption{Histograms of order sizes for 4 CAC40 stocks - ACAp and BNPp (top), SANp and VIVp (bottom) }
\label{fig:ordersizes}
\end{center}
\end{figure}

In our basic reference model, we assumed that orders sizes are constant, i.e. all limit order submissions, cancellations and executions were from an equal number of shares. This is similar to the model of \cite{cont2010stochastic}, which assumed that all orders are of unit size, which they set to correspond to the average size of limit orders observed for the asset. Abstracting away the order size aspect is an approximation one can make in order to simplify the model. However, such a simplifying assumption is not likely to be supported by the data, as we illustrate in Figure \ref{fig:ordersizes}. Clearly, one observes that there is a range of distribution shapes for the order sizes of different assets.  

It is clear that the distribution of order sizes will be affected by features such as minimum order sizes on an exchange (in number of shares, lots, or weight, depending on what is being traded). We observe empirically that for a range of equities traded in a number of countries, the distribution of order sizes has clear peaks at round figures - see Figure \ref{fig:ordersizes} for evidence of clustering order volumes at multiples of 100 shares, for example. This seems to be independent of the level at which they are submitted, whether it is a buy or a sell order, as well as the intensity of the order submissions in that period. 

Therefore, we present a case study where we relax the assumption of a fixed order size, by considering instead a stochastic model where we assume that the order size is drawn from a mixture of distributions. In this case, we assume that both the limit and market order sizes are obtained by sampling from the following Gamma mixture
\begin{equation}
O_{i,t}^{LO,k,s} \sim w \, Gamma(\kappa_1,\theta_1) + (1-w)\, Gamma(\kappa_2,\theta_2), \forall i,t,k,s 
\end{equation} 
where
\begin{equation}
Gamma(O;\kappa,\theta)=\frac{1}{\Gamma(\kappa,\theta^{\kappa})}O^{\kappa-1} \exp\left [ {-\frac{O}{\theta}} \right ]; \;\; O \in \mathbb{R}^+,
\end{equation}
with positive shape parameters $\kappa_1,\kappa_2$ and positive scale parameters $\theta_1,\theta_2$. We set $\kappa_1=1,\kappa_2=2$ as we observed there was a mode present in the empirical distributions of order sizes and we estimated the scale parameters for each mixture component to place the mode in the appropriate locations. Hence, we additionally estimate the parameters $\theta_1,\theta_2$ and the mixture weight $w$. 

\begin{figure}
\begin{center}
\includegraphics[width=0.5\textwidth]{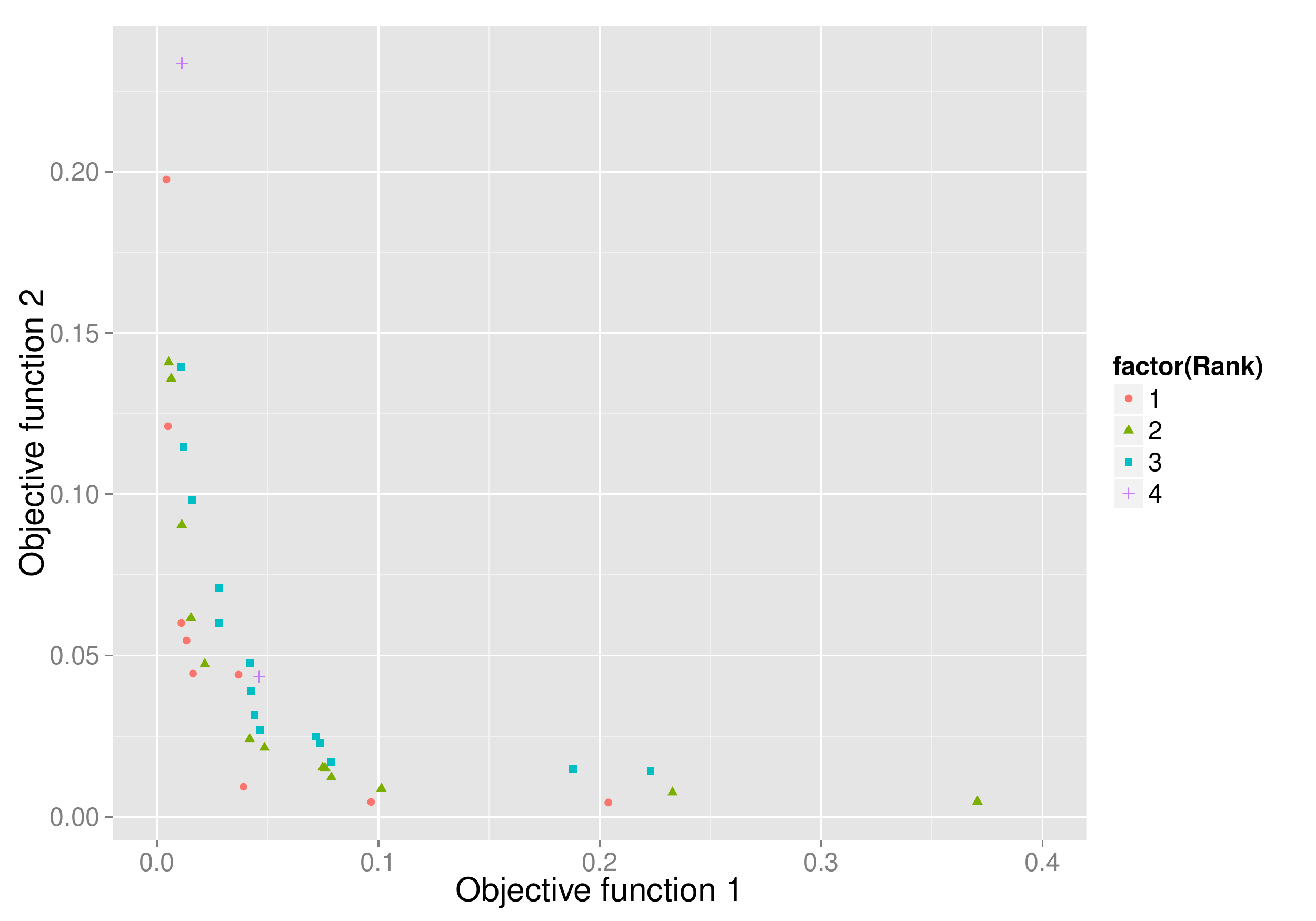}
\caption{Objective function values for the parameter vectors produced by the multi-objective II method, in the case where we assume that order sizes are follow a mixture of Gamma distributions.}
\label{fig:gammaobjvalue}
\end{center}
\end{figure}

We run the stochastic optimization framework using the same settings (a parameter population of 40 candidate solutions and an evolution over 40 generations) and calibrate the relaxed reference model with the stochastic model for the order sizes to the same data set used in the reference model fit, i.e. the LOB data for BNP Paribas over an entire day.
We obtain a Pareto optimal front which again contained multiple parameter vector solutions which were spread out over the Pareto front, indicating a successful exploratory search by the genetic search framework. Importantly, as shown in Figure \ref{fig:gammaobjvalue} we observe the realized objective function values for the relaxed reference model, which we observe are clear improvements on the objectives achieved by the comparison basic reference model case in which the order sizes were fixed. 

\begin{figure}
\begin{center}
\includegraphics[width=0.45\textwidth]{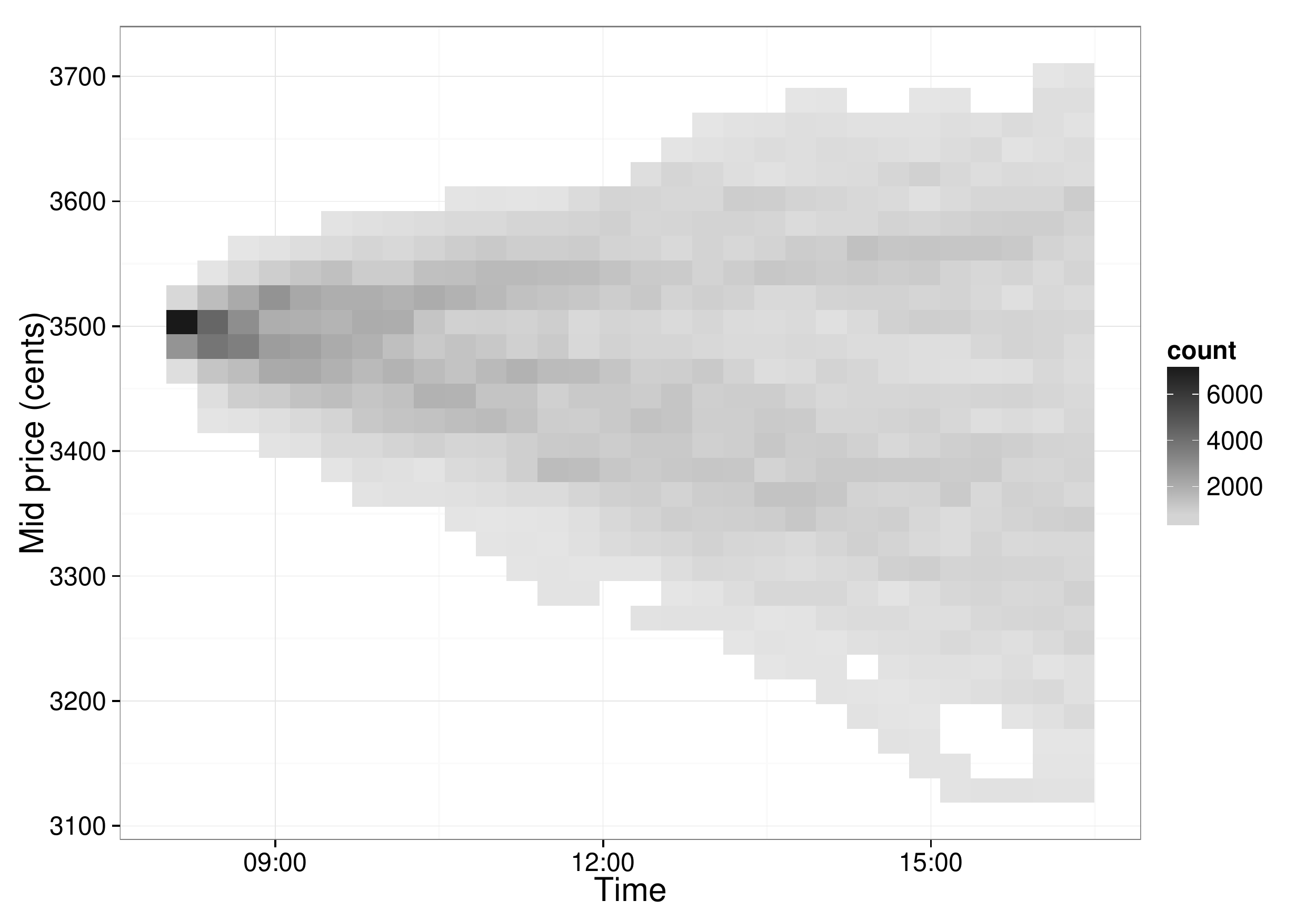}
\includegraphics[width=0.45\textwidth]{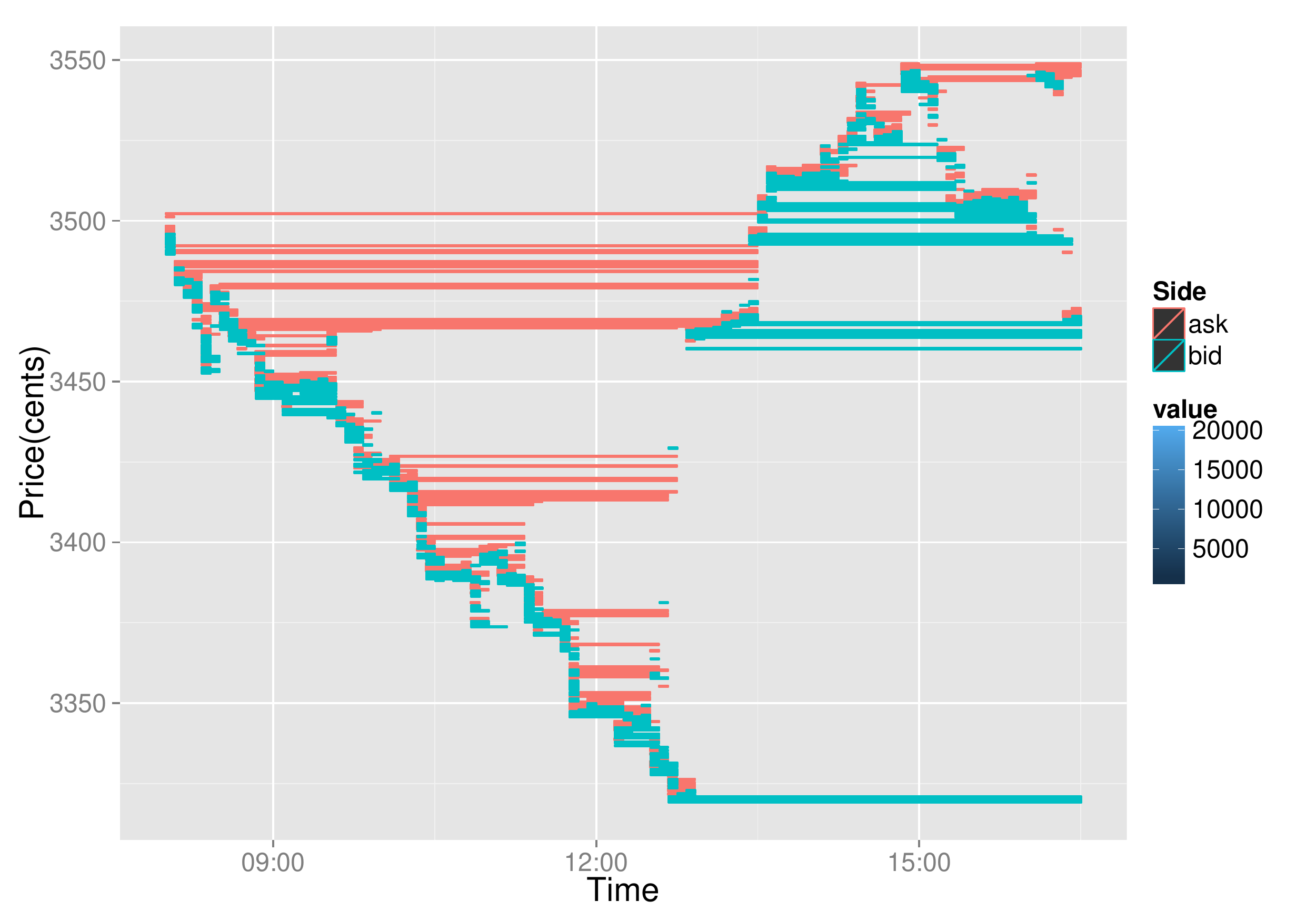}
\includegraphics[width=0.45\textwidth]{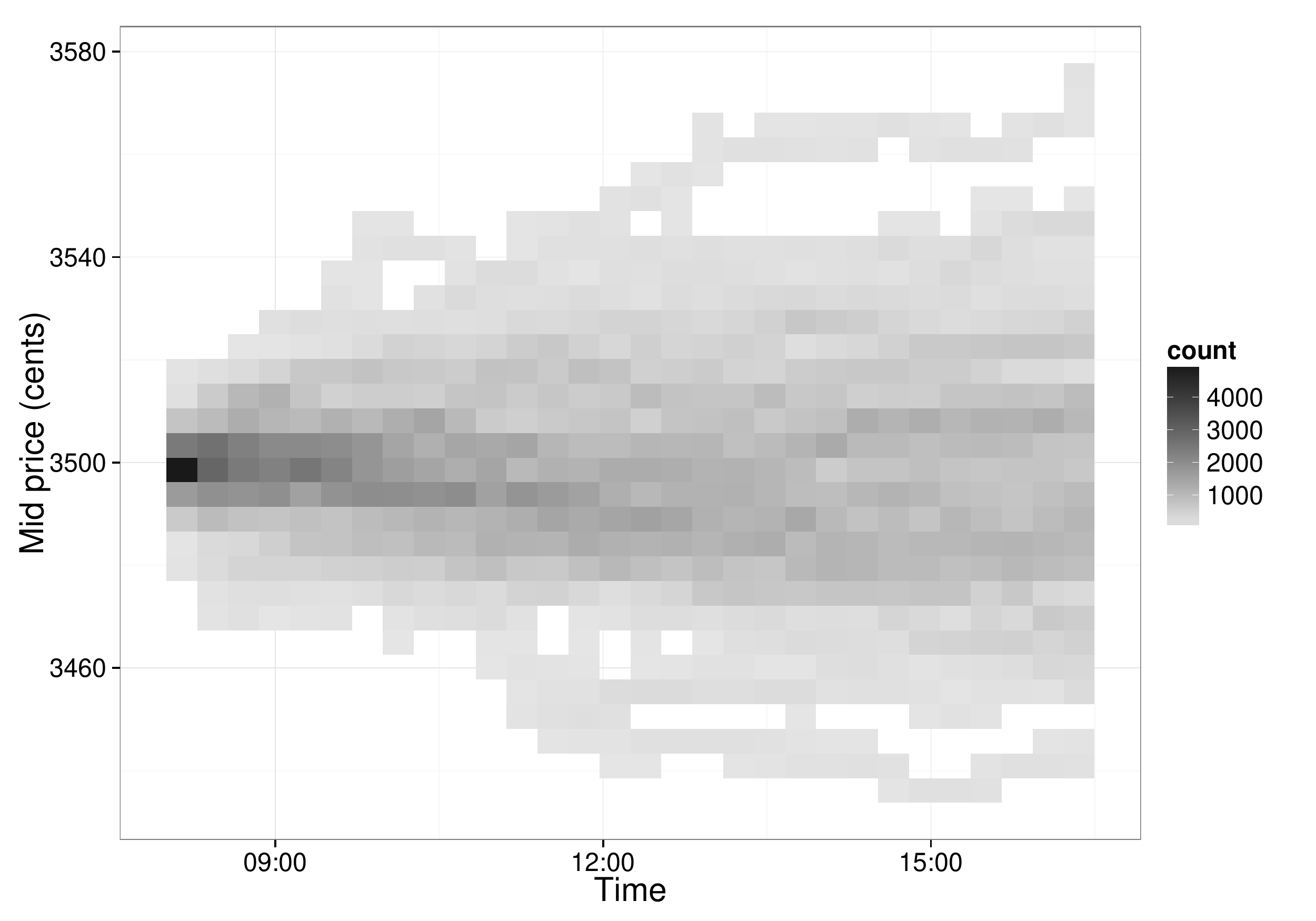}
\includegraphics[width=0.45\textwidth]{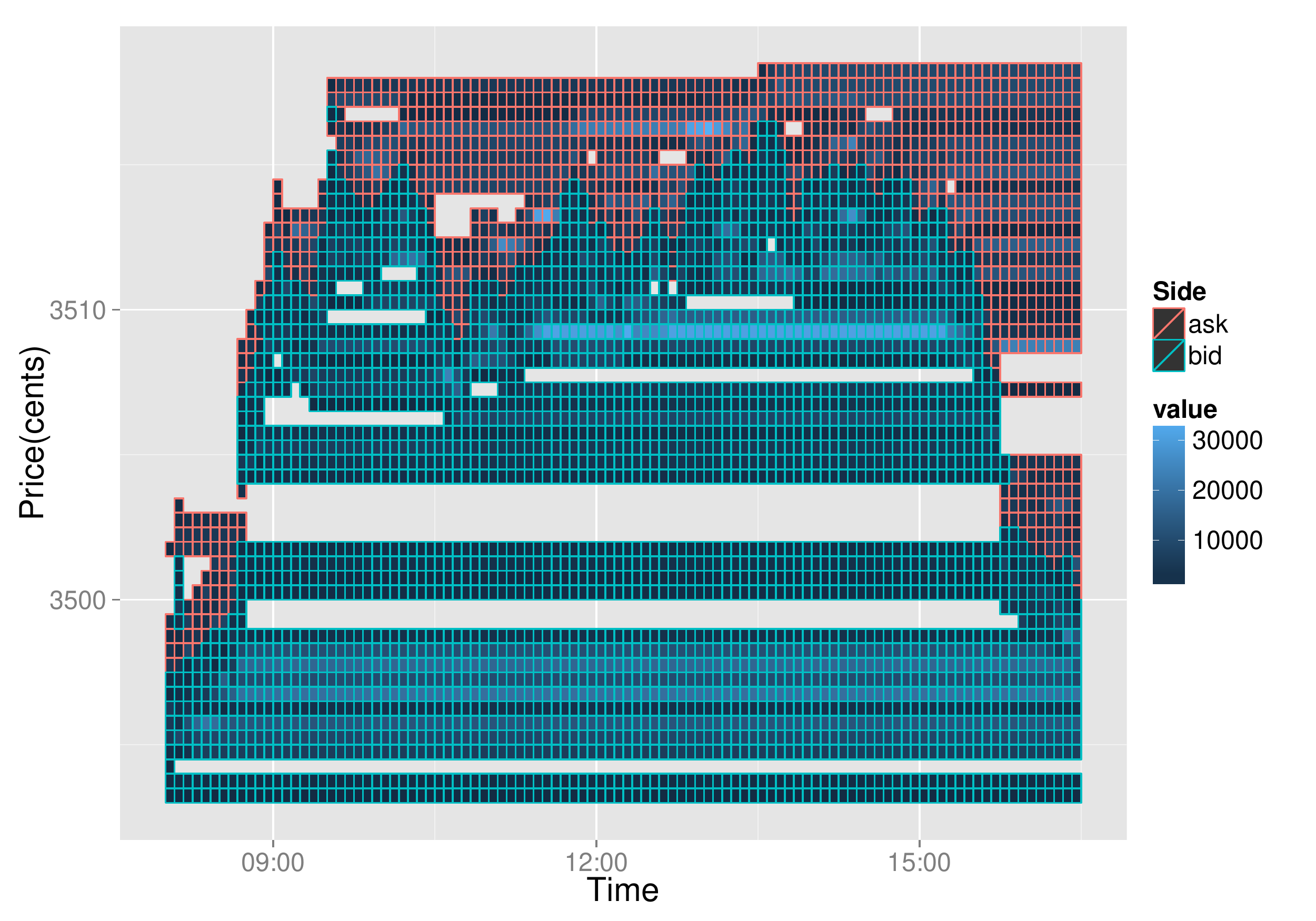}
\caption{Simulations using 2 of the non-dominated parameter vectors resulting from estimating the basic model with NSGA-II, but assume that order sizes are drawn from a mixture of gamma distributions.}
\label{fig:simsmixgamma}
\end{center}
\end{figure}

Figure \ref{fig:simsmixgamma} shows the intensity of the volume process and the evolution of the spread for a simulated trading day for 2 of these parameter vectors selected from the Pareto optimal front. Similarly to the reference model, the price and volume trajectories are still quite flexible between the different feasible, Pareto optimal solutions obtained for this calibration. 

\subsubsection{Introducing asymmetry and skewness to Limit Order intensity by depth}
In the reference model, we assumed that the skewness parameter vector $\bm{\gamma}$ for the multivariate skew-t distribution assumed for the number of limit orders and cancellations at each level of the LOB were fixed to a common skew. This parsimonious choice was encoded in the model by the reference model assumption $\boldsymbol{\gamma}^{LO,a}=\boldsymbol{\gamma}^{LO,b}=\boldsymbol{\gamma}=\gamma_0 \bm{1}$ and $\gamma^{MO}=\gamma_0$, i.e. there was only one skewness parameter which was common to all levels on both the bid and ask. The effect of this assumption on the price and volume dynamics in the reference model is now assessed by relaxing this feature and performing calibration of a relaxed version of the reference model to the same day of data from BNP Paribas. 

We now allow $\bm{\gamma}^{LO,a}=\bm{\gamma}^{LO,b}=\bm{\gamma}=\left \{  \right. \gamma^{LO,-l_d+1},\ldots, \gamma^{LO,l_p} \left.  \right \}=\bm{\gamma}^{C,a}=\bm{\gamma}^{C,b}$, in order to gain additional flexibility in modelling the skewness in the multivariate counts for limit order and cancellation data. We also allow $\gamma^{MO,a}=\gamma^{MO,b}=\gamma_0^{MO}$ to enable the skewness of the market order data to be modelled separately. This will entail estimating an additional $l_d + l_p$ parameters. Again, we assess whether the Pareto optimal solutions improve in minimizing the objective functions under this relaxation of the constraints in the reference model assumptions. 

\begin{table}[ht]
\caption{Non-dominated solutions for the model where the elements of the skewness vector are allowed to vary.}
\centering
\resizebox{0.99\linewidth}{!}{
\begin{tabular}{rrrrrrrrrrrrrrrr}
  \hline
 & $\mu_0^{LO,p}$ & $\mu_0^{LO,d}$ & $\mu_0^{MO}$ & $\gamma_0^{MO}$ & $\nu_0$ & $\sigma^{MO}_0$ & $\gamma_0^{LO,-2}$ & $\gamma_0^{LO,-1}$ & $\gamma_0^{LO,0}$ & $\gamma_0^{LO,1}$ & $\gamma_0^{LO,2}$ & $\gamma_0^{LO,3}$ & $\gamma_0^{LO,4}$ & $\gamma_0^{LO,5}$ & $\mathrm{Tr}(\Sigma)$ \\ 
  \hline
1 & 39.35 & 4.00 & 0.54 & -7.36 & 46.24 & 7.89 & 4.32 & -7.30 & 1.89 & -4.49 & -7.86 & 4.51 & 4.78 & -6.72  & 7.97 \\ 
  2 & 38.48 & 3.98 & 5.81 & -1.35 & 8.63 & 8.21 & 7.41 & 4.35 & 7.47 & -6.86 & -2.29 & 1.16 & 4.74 & -5.77  & 5.82 \\ 
  3 & 39.54 & 3.39 & 0.54 & -6.46 & 46.24 & 7.89 & 4.32 & -7.30 & 3.13 & -4.49 & -7.86 & 2.67 & 4.78 & -6.72  & 6.41 \\ 
  4 & 11.33 & 2.56 & 2.53 & 6.90 & 3.14 & 1.98 & -3.32 & -3.55 & -8.42 & -3.32 & -5.53 & -4.10 & -4.58 & 4.40  & 5.53 \\ 
  5 & 37.61 & 3.98 & 1.16 & -1.35 & 2.59 & 8.21 & 4.40 & -7.50 & -6.64 & -7.61 & -7.56 & 1.20 & 4.78 & -5.11  & 7.41 \\ 
  6 & 18.25 & 4.00 & 1.05 & -2.34 & 2.52 & 8.21 & -0.05 & -8.93 & -3.35 & -7.37 & -7.43 & 3.67 & 2.81 & -2.61  & 5.67 \\ 
  7 & 13.40 & 5.97 & 6.14 & -1.25 & 22.02 & 8.69 & 6.11 & -1.65 & -6.36 & -8.16 & -2.75 & 3.34 & 8.76 & 6.81  & 6.42 \\ 
  8 & 39.35 & 4.00 & 0.71 & -6.44 & 5.67 & 2.25 & -3.25 & -7.34 & 1.89 & -4.47 & -7.86 & 4.51 & 4.78 & -7.18  & 4.42 \\ 
   \hline
\end{tabular}
}
\label{tab:estskew}
\end{table} 

Table \ref{tab:estskew} shows that in none of the parameter vectors produced by the multi-objective II estimation method are the elements of the skewness vector close to being equal to one another, which indicates that the use of the skew vectors with different skew at each level of the LOB for the bid and ask, in the Multivariate Skew-t distribution, is appropriate for the calibration to real data. As expected, incorporating these features improves the model power and suitability, measured by the objective function values achieved by the solutions in the Pareto optimal front, for the simulated stochastic agent LOB model realizations, when compared to the reference model. 

\begin{figure}
\begin{center}
\includegraphics[width=0.5\textwidth]{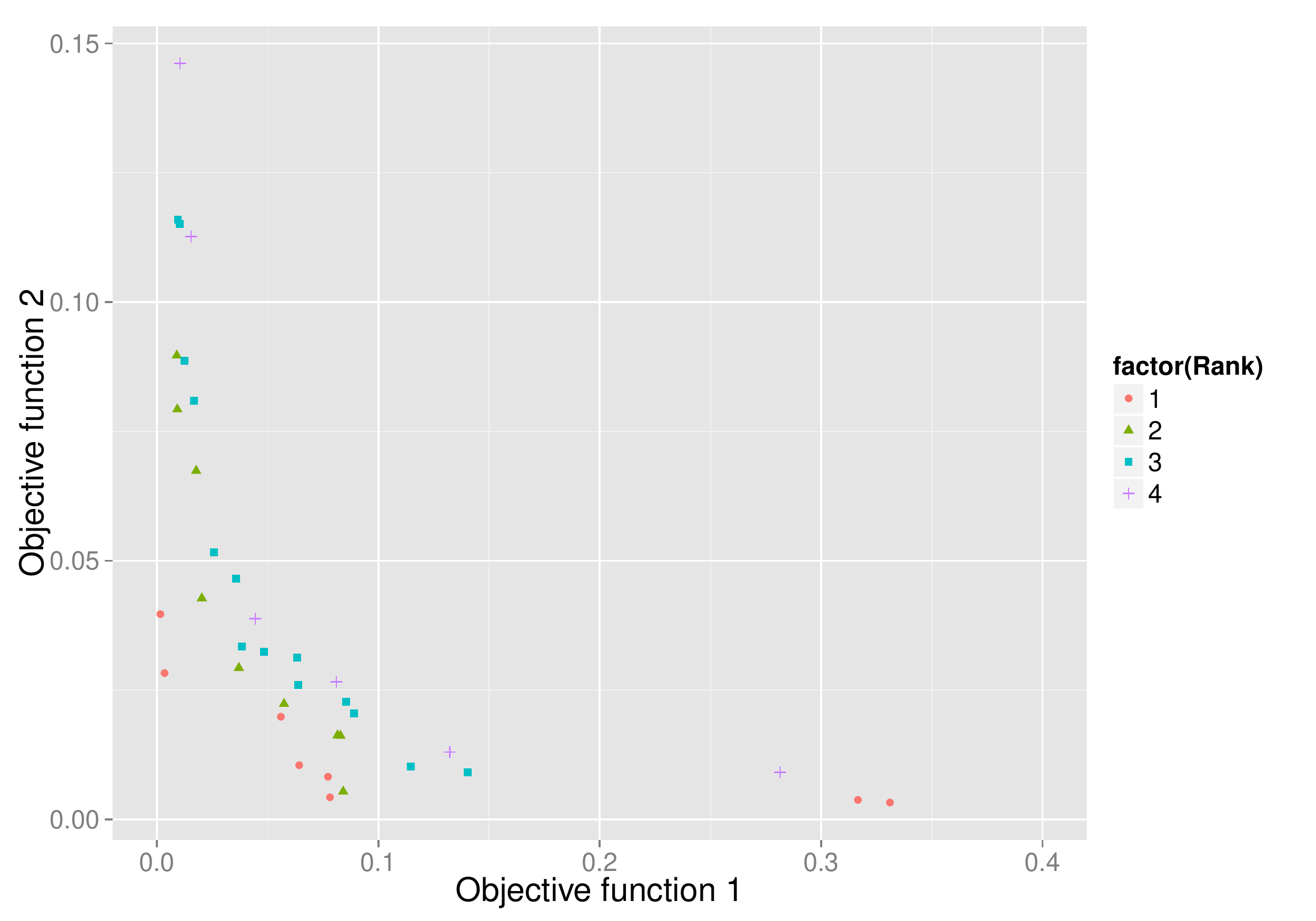}
\caption{Objective function values for the parameter vectors produced by the multi-objective II method, in the case where we relax the assumption that the elements of the skewness vector in the Multivariate Skew-t distribution are equal.}
\label{fig:skewobjvalue}
\end{center}
\end{figure}

\begin{figure}
\begin{center}
\includegraphics[width=0.45\textwidth]{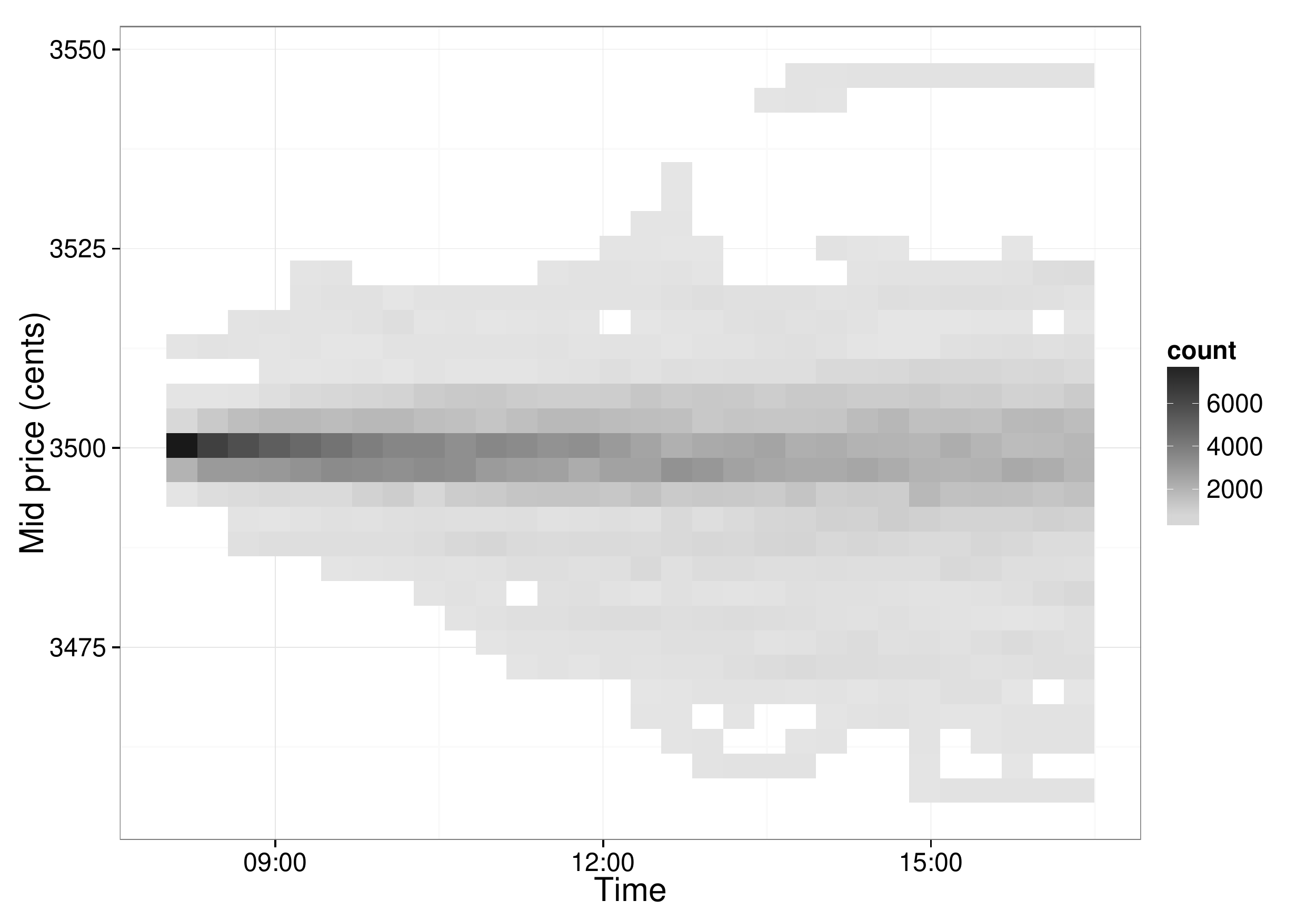}
\includegraphics[width=0.45\textwidth]{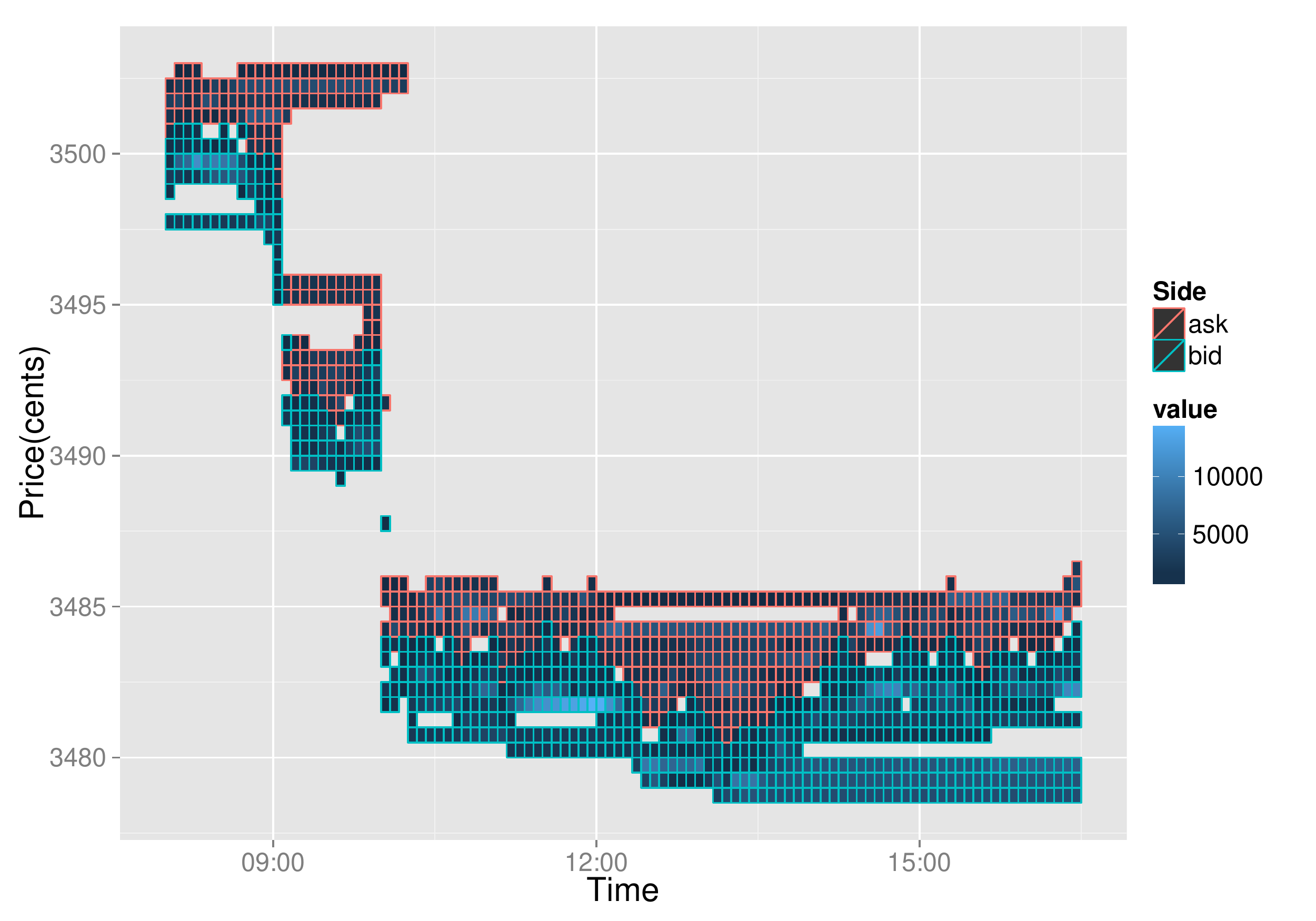}
\includegraphics[width=0.45\textwidth]{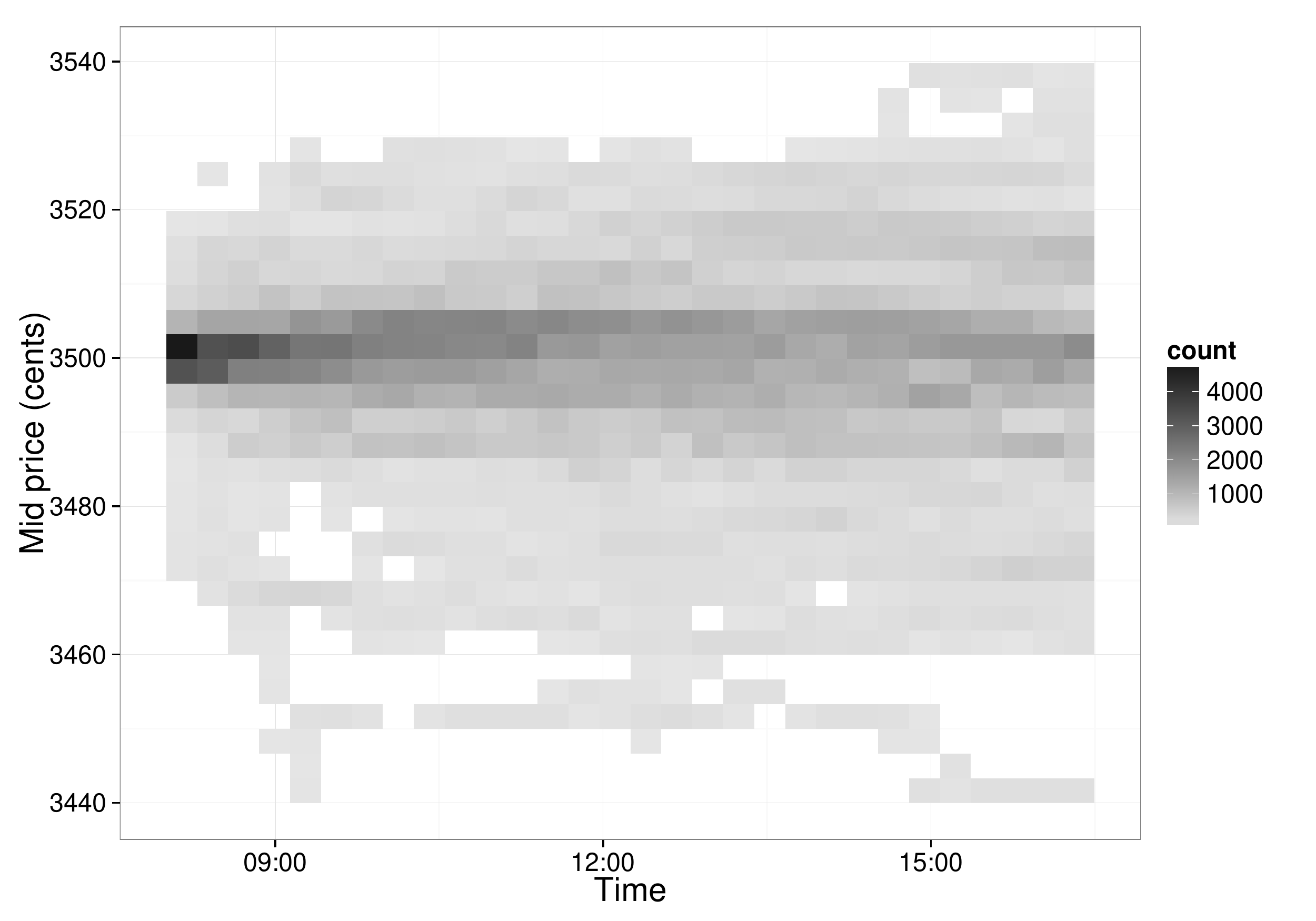}
\includegraphics[width=0.45\textwidth]{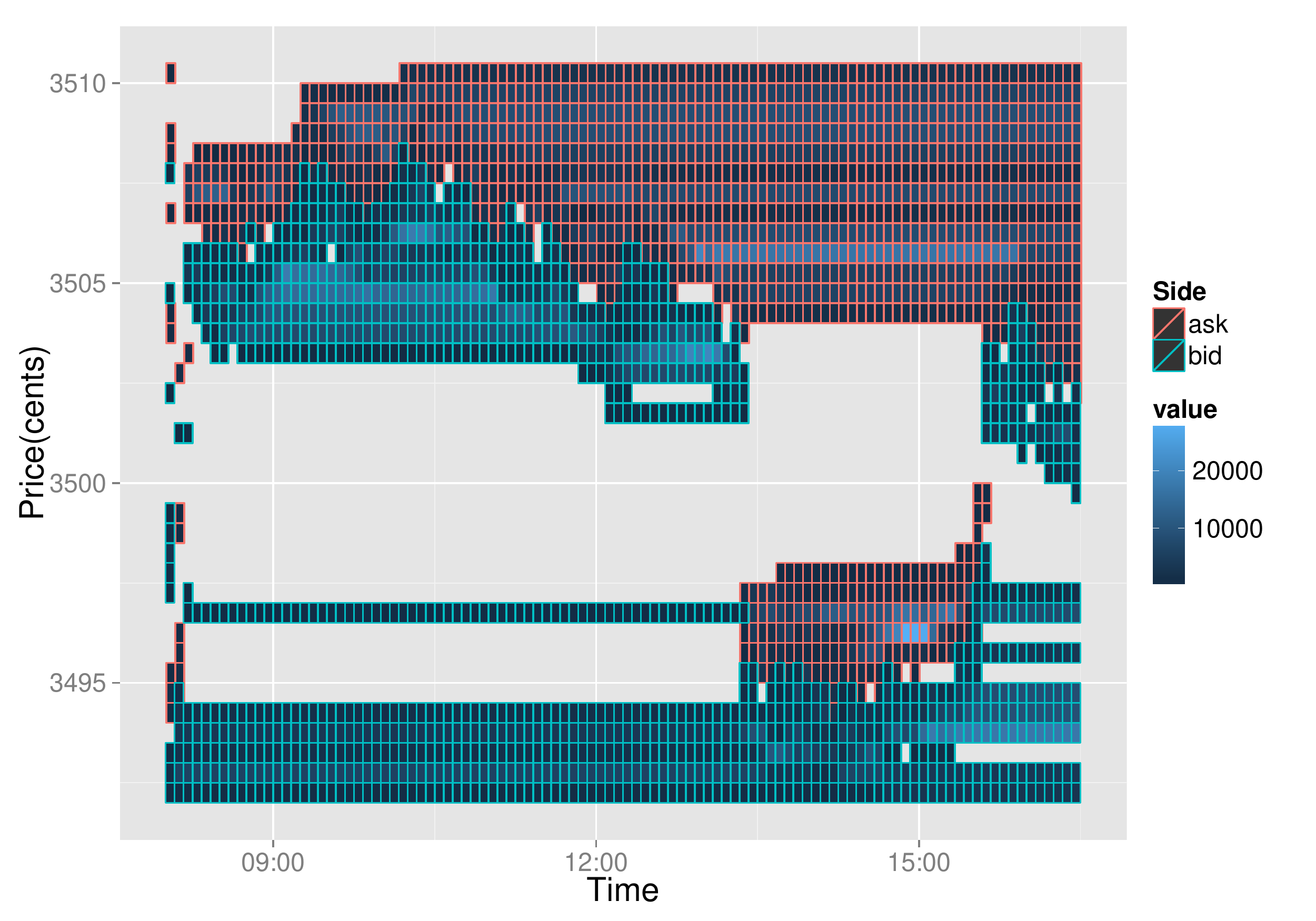}
\caption{Simulations using 2 of the non-dominated parameter vectors resulting from estimating the basic model with NSGA-II, but relaxing the assumption that the elements of the skewness vector in the Multivariate Skew-t distribution are equal.}
\label{fig:simskew}
\end{center}
\end{figure}

\section{Regulatory interventions via the SR-ABM stochastic LOB agent model}
\label{sec:reg}

In building our stochastic agent-based LOB simulation model, we were motivated by the increasing desire of regulators, exchanges and brokerage houses to better understand the role of intervention in electronic exchanges. In this regard, there have been a sequence of new regulations being instigated throughout Europe and the US to further manage the processing, placement and clearing of trades in electronic exchanges\footnote{These regulations, which in Europe fall under the `Lamfalussy Directives' include the Prospectus Directive, the Market Abuse Directive, the Transparency Directive and the Markets in Financial Instruments Directive (MiFID)}. 

The the Markets in Financial Instruments Directive (MiFID) aims to develop a harmonised regulation for investment services across the 31 member states of the European Economic Area. 
Several components of MiFID can be better understood by the type of analysis we undertake in this paper. For instance, one aspect pertaining to the brokerage hoses involves the key aspect of this directive known generically as `Best Execution' practice\footnote{MiFID’s best execution regime is set out as follows in the Directives. Article 21 of Level 1  and Articles 44 and 46 of Level 2 set out the requirements for investment firms that provide the  service of executing orders on behalf of clients for MiFID financial instruments and, indirectly via  Article 45(7), for investment firms that provide the service of portfolio management, when  executing decisions to deal on behalf of client portfolios.}. Under this feature of the directive, MiFID will require that firms take all reasonable steps to obtain the best possible result in the execution of an order for a client. The best possible result is not limited to execution price but also includes cost, speed, likelihood of execution and likelihood of settlement and any other factors deemed relevant. As is clearly evident, this directive therefore speaks directly to liquidity in the LOB and the need to develop a better understanding of which features and market behaviours by agents in the market affect liquidity either in volume or price. An intrinsic part of this process is the consideration of volumes at different levels of the LOB.

In addition to developing a better understanding of the LOB stochastic process, regulators also have an important role to play in trying to determine how best to manage certain types of potentially undesirable market behaviours by agents. In this regard, we refer to behaviours that may be disruptive, cause excess volatility in price or illiquidity throughout the trading day in given asset's LOB. 

\subsection{Related ABM studies of regulatory interventions}

The introduction of MiFID has increased competition and allowed for the trading of stocks in pan-European multilateral trading facilities (MTFs). The trading on one venue will undoubtedly affect the trading interest in another, through the activity of cross-market arbitrageurs. In addition, there is the possibility that regulation can be imposed on one market, but not another, which will have implications for the efficacy of the regulation itself. Both \cite{mannaro2008using} and \cite{westerhoff2006effectiveness} have considered this in the context of an ABM, but with simpler models than the one considered here, which do not take into account the liquidity considerations of the agents. We extend these studies using the stochastic agent-based LOB simulation model developed in this manuscript.

In contrast to the ABM model \cite{westerhoff2003heterogeneous}, set up to study the effect of a transaction tax in a financial market, in our model the agents' strategy is not dependent on profitability. This is because of the division of our trading agents according to their liquidity considerations: Traders often consume liquidity due to considerations other than profit, such as rebalancing the weights of their holdings in a fund. They cannot simply choose to become liquidity providers because of the superior profitability of these agents, for a number of reasons. These include the investment in technology required to be able to carry out such a strategy in the millisecond environment, the inventory they will be required to hold, and, possibly, regulatory or exchange obligations they will have to adhere to.

\subsection{Quote-to-trade ratio}
The intervention we will consider here, as an example of the type of experiment that can be performed using our model, is the imposition of a quote-to-trade ratio. This ratio is already considered in certain exchanges, such as the LSE, which allows for 500 quotes per trade. Further quotes are allowed in the case of the LSE, but are subject to a 5 pence surcharge for every order\footnote{\url{http://www.londonstockexchange.com/products-and-services/trading-services/pricespolicies/tradingservicespricelisteffective2december2013.pdf}}. In our model, we have made the assumption that the baseline limit order submission (or quote) intensity at every level $\mu^{LO,a,i}_0$ is equal to the baseline cancellation intensity $\mu^{C,a,i}_0$. That is, potentially all orders submitted in an interval can be cancelled prior to execution. 

Given the setup of our model, it is more convenient to enforce a stochastic limitation for excessive trading, rather than a hard limit of (say) 100 limit orders to 1 market order. For a quote-to-trade ratio $q=\frac{100}{1}$, we impose the limit by specifying that for the cancellation activity $\mu^{C,a,i}_t = (1-\frac{1}{q})\mu^{LO,a,i}_t$. This is an approach similar in concept to that taken by \cite{ait2013high}, who, rather than enforcing a strict minimum resting time of 500 milliseconds, instead subject every order to a random minimum resting time that is exponentially distributed, but with the same mean. 

\begin{figure}
\begin{center}
\includegraphics[width=0.42\textwidth]{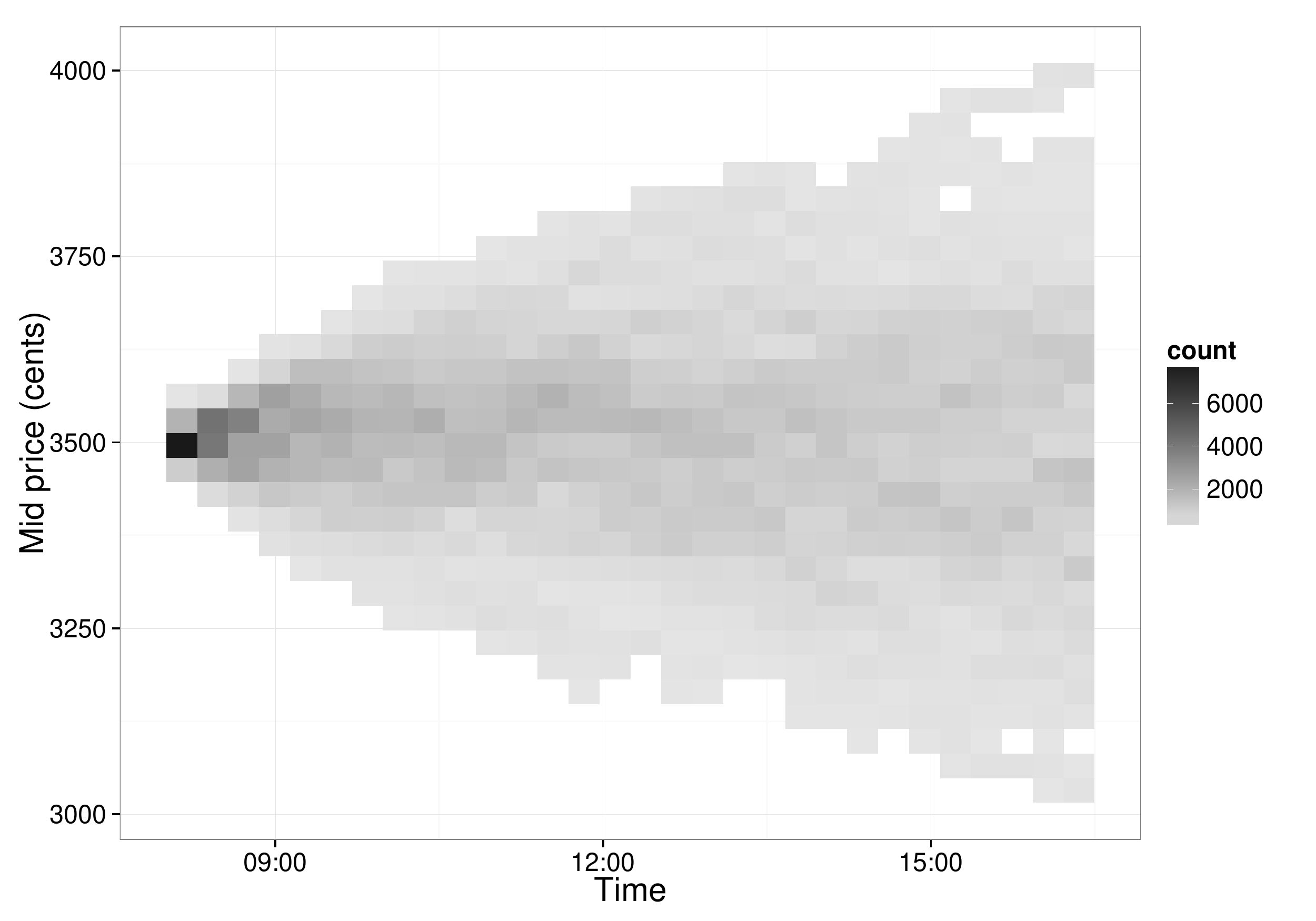}
\includegraphics[width=0.42\textwidth]{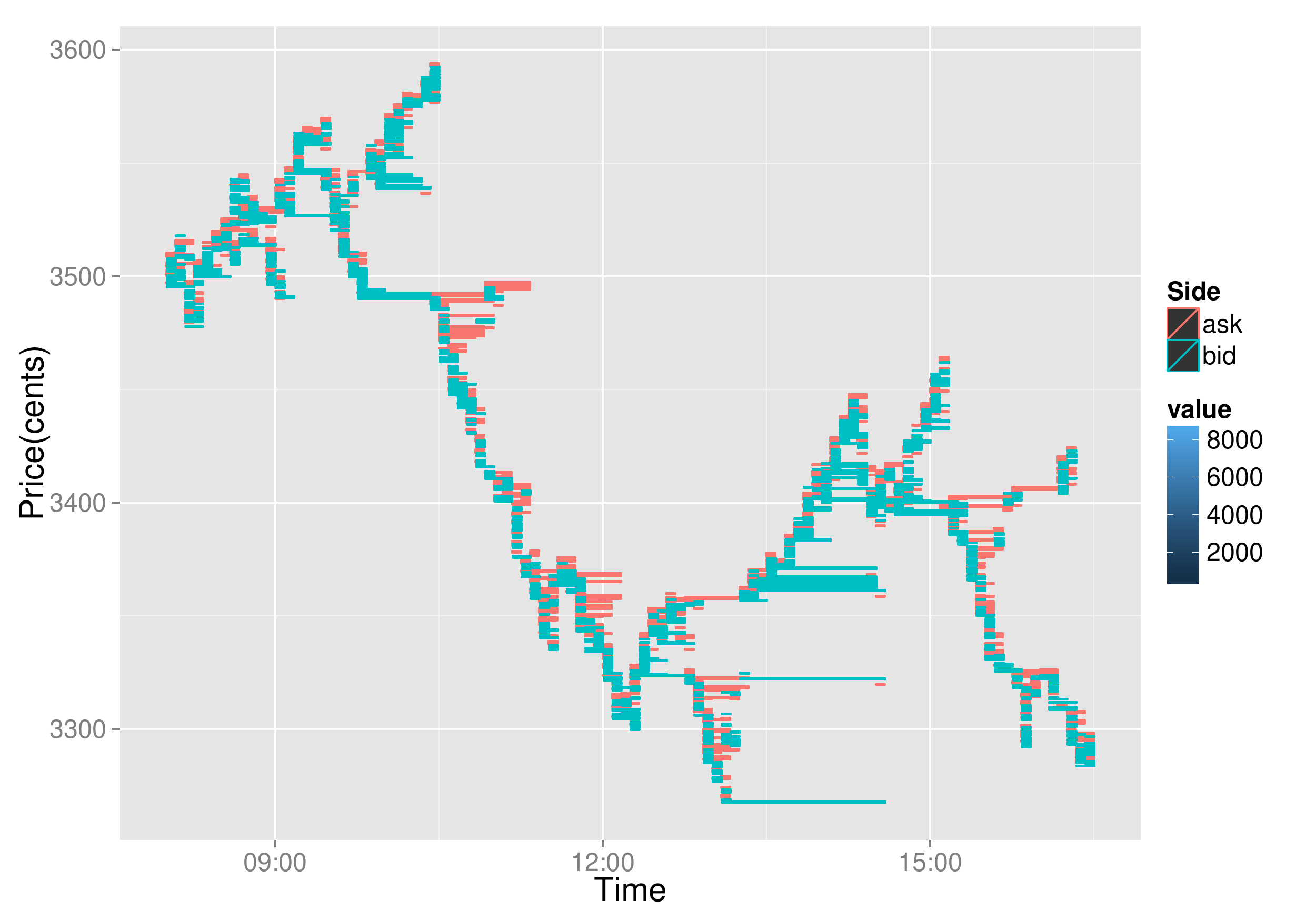}
\includegraphics[width=0.42\textwidth]{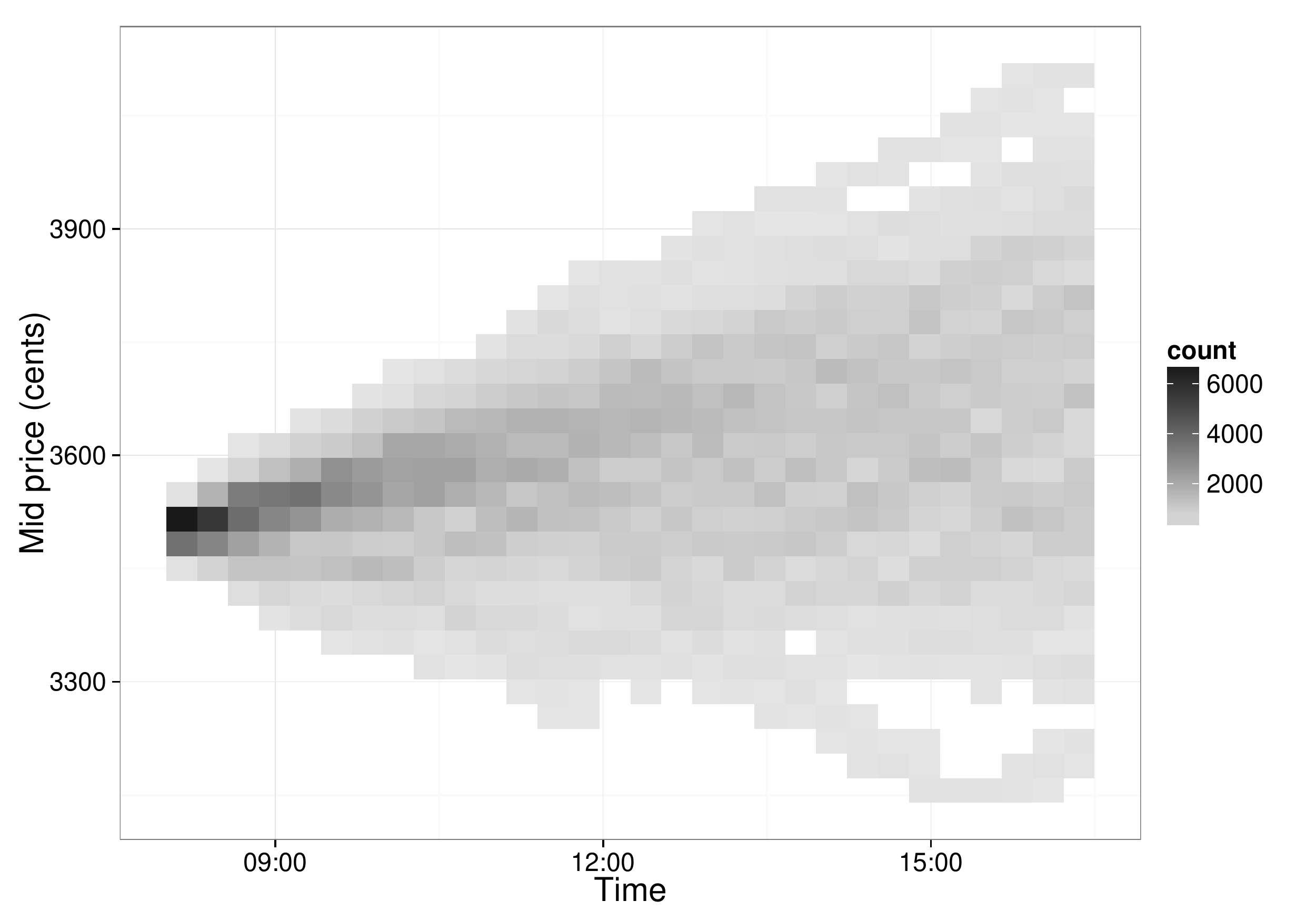}
\includegraphics[width=0.42\textwidth]{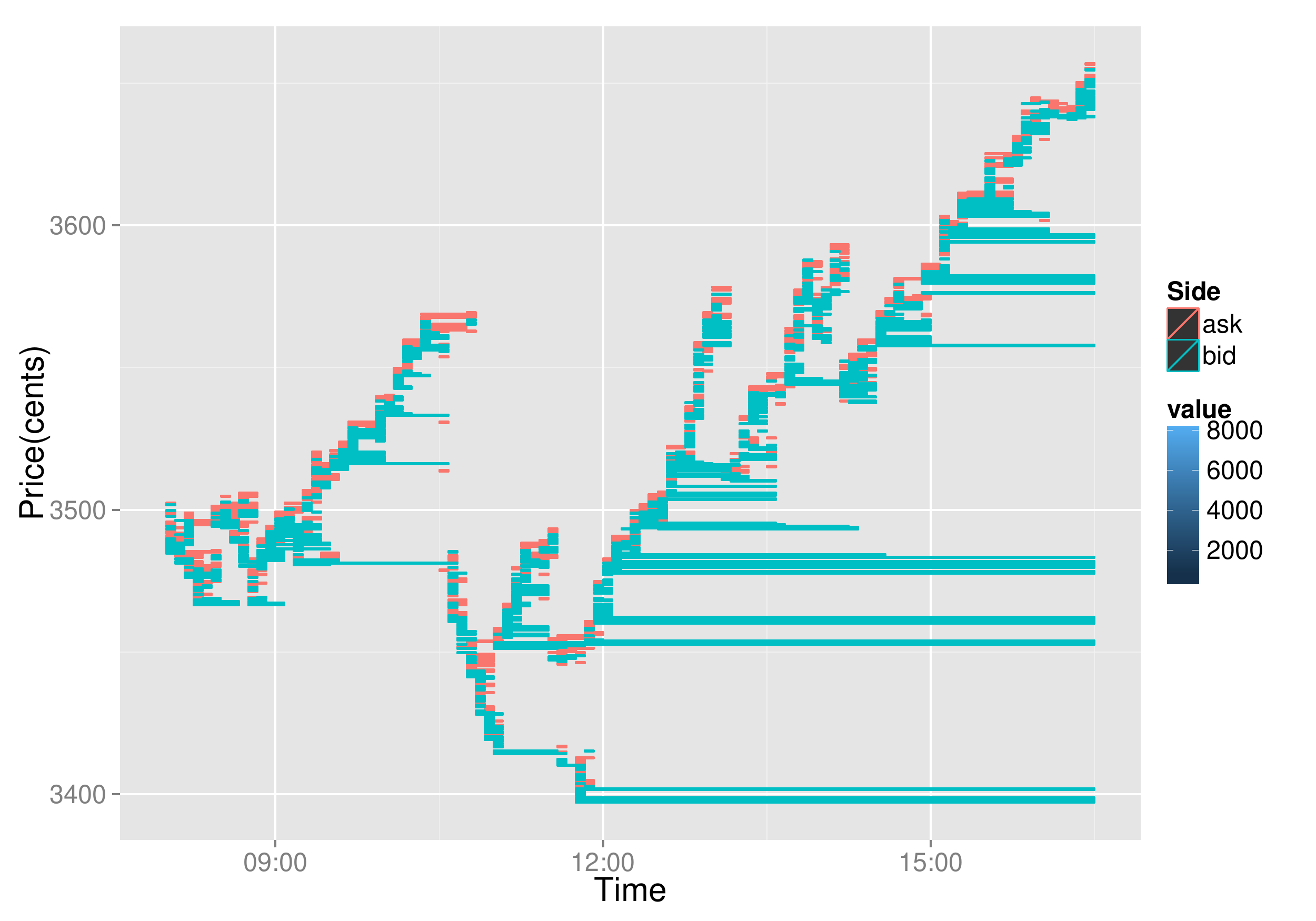}
\includegraphics[width=0.42\textwidth]{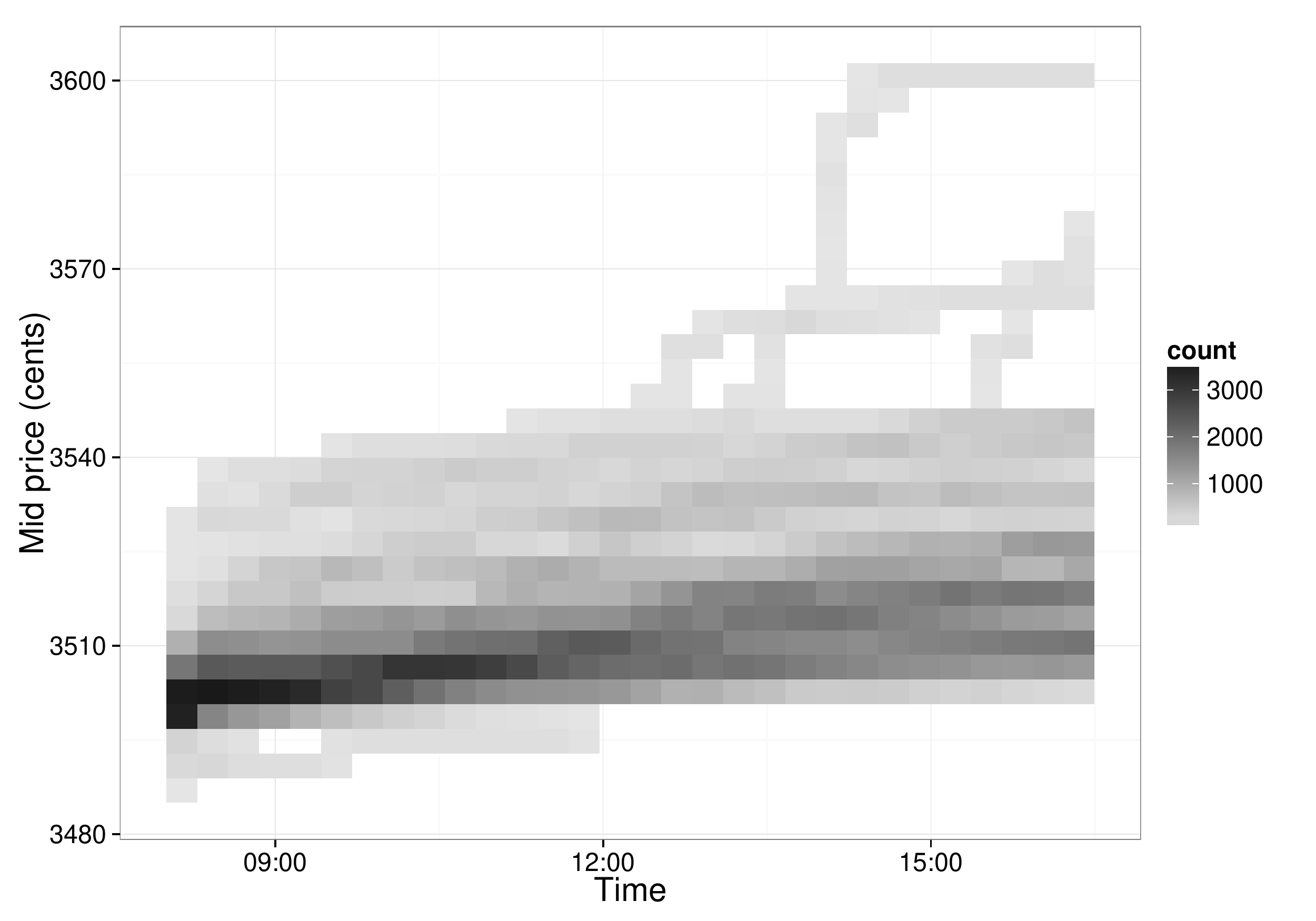}
\includegraphics[width=0.42\textwidth]{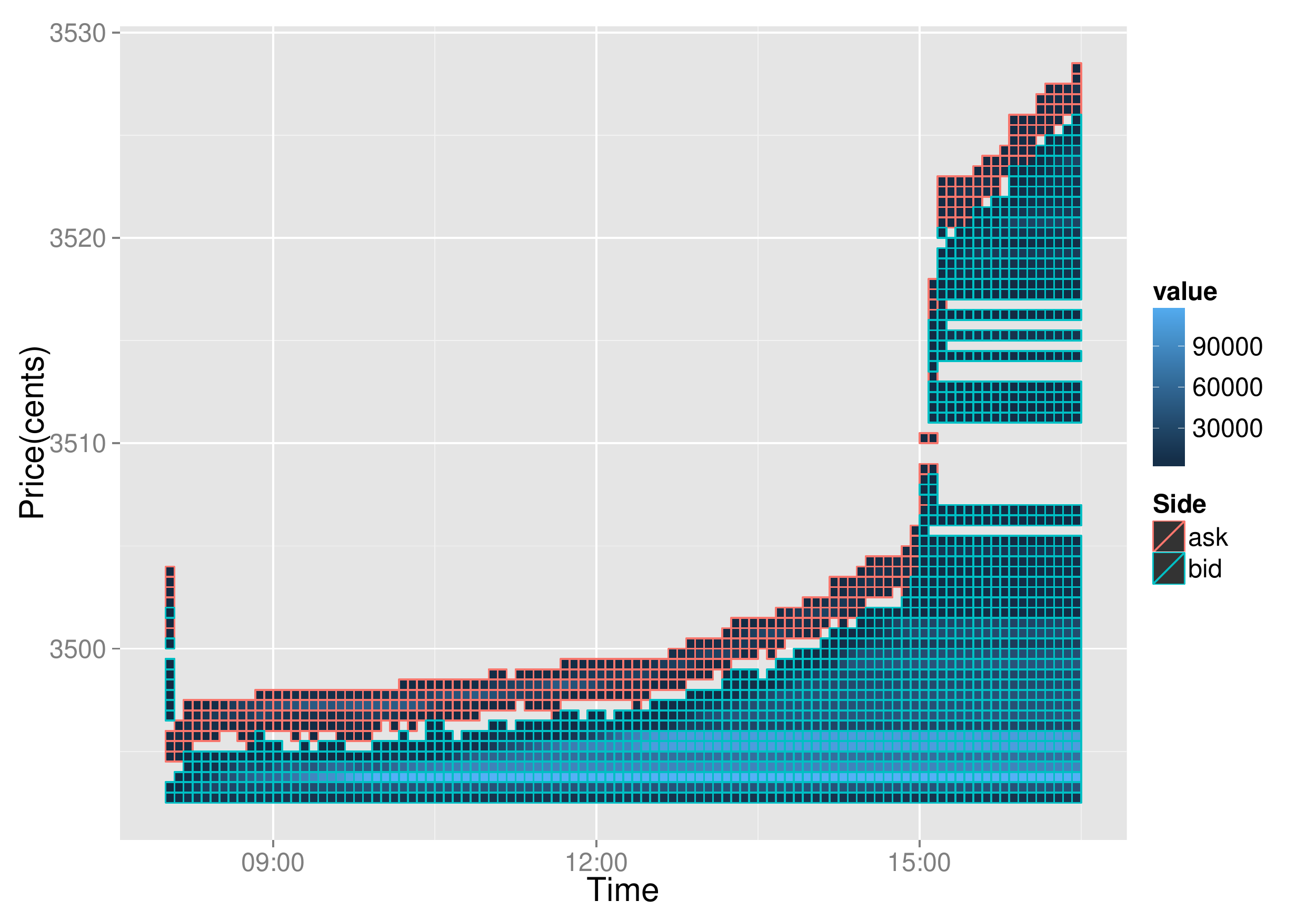}
\caption{Simulations of the basic model, with the addition of a `quote-to-trade ratio' regulatory intervention. The mid-price process and daily LOB volumes with a quote-to-trade ratio of $q=\frac{500}{1}$ (top), $\frac{100}{1}$ and $\frac{20}{1}$ (bottom). }
\label{fig:simsreg}
\end{center}
\end{figure}

We evaluate the outcome of such an intervention in our simulated LOB for 3 different quote-to-trade ratios, i.e. $q \in \left\{ \frac{500}{1}, \frac{100}{1}, \frac{20}{1} \right\}$. Figure \ref{fig:simsreg} shows the effect of the regulation on individual realisations of daily activity, as well as the price process in repeated realisations. We have chosen one of the parameter vectors from the estimation of the basic model which generally showed excessive volatility. We note that, in our model, increasing $q$ (and thus, reducing the relative number of cancellations) has the effect of constraining the mid-price process, and thus, curbing excess volatility. 

While one cannot draw definite conclusions about the effect of such an intervention through an ABM simulation, it is a step a regulator may consider, particularly when comparing different approaches. For example, even in the implementation of a quote-to-trade ratio, the regulator may have a number of choices, for example, regarding the period over which they consider the ratio. We argue that our model can be informative for such considerations, and, given its flexibility, can give rise to a large number of computational experiments and scenario analysis studies that will better inform policy makers of the impact their policies may have on the market behaviours of traders.   

\section{Conclusion}
\label{sec:conc}
We have presented a new form of agent-based model, in order to capture features of the complex stochastic process that is the Limit Order Book. The agent types we considered are representative of the classes of market participants in modern financial markets: In electronic LOBs, traders can be broadly separated according to their liquidity requirements, into liquidity providers and liquidity demanders. This is certainly more representative of the motivation for trading activity, compared to the chartist and fundamentalist models considered in the past (e.g. \cite{frankel1988chartists,kirman1993ants,farmer2002price,westerhoff2003nonlinearities,manzan2007heterogeneous}). 

We have modelled the activity resulting from the entire class of agents, which has enabled us to directly model the dependence in event (limit order submission, cancellation and market order) activity between the different levels of the LOB, which would not have been possible by considering simpler formulations for individual agent strategies. We have employed a flexible Multivariate Skew-t model for the event intensities, which is unique for its ability to capture asymmetric and heterogeneous dependence, and its scalability in high dimensions \citep{demarta2005t,fung2010tail}. This has resulted in a very general formulation of the ABM, which also enables one to model the heterogeneity in order sizes. 


In our estimation of the model, we proposed an extension to standard simulation-based approaches, considering multiple auxiliary models (relating to the price and volume processes) in a multi-objective problem. 
We developed a novel Indirect Inference multi-objective optimisation method which uses the concepts of stochastic ordering and Pareto optimality to select most suitable candidate parameter vector solutions when calibrating the stochastic agent LOB model.

We have shown that even a parsimonious, baseline version of the model, which assumes fixed order sizes and no heterogeneity in the skewness of the distribution intensities for limit order placements and cancellations, is still able to generate produce a range of plausible LOB stochastic dynamic behaviours. Relaxing the baseline model assumptions, however, generally leads to an improvement in the model estimates, in terms of their ability to produce simulations that closely reflect the price and volume dynamics observed in real data on a typical, i.e. non-eventful in terms of shocks or liquidity droughts, trading day.

The flexible LOB framework presented here, coupled with the proposal of a new simulation-based estimation method is an important contribution towards LOB modelling. We have proposed a model that can capture rarely studied LOB features, such as the dependence in the intensity of LOB activity at different levels. In addition, we have shown that the model is sufficiently detailed, such that one can use it as a testbed for proposed regulation. We demonstrated that a sufficiently high stochastic limitation on the number of cancellations, which would be similar to the imposition of a quote-to-trade ratio, can reduce excessive volatility in our simulated LOB.

\vspace{2em}

\section*{Acknowledgements}

EP acknowledges the support of Prof. Mark Harman for discussions on agent-based modelling and initial work on calibration through multi-objective optimisation. EP and GWP acknowledge the support of the Institute of Statistical Mathematics, Tokyo, Japan and Prof. Tomoko Matsui for support during a work on this paper during a research visit. EP and GWP also acknowledge the Systemic Risk Center, LSE for extended discussions during this manuscript. GWP acknowledges the support of the Oxford Mann Institute, Oxford University.


 

\bibliography{all.bib}

\newpage
\appendix

\section{Adaptive genetic evolutionary search for multi-objective optimisation}
\label{app:evosearch}

A search strategy is also required to explore the parameter space in seeking Pareto optimal sets of parameters for the agents, i.e. liquidity provider and liquidity demander parameter vectors in the stochastic LOB model. In this regard, one may consider a multi-objective evolutionary algorithm (MOEA) framework. Such approaches have been the focus of extensive study over the past 15 years (see, e.g. \cite{zhou2011multiobjective}, \cite{eiben2003introduction}, and references within) and would be particularly applicable to the problem at hand. There are several reasons for their popularity: they are inherently parallel, they feature operators to combine and mutate candidate solutions to rapidly arrive at improved solutions and are able to capture multiple Pareto-optimal solutions during the optimisation \citep{zitzler2000comparison}, which can be spread out across the Pareto front.  In addition, there has been recent advances to better understand the relationship between such optimisation search frameworks and stochastic genetic search methods, see for instance discussions in \cite{emmerich2013evolve}. In this paper, we explore the utilisation of adaptive mutation kernels in the simulation based Multi-objective-II framework to efficiently explore the parameter space, where our approach merges traditional genetic search algorithms with adaptive Markov kernels utilised in adaptive MCMC methods, such as those studied in \cite{haario2006dram}, \cite{roberts2009examples} and \cite{andrieu2006ergodicity}.

The MOEA used in this paper is based on the NSGA-II (Non-dominated Sorting Genetic Algorithm II), developed by \cite{deb2002fast}. This is an \textbf{elitist} MOEA, and in every iteration, combines the best parent solutions with the best offspring to produce a new family of candidate solutions. It produces a diverse \textbf{Pareto-optimal} front (i.e. the solutions are well-spread out across the front, due to the algorithm's use of a crowding distance operator) with low computational requirements ($O(mN^{2})$ computational complexity, where $m$ is the number of objectives, and $N$ is the population size). 

The algorithm is perhaps the most popular MOEA and is frequently used as a performance benchmark for other algorithms \citep{coello2007evolutionary}. It has been used in various applications, including the generation expansion planning problem in power systems \citep{kannan2009application} and for balancing objectives in groundwater monitoring designs \cite{reed2004striking}. In addition, it has been been further developed in a Bayesian setting, in order to solve discrete multi-objective decomposable problems (see, e.g. \cite{khan2003bayesian,laumanns2002bayesian,khan2002multi}). Within this algorithm, we extend the features by also incorporating an adaptive global and local mutation kernel for a subset of the stochastic agent-based LOB model parameters $\bm{\theta}$. We first present an overview of the optimisation algorithm structure:
\begin{enumerate}
\item{First, a family, or population, of $N$ candidate solutions is initialised randomly from the feasible region.
}
\item{For each solution, the objective functions are calculated and a rank is obtained reflecting Pareto dominance. That is, solutions are sorted into fronts, with the first front consisting of solutions that are not dominated by any other solutions, the second consisting of solutions that are only dominated by a single solution, and so on.  Solutions are also assigned a crowding distance value, indicating the Euclidean distance from other solutions on the same front. 
}
\item{From this family of solutions, the crowding comparison operator is applied, and chooses the best solutions according to their rank, and in the case of ties, according to the crowding distance value.
}
\item{Then, one or more evolutionary operators (detailed in the following section) are applied to evolve the selected set of solutions.
}
\item{The new solutions are combined with the current family of solutions and the process is repeated from the second step, for a set number of iterations.
}
\end{enumerate}

The algorithm outputs the non-dominated set of solutions with the highest ranking. We provide details about the operators used in multi-objective Indirect Inference procedure in the following section. \\

\subsection{Algorithm settings and evolutionary operators} 
Details of a large number of evolutionary operators used in MOEAs can be found in \cite{coello2007evolutionary}. In NSGA-II, one has to first select the size of the population of candidate solutions for every iteration of the algorithm, in addition to the number of iterations (called generations in  the MOEA nomenclature). In our optimisation, we use a population size of $N=40$ parameter sets, and run the optimisation for a total of 40 generations. 

We referred to a number of operators used to evolve and choose amongst the set of solutions, and we provide further information here about their function:
\begin{itemize}
\item{\textbf{Selection operator:} From the second iteration of the algorithm onwards, there will be $2N$ sets of candidate solutions in step 3. The best $N$ solutions are chosen based on a) dominance and b) crowding distance, or the distance of the solution from its neighbours. If the number of solutions on the first front is less than $N$, they are all selected, and the remainder are taken from further fronts. In the case where one must select fewer solutions than the number of solutions on a particular front, the solutions with the highest crowding distance value are chosen.  
}
\item{\textbf{Crossover operator:} The Simulated Binary Crossover (SBX) operator is used. From two candidate solutions $\theta_1,\theta_2$, two new solutions $\theta'_1,\theta'_2$ are formed, where the $k$-th elements are as follows:
\begin{align}
\theta'_{1,k} &=\frac{1}{2}[(1-\alpha_k)\theta_{1,k}+(1+\alpha_k)\theta_{2,k}]\\
\theta'_{2,k} &=\frac{1}{2}[(1+\alpha_k)\theta_{1,k}+(1-\alpha_k)\theta_{2,k}]
\end{align}
Here, $\alpha_k$ is a random sample from a distribution with density 
\begin{equation*}
p(\alpha) = \begin{cases} \frac{1}{2}(\eta+1)\alpha^{\eta_c} &\mbox{if }0<\alpha\leq 1  \\ 
\frac{1}{2}(\eta+1)\frac{1}{\alpha^{\eta_c+2}} & \mbox{if } \alpha>1 \end{cases} 
\end{equation*}
We use the crossover operator with probability 0.7 and a distribution index $\eta_c=5$.
}
\item{\textbf{Mutation operator:} The polynomial mutation operator is used. The mutation operator perturbs elements of the solution, according to the distance from the boundaries. 
\begin{equation*}
\theta'_{k}=\theta_{k}+\delta_k(\theta_{k_{U}}-\theta_{k_{L}})
\end{equation*}
where we have for $\delta_k$
\begin{equation*}
\delta_k=\begin{cases} (2\gamma_k)^{\frac{1}{\eta_m+1}}-1 &\mbox{if }\gamma_k<0.5  \\ 
1-[2(1-\gamma_k)]^{\frac{1}{\eta_m+1}} & \mbox{if } \gamma_k \geq 0.5 \end{cases} 
\end{equation*}
Here, $\gamma_k$ is uniformly distributed on $(0,1)$ and the distribution index $\eta_m=10$. The polynomial mutation operator is used with probability 0.2.
}
\end{itemize}
\noindent \textbf{Covariance matrix mutation and sampling:} 
The NSGA-II algorithm discussed above is only able to produce binary, integer, or real encodings for the output solution vectors. However, the stochastic process for the limit order submission activity by liquidity providers requires the specification of a positive definite and symmetric covariance matrix for the generation of intensities from a multivariate skew-t distribution. We cannot naively extend the evolutionary operators above (crossover and mutation) to produce new sets of covariance matrix candidate solutions which guarantee that the positive definiteness and symmetry constraints of the covariance matrix are preserved.  We thus propose an extension to the MOEA, effectively another operator that will generate candidate solutions for the covariance matrices, such that every new generation remains in the manifold of positive definite matrices. This operator will generate new candidate covariance matrices once the evolutionary operators discussed previously have been applied. 

To ensure that the optimisation algorithm searches the space of feasible solutions efficiently and does not get stuck in a suboptimal region of the space of possible solutions, our covariance matrix sampling operator has two components to undertake exloration and exploitation type moves. The mutation kernel is comprised of a mixture of Inverse Wishart distributions with different parameters, as per the proposal of \cite{peters2012copula}, one mixture component to provide global search (exploration) and a second mixture component to provide local searches (exploitation). To do this efficiently, it is based on an adaptive learning strategy for the specification of the local mixture component. In this case, the algorithm will explore the local region with high probability, but make potentially larger moves with smaller probability. 

We now describe one complete covariance mutation step. In the $n$-th generation of the MOEA, we generate $\left \{ \Sigma_{n,i} \right \},i=1\ldots N$ from a mixture distribution $q(\Sigma_{n,i})$ defined as follows:
\begin{equation*}
q(\Sigma_{n,i})=(1-w_1)\mathcal{IW}(\Psi_n,p_1)+w_1 \mathcal{IW}(\Psi,p_2)
\end{equation*} 
where $p_1,p_2$ are degrees of freedom parameters with $p_2<p_1$, and where $w_1$ is small so that sampling from the second distribution happens infrequently. Here $\Psi$ denotes an uninformative positive definite matrix, with the effect that sampling from the second distribution leads to moves away from the local region being explored. $\Psi_n$ is also a positive definite matrix, fitted based on moment matching to the sample mean of the successfully proposed candidate solutions in the previous stage of the Multi-Objective optimisation as follows:
\begin{equation*}
\Psi_n=\frac{1}{\sum_{t=1}^n w^{t}} \sum_{t=1}^n w^{t} \frac{1}{\sum_{i=1}^N \frac{1}{r_{t,i}}} \sum_{i=1}^N \frac{1}{r_{t,i}}  \tilde{\Sigma}_{t,i}
\end{equation*} 
where $r_{t,i}$ is the non-domination rank of the $i$-th solution in the $t$-th generation, and $w^t$ with $w<1$ is an exponential weighting factor. 

\section{Further results}
\label{app:results}

In Sections \ref{subsec:calib} and \ref{subsec:calib} we presented results for the calibration of the reference model and models where we relaxed certain assumptions, respectively. This calibration was performed using the data from a single asset (BNP Paribas) over one day, in order to be able to present detailed results regarding objective function values, LOB evolution over individual simulations using individual solutions on the Pareto front, as well as summaries of repeated simulations. In this section, we repeat the calibration of the reference model for 5 assets (BNP Paribas, Credit Agricole, Total SA, Technip SA and Sanofi) every trading day between 01/02/2012 and 21/02/2012. The stocks were chosen from the French CAC40 stocks, and are therefore amongst the most liquid stocks in the country. Specifically, we chose assets that are representative of different industries (banking, energy and pharmaceutical) and have different ticksizes (minimum price increments) and market capitalisations, as these are factors that affect daily trading activity.  

We summarise the results as follows: We first calibrate the reference model for each day and each asset individually, from which we obtain a set of $J$ solutions (i.e. non-dominated solutions on the Pareto front) every time. For each solution (parameter vector $\hat{\bm{\theta}}_j, j \in 1 \ldots J$), we simulate the LOB model $N$=50 times and fit the auxiliary models to the simulated data to obtain $N$ auxiliary model parameter vectors $\beta^{i,j,\ast}_1$ and $\beta^{i,j,\ast}_2,i \in 1 \ldots N$. The former are the ARIMA model parameters fit to the volume process on the bid and ask side, and the latter are the GARCH model parameters fit to the log returns.   
   
\begin{figure}
\begin{center}
\includegraphics[width=0.49\textwidth]{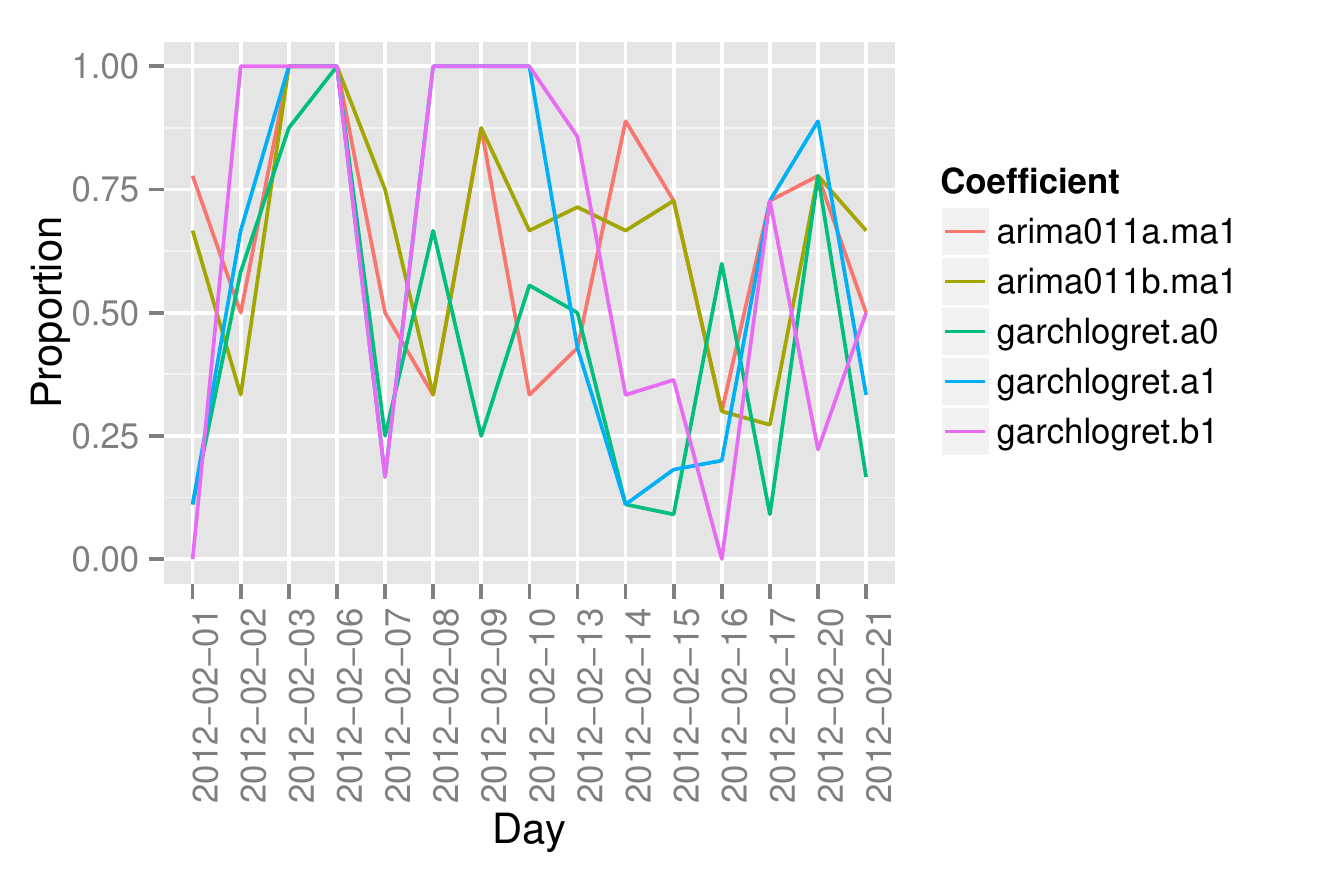}
\includegraphics[width=0.49\textwidth]{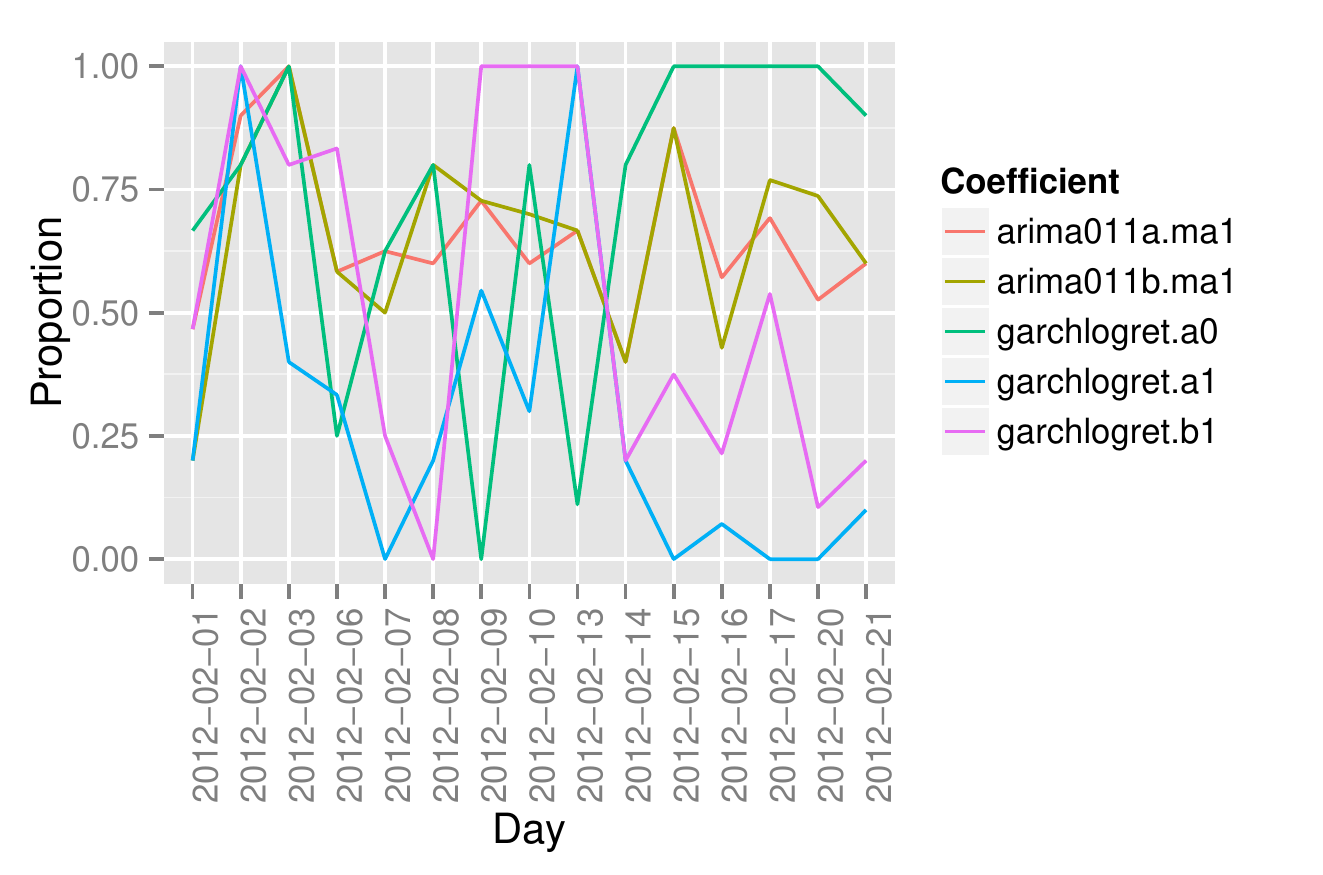}
\caption{The proportion of solutions on the Pareto front for which the coefficients of the auxiliary model fit to the real data lie within the 95\% confidence interval of the coefficients of the auxiliary model fit to the simulated data, for each trading day between 01/02/2012 and 21/02/2012 for 5 different stocks. (Left): BNP Paribas. (Right): Credit Agricole. }
\label{fig:simsprop1}
\end{center}
\end{figure}

\begin{figure}
\begin{center}
\includegraphics[width=0.49\textwidth]{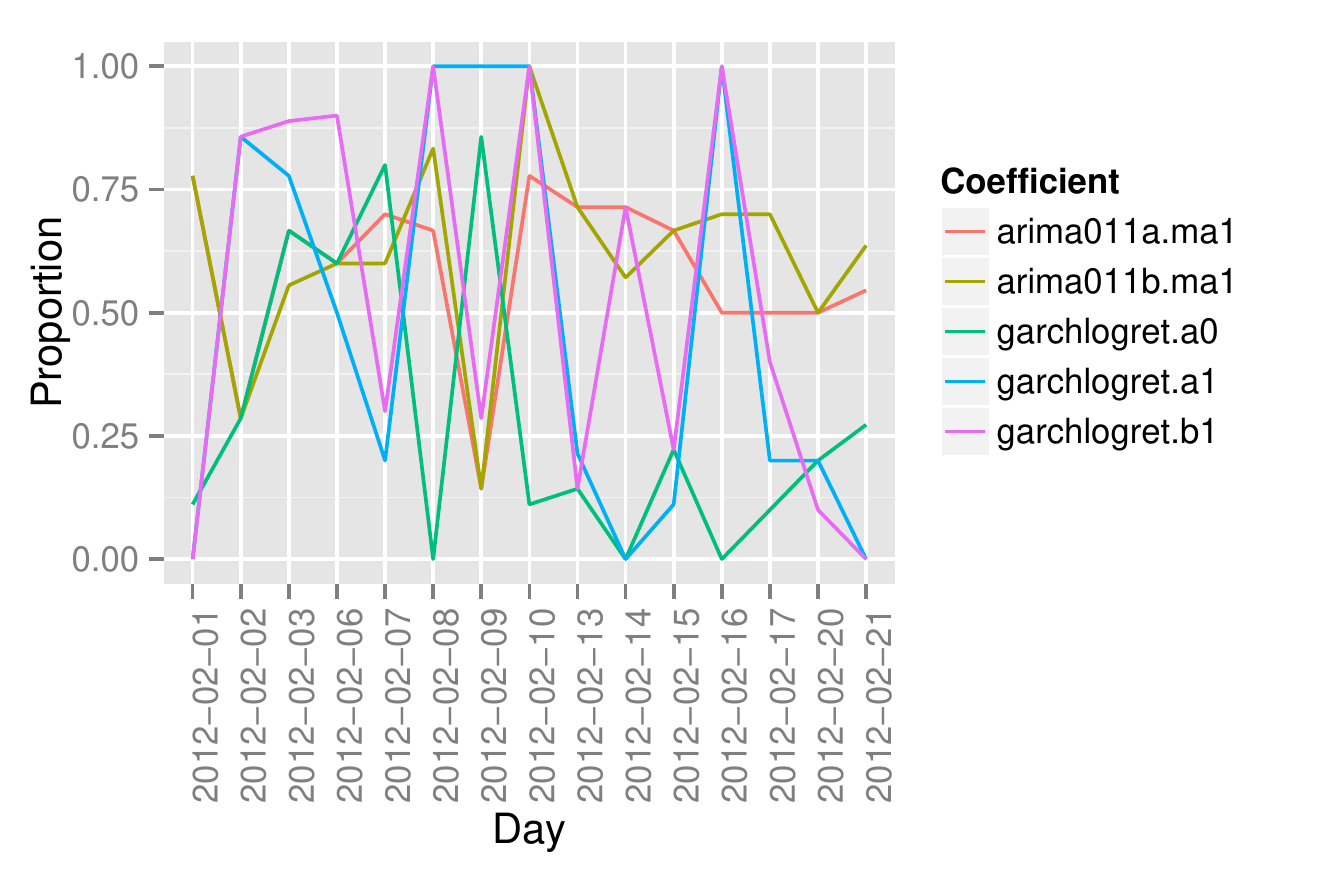}
\includegraphics[width=0.49\textwidth]{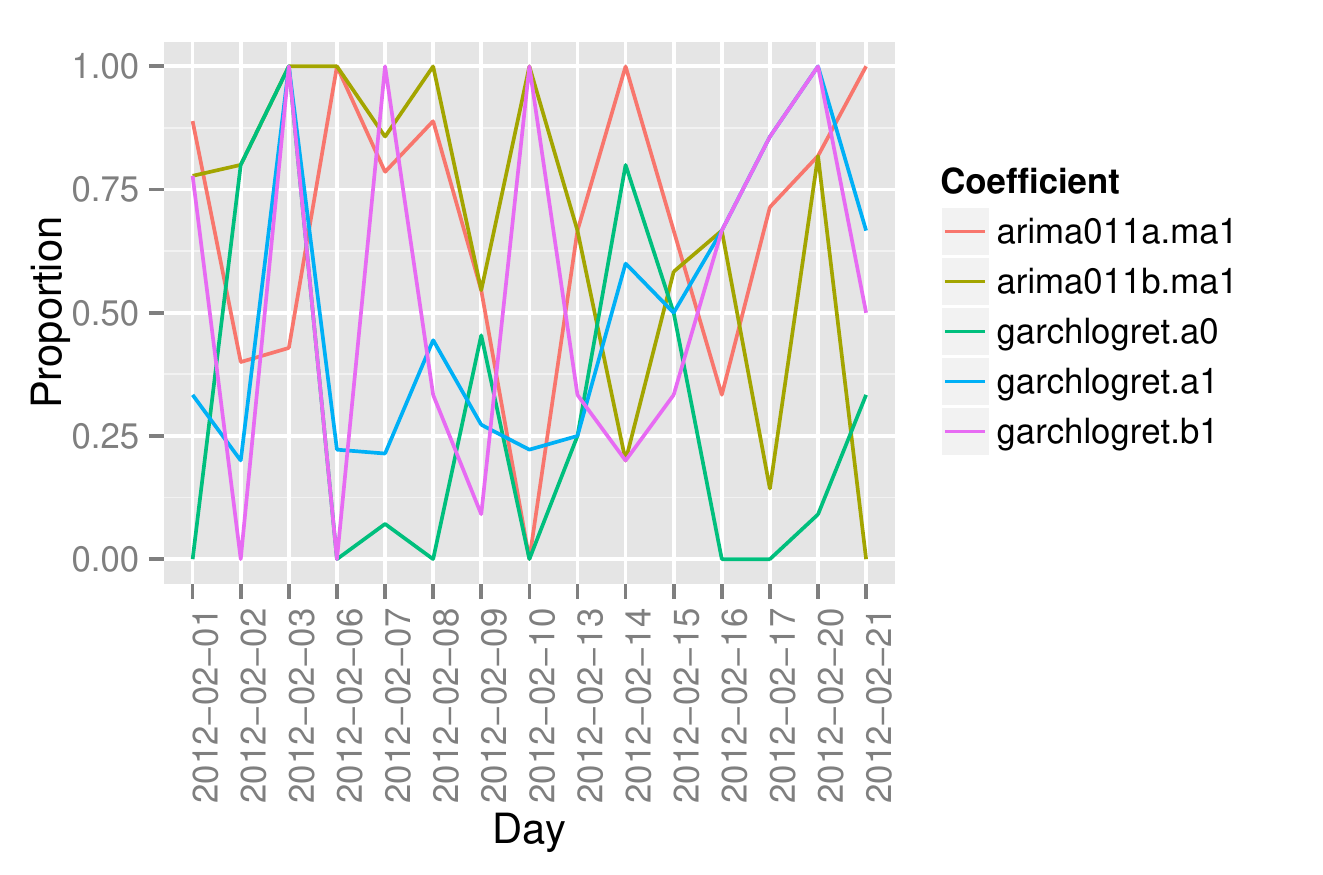}
\includegraphics[width=0.49\textwidth]{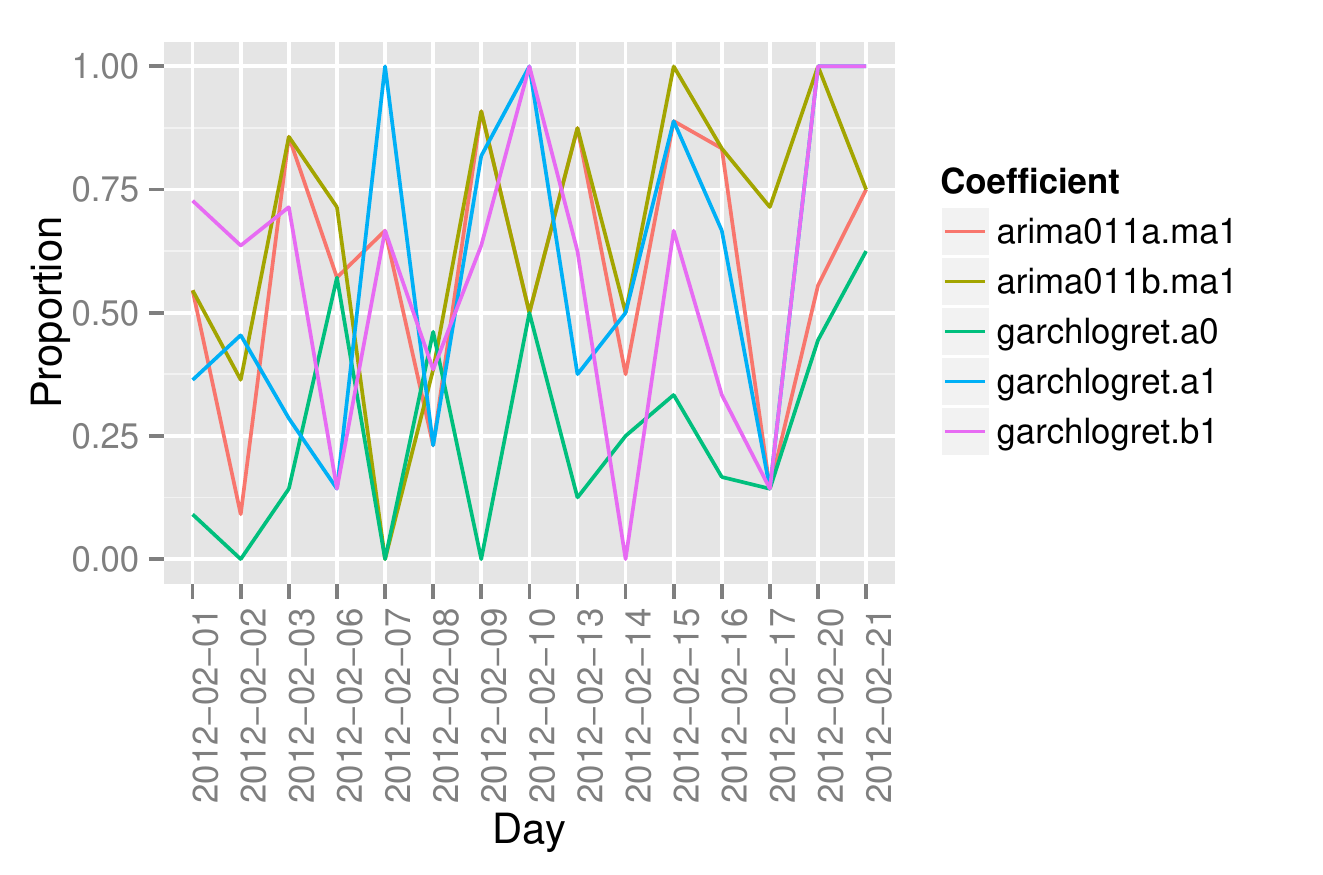}
\caption{The proportion of solutions on the Pareto front for which the coefficients of the auxiliary model fit to the real data lie within the 95\% confidence interval of the coefficients of the auxiliary model fit to the simulated data, for each trading day between 01/02/2012 and 21/02/2012 for 5 different stocks. (Left) Total SA. (Right) Technip SA. (Bottom): Sanofi.}
\label{fig:simsprop2}
\end{center}
\end{figure}

We can then construct the empirical distribution for each parameter in these vectors, and determine the 95\% confidence interval. From this, we can determine whether the parameter coefficients of the auxiliary model fit to the real data lie within this range, for each asset on the Pareto front. In Figures \ref{fig:simsprop1} and \ref{fig:simsprop2}, we show for each day, each asset and each auxiliary model parameter, the proportion of solutions on the Pareto front for which the coefficients of the auxiliary model fit to the real data lie within the 95\% confidence interval of the coefficients of the auxiliary model fit to the simulated data. 

We note that the proportion varies over time, as one would expect, as not all solutions on the Pareto front will give rise to LOB dynamics that closely reflect those observed in real data. However, we note that this proportion is generally more than 25\% for most parameters and most days. Thus, within the set of solutions produced by our estimation procedure, there is a subset which produce simulations which are similar to real trading observations in terms of their price and volume behaviour, which are the summaries of the LOB which our auxiliary models related to.

\end{document}